\documentclass{aa}

\renewcommand{\rm}{\mathrm}

\def\({\left(}
\def\r){\right)}

\newcommand*{\nt}{\textrm}
\newcommand*{\dif}{\textrm{d}}
\usepackage{physics}
\usepackage{ulem}
\usepackage{cancel}

\usepackage{bm}
\usepackage{verbatim}
\usepackage{subcaption}
\usepackage{graphicx}
\usepackage{txfonts}
\usepackage{subcaption}
\usepackage{acronym}
\usepackage[dvipsnames]{xcolor}
\usepackage[font=footnotesize,labelfont=bf]{caption}
\usepackage[colorlinks = true,
            linkcolor = blue,
            urlcolor  = blue,
            citecolor = blue,
            anchorcolor = blue]{hyperref}

\usepackage[noabbrev]{cleveref}
\usepackage{todonotes}

\begin{document} 

\title{A revised density split statistic model for general filters}

   \author{Pierre Burger\inst{1}
   		  \and Oliver Friedrich\inst{2,3}
   		  \and Joachim Harnois-D\'eraps\inst{4,5}
          \and Peter Schneider\inst{1} 
          }
   \institute{Argelander-Institut f\"ur Astronomie, Auf dem H\"ugel 71, 53121 Bonn, Germany 
   \and 
   Kavli Institute for Cosmology, University of Cambridge, CB3 0HA Cambridge, UK
   \and
   Churchill College, University of Cambridge, CB3 0DS Cambridge, UK
   \and  
   School of Mathematics, Statistics and Physics, Newcastle University, Newcastle upon Tyne, NE1 7RU, UK
   \and  
   Astrophysics Research Institute, Liverpool John Moores University, 146 Brownlow Hill, Liverpool L3 5RF, UK
\\ \email{pburger@astro.uni-bonn.de}
             }

   \date{Received 24 June 2021/ Accepted 02 March 2022}

  \abstract
  {Studying the statistical properties of the large-scale
structure in the Universe with weak gravitational lensing is a
prime goal of several current and forthcoming galaxy surveys. The power that weak lensing has to constrain cosmological parameters can
be enhanced by considering statistics beyond
second-order shear correlation functions or power spectra. One such higher-order probe that has proven successful in observational data is density split statistics (DSS), in which one analyses the mean shear profiles around points that are classified according to their foreground galaxy density.}
{In this paper, we generalise the most accurate DSS model to allow for a broad class of angular filter functions used for the classification of the different local density regions. This approach is motivated by earlier findings showing that an optimised filter can provide tighter constraints on model parameters compared to the standard top-hat case. }
{As in the previous DSS model we built on large deviation theory approaches and approximations thereof to model the matter density probability distribution function, and on perturbative calculations of higher-order moments of the density field. The novel addition relies on the generalisation of these previously employed calculations to allow for general filter functions and is validated on several sets of numerical simulations.}
{It is shown that the revised model fits the simulation measurements well for many filter choices, with a residual systematic offset that is small compared to the statistical accuracy of current weak lensing surveys. However, by use of a simple calibration method and a Markov chain Monte Carlo analysis, we studied the expected sensitivity of the DSS to cosmological parameters and find unbiased results and constraints comparable to the commonly used two-point cosmic shear measures. Hence, our DSS model can be used in competitive analyses of current cosmic shear data, while it may need refinements for forthcoming lensing surveys.}
   {}

   \keywords{gravitational lensing: weak --methods: statistical -- surveys -- Galaxy: abundances -- (cosmology:) large-scale structure of Universe
               }

   \maketitle

%
\section{Introduction}
Studying the matter distribution of the present large-scale structure reveals a wealth of information about the evolution of the Universe. In particular, its distorting effect on the propagation of light from distant galaxies, known as cosmic shear, can be captured by analysing weak lensing surveys. By comparing the results of cosmological models with the observed signal, one can constrain cosmological parameters \citep[see e.g.][]{Asgari2021,DES2021,Hamana:2020}. 

The preferred methods used to infer statistical properties of the matter and galaxy distribution concentrate on second-order statistics, such as the two-point correlation functions or their Fourier counterparts, the power spectra. Although these statistics have an impressive accuracy to describe for instance primordial perturbations visible in the cosmic microwave background \citep[CMB; e.g.][]{Aghanim:2020} they probe only the Gaussian information present in the density fluctuations. However, these initial conditions developed significant non-Gaussian features by means of non-linear gravitational instability, which can only be investigated with higher-order statistics. Although they are typically more time consuming to model and measure, these higher-order statistics scale differently with cosmological parameters, and are not affected in the same way by residual systematics. Hence, by jointly investigating second- and higher-order statistics, the constraining power on cosmological parameters increases \citep[see e.g.][]{Berge2010,Pyne2021,Pires:2012,Fu:2014,Kilbinger2005}.

A large number of analytical models for the two-point statistics exists in the literature \citep{Takahashi2012,Heitmann:Lawrence:2014,EuclidEmulator2020,Mead2020,Nishimichi2021}; however, the analysis of higher-order statistics is usually based on simulations. Analytical models for higher-order lensing statistics are rare, although they are important not only for 
scientists to understand physical processes, but also to cross-check simulations, which are usually only tested against Gaussian statistics. For example, \citet{Reimberg2018} and \citet{Barthelemy2020} used large deviation theory (LDT) to compute the reduced-shear correction to the aperture mass probability distribution function (PDF); \citet{Munshi2020} and \citet{Halder2021} analytically modelled the integrated shear three-point function; the lensing peak count function was modelled in \citet{2010ApJ...719.1408F,2015A&A...576A..24L} and \citet{2018MNRAS.474.1116S}, while the lensing PDF is modelled in \citet{boyle2020}.

The examples mentioned above all pertain to the analysis of cosmic shear data. However, it has been established in recent analyses that the addition of foreground  clustering data, and their cross-correlation with the background source galaxies, yield significantly better constraints \citep{DES:2018,Heymans:2021}. While the central analyses focused again on two-point statistics, \citet[][hereafter F18]{Friedrich:Gruen:2018} developed a competitive model based on density split statistics (hereafter DSS). The idea is to measure the mean tangential shear around small sub-areas of the survey, and to stack the signal according to the foreground galaxy density in these sub-areas. We expect the tangential shear to be larger around points with a high density of foreground galaxies, given they correspond to a large matter overdensity on average. The model derived in F18 is based on non-perturbative calculations of the matter density PDF, which predicts the shear profiles and the probability density of galaxy counts in the sub-areas, for a given  cosmological model, a redshift distribution for the source and lens galaxy populations, and a mean galaxy density. In \citet[][hereafter G18]{Gruen:Friedrich:2018}, 
the F18 model is used to constrain cosmological parameters from DSS measurements from the Dark Energy Survey (DES) First Year and Sloan Digital Sky Survey (SDSS) data, which yields constraints on the matter density $\Omega_\nt{m}= 0.26^{+0.04}_{-0.03}$ that agree and are competitive with the DES analysis of galaxy and shear two-point functions \citep[see][]{DES:2018}.

One of the motivations of this work is based on \citet[][hereafter B20]{Burger:2020}, who use a suite of numerical simulations to show that using matched filter functions for searching peaks and troughs in the galaxy and matter density contrast has clear advantages compared to the top-hat filter used in the F18 model, both in terms of the overall signal S/N and in recovering accurately the galaxy bias term. Another motivation of using compensated filters is that these filters are more confined in Fourier space and are therefore better at smoothing out large $\ell$-modes where baryonic effects play an important role \citep{Asgari2020}. Therefore, it is of interest to generalise the DSS to a broader set of filter functions.
Smoothing cosmic density fields with filters other than top-hat ones significantly complicates the LDT-like calculations used by F18 and G18 \citep[cf.][]{Barthelemy2020} because for top-hat filters the Lagrangian to Eulerian mapping inherent in LDT is particularly simple. However, we find here that density split statistics with non-top-hat filters that are sufficiently concentrated around their centres can still be accurately modelled with computationally feasible extensions of approximations made by F18. This paper describes our modifications to the F18 model that will allow us to optimise filtering strategies when applying density split statistics to Stage III weak lensing surveys such as KiDS. Throughout this paper, if not otherwise stated, we assume a spatially flat universe.

This work is structured as follows. In Sect.~\ref{Aperture_stat} we review the basics of the aperture statistics; we then detail our changes to the F18 model in Sect.~\ref{new_model}. In Sect.~\ref{Sect:Data} we describe the simulations, and the construction of our mock data  used to validate the revised model. In Sect.~\ref{sec_model_test} we compare the model predictions with simulations, and establish the model's limitations. We summarise our work in Sect.~\ref{sec:Conclusion}.

\section{Aperture statistics}
\label{Aperture_stat}
The lensing convergence $\kappa$ and shear $\gamma$ are related via the lensing potential $\psi$ \citep{Schneider1992} as
\begin{equation}
    \kappa(\boldsymbol{\theta}) = \frac{1}{2} (\partial_1^2+\partial_2^2)  \psi(\boldsymbol{\theta}) ~,~ \gamma(\boldsymbol{\theta}) = \frac{1}{2}(\partial_1^2-\partial_2^2+2\mathrm{i}\partial_1\partial_2)\psi(\boldsymbol{\theta})\, ,
\end{equation}
with $\partial_i = \frac{\partial}{\partial \theta_i}$ and $\boldsymbol{\theta}$ the angular position on the sky; we employ the flat-sky approximation. Given a reference point in a Cartesian coordinate system on the sky and a second point whose separation to the first is oriented at an angle $\phi$ with respect to that coordinate system, we can express the shear at the second point in terms of the tangential and cross-shear with respect to the first point as
\begin{equation}
   \gamma_{\rm t} = - \mathrm{Re}(\gamma\, \mathrm{e}^{-2\mathrm{i}\phi}) \quad , \quad \gamma_\times = - \mathrm{Im}(\gamma\, \mathrm{e}^{-2\mathrm{i}\phi})\, ,
\end{equation}
where the factor $2$ in the exponent is due to the polar nature of the shear. Given a convergence field $\kappa(\boldsymbol{\theta})$, the aperture mass at position $\boldsymbol{\theta}$ is defined as
\begin{equation}
    M _{\nt{ap}}\left(\boldsymbol{\theta}\right) \coloneqq \int \dd^2\theta' \,\kappa(\boldsymbol{\theta}+\boldsymbol{\theta}')\,U(|\boldsymbol{\theta'}|) \, ,	
    \label{Map}
\end{equation}
where $ U(\vartheta)$ is a compensated axisymmetric filter function, such that $ \int\vartheta\, U(\vartheta)~ \dif\vartheta=0$. As shown in
\citet{Schneider:1996}, if $U$ is compensated, $M_{ \nt{ap}}$ can also be expressed in terms of the tangential shear $\gamma_{\nt{t}}$ and a related filter function $Q$ as
\begin{equation}
     M_{ \nt{ap}}(\boldsymbol{\theta}) = \int\dd^2\theta'\,\gamma_{\nt{t}}(\boldsymbol{\theta}+\boldsymbol{\theta}')\,Q(|\boldsymbol{\theta}'|)\, ,
     \label{eq:MapQ}
\end{equation}
where
\begin{equation}
 Q(\vartheta) = \frac{2}{\vartheta^2} \int\limits_0^{\vartheta}\dd \vartheta'\,\vartheta'\,U(\vartheta') - U(\vartheta) \, ,
 \label{NewQ}
\end{equation}
which can be inverted, yielding
\begin{equation}
 U(\vartheta) = 2\int\limits_{\vartheta}^{\infty} \dd\vartheta'\, \frac{Q(\vartheta')}{\vartheta'} - Q(\vartheta) \, .
\label{NewU}
\end{equation}
 
In analogy to $M_{ \nt{ap}}$, we define, as done in B20, the aperture number counts 
\citep[][]{Schneider:1998}, or aperture number, as
\begin{equation}
    N_{ \nt{ap}}(\boldsymbol{\theta}) \coloneqq \int\dd^2\theta' \,n(\boldsymbol{\theta}+\boldsymbol{\theta}')\,U(|\boldsymbol{\theta}'|)\, ,
    \label{Nap}
\end{equation}
where $ U(\boldsymbol{\vartheta})$ is the same filter function as in
Eq.~\eqref{Map} and $ n(\boldsymbol{\vartheta})$ is the (foreground) galaxy number density on the sky. This definition of the aperture number is equivalent to the `Counts-in-Cell' (CiC) from \citet{Gruen:2015} if a top-hat filter of the form
\begin{equation}
U_\nt{th}(\vartheta) = \frac{1}{A}\mathcal{H}(\vartheta_\nt{th}-\vartheta) \, ,
\label{U_th_filter}
\end{equation}
is used, where $\mathcal{H}$ is the Heaviside step function and $A$ is the area of the filter. However, B20 demonstrated that top-hat filters are not optimal, and that a better performance is achieved by an adapted filter in terms of signal-to-noise-ratio (S/N) and in recovering accurately the galaxy bias term. In this paper we compute aperture mass statistics with Eq.~\eqref{eq:MapQ} using simulated weak lensing catalogues of background source galaxies, notably regarding positions and ellipticities, and aperture number statistics with Eq.~\eqref{Nap} from the position of simulated foreground lens galaxies (see Sect.~\ref{Sect:Data}).

\section{Revised model}
\label{new_model}
In this section we describe our modifications of the original F18 model. Although the derivations shown here are self-contained, we recommend the interested reader to consult the original F18 paper. In particular, it is shown there that the full non-perturbative calculation of the PDF within large deviation theory (LDT) can be well approximated with a log-normal PDF that matches variance and skewness of the LDT result. This allowed F18 and G18 to replace the expensive LDT computation with a faster one, hence making the calculation of full Markov chain Monte Carlo (MCMC) functions feasible. The reason why this approximation works well is that, for top-hat filters, the scaling of variance and higher-order cumulants in LDT is similar to that found in log-normal distributions. This cannot be expected a priori for other filter functions. However, through comparison with N-body simulations we find here (cf.\ Sect.~\ref{sec_model_test}) that either a simple log-normal or a combination of two log-normal distributions still accurately describes the density PDFs required to analyse density split statistics with more general classes of filters. The following section describes these calculations. In order to reduce the mathematical calculations in this section, some derivations are detailed in Appendix~\ref{sec:deatiled_der}. 

We start by defining the line-of-sight projection of the 3D matter density contrast $\delta_{\mathrm{m,3D}}$, weighted by a foreground (lens) galaxy redshift probability distribution $n_\mathrm{f}(z)$ as
\begin{equation}
\delta_{\mathrm{m,2D}}(\boldsymbol{\theta}) = \int \dd \chi \, q_\mathrm{f}(\chi)\, \delta_{\mathrm{m,3D}}(\chi\boldsymbol{\theta},\chi) \, ,
\label{eq:mattercontrast_2d}
\end{equation}
where $\chi$ is the co-moving distance and the projection kernel $q_\mathrm{f}(\chi)$ is
\begin{equation}
q_\mathrm{f}(\chi) = n_\mathrm{f}(z[\chi])\frac{\dd z[\chi]}{\dd \chi} \, .
\label{eq:projection_kernel}
\end{equation}

This 2D matter density contrast can then be used together with a linear bias term to represent a tracer density contrast (see Sect.~\ref{Sec:Characteriscfunction} or Sect.~\ref{Sect:Data}). Following F18, the next step consists of smoothing the results
with a filter $U$ of size $\Theta$:
\begin{align}
\delta_{\mathrm{m},U}(\boldsymbol{\theta}) &\equiv \hspace{-0.2cm}\int\limits_{|\boldsymbol{\theta}'|<\Theta}\hspace{-0.2cm} \dd^2 \theta' \,\delta_{\mathrm{m,2D}}(\boldsymbol{\theta}+\boldsymbol{\theta}') \, U(|\boldsymbol{\theta}'|) \, .
\end{align} 
This simplifies in the case of a top-hat filter of size $\Theta$ to
\begin{equation}
\delta_{\mathrm{m},\mathrm{th}}^{\Theta}(\boldsymbol{\theta}) = \frac{1}{A} \hspace{-0.1cm} \int\limits_{|\boldsymbol{\theta}'|<\Theta} \hspace{-0.1cm} \dd^2\theta' \;\delta_{\mathrm{m,2D}}(\boldsymbol{\theta}+\boldsymbol{\theta}') \ .
\label{eq:delta_th}
\end{equation}

Similar to the 2D density contrast, the convergence, which is needed to describe the DSS signal, is given by
\begin{equation}
\kappa(\boldsymbol{\theta}) = \int \dd \chi \, W_\mathrm{s}(\chi)\, \delta_{\mathrm{m,3D}}(\chi\boldsymbol{\theta},\chi) \, ,
\label{eq:kappa}
\end{equation}
where $W_\mathrm{s}(\chi)$ is the lensing efficiency defined as
\begin{equation}
    W_\mathrm{s}(\chi) = \frac{3\Omega_\mathrm{m}H_0^2}{2c^2}\int_\chi^\infty \dd \chi'\, \frac{\chi(\chi'-\chi)}{\chi'a(\chi)}\,q_\mathrm{s}(\chi') \;,
    \label{eq:lensing_efficiency}
\end{equation}
with $q_\mathrm{s}(\chi)=n_\mathrm{s}(z[\chi])\frac{\dd z[\chi]}{\dd \chi}$ being the line-of-sight probability density of the sources, $\Omega_\mathrm{m}$ the matter density parameter, $H_0$ the Hubble parameter, and $c$ the speed of light. The mean convergence inside an angular separation $\vartheta$, $\kappa_{<\vartheta}$, follows then in analogy to Eq.~\eqref{eq:delta_th} by substituting $\delta_{\mathrm{m,2D}}(\boldsymbol{\theta})$ with $\kappa(\boldsymbol{\theta})$.

The aim of our model is to predict the tangential shear profiles $\gamma_\textrm{t}$ given a quantile $\mathcal{Q}$ of the foreground aperture number $N_{\textrm{ap}}$, $\langle \gamma_\textrm{t} |  \mathcal{Q} \rangle$, where for instance the highest quantile is the set of lines of sight of the sky that have the highest values of $N_{\textrm{ap}}$. Therefore, to determine $\langle \gamma_\textrm{t} |  \mathcal{Q} \rangle$ the model calculates $\langle \gamma_\textrm{t} |  N_{\textrm{ap}} \rangle$ and sums up all that belong to the corresponding quantile $\mathcal{Q}$.
The expectation value of $\langle \gamma_\textrm{t}|  N_{\textrm{ap}} \rangle$ is computed from the convergence profile as
\begin{align}
\langle \gamma_\textrm{t}(\vartheta) |  N_{\textrm{ap}} \rangle &= \langle \kappa_{<\vartheta} |  N_{\textrm{ap}} \rangle - \langle \kappa_{\vartheta} |  N_{\textrm{ap}} \rangle = -\frac{\vartheta}{2}\frac{\dd}{\dd \vartheta}\langle \kappa_{<\vartheta} |  N_{\textrm{ap}} \rangle \, ,
\end{align}
where $\kappa_{\vartheta}$ is the azimuthally averaged convergence at angular separation  $\vartheta$ from the centre of the filter, and $\kappa_{<\vartheta} $ is the average convergence inside that radius. The latter quantity, conditioned on a given $N_{\textrm{ap}}$, can be  specified by
\begin{align}
\langle \kappa_{<\vartheta} |  N_{\textrm{ap}} \rangle &= \int \dd \delta_{\mathrm{m},U} \;\langle \kappa_{<\vartheta} | \delta_{\mathrm{m},U} ,  N_{\textrm{ap}}\rangle p(\delta_{\mathrm{m},U} |  N_{\textrm{ap}})\\
&\approx  \int \dd \delta_{\mathrm{m},U} \;\langle \kappa_{<\vartheta} | \delta_{\mathrm{m},U}  \rangle\, p(\delta_{\mathrm{m},U} |  N_{\textrm{ap}})\,,
\label{Eq:Kappa_given_Nap}
\end{align}
where in the second step we assumed that the expected convergence within $\vartheta$ only depends on the projected matter density contrast $\delta_{\mathrm{m},U}$ and not on the particular realisation of shot-noise in $N_{\textrm{ap}}$ found within that fixed matter density contrast\footnote{This assumption is not evident per se, since via mode coupling the large-scale profile of a given density perturbation may well be correlated to the shot-noise (i.e. small-scale fluctuations) of galaxy formation in the centre of that perturbation. F18 have found the approximation $\langle \kappa_{<\vartheta} | \delta_{\mathrm{m},U} ,  N_{\textrm{ap}} \rangle \approx \langle \kappa_{<\vartheta} | \delta_{\mathrm{m},U} \rangle$ to be accurate in the Buzzard N-body simulations \citep{DeRose2019}, but a more stringent investigation of this assumption is left for future work.}.

By use of Bayes' theorem, we can express the conditional PDF as
\begin{equation}
p(\delta_{\mathrm{m},U} |  N_{\textrm{ap}}) = \frac{ p(N_{\textrm{ap}} | \delta_{\mathrm{m},U}) p(\delta_{\mathrm{m},U})}{p(N_{\textrm{ap}})}\;,
\label{Bayes}
\end{equation}
where $p(N_{\textrm{ap}} | \delta_{\mathrm{m},U})$ is the probability of finding $N_{\textrm{ap}}$ given the smoothed density contrast $\delta_{\mathrm{m},U}$. The normalisation in the denominator of Eq.~(\ref{Bayes}) follows by integrating over the numerator,
\begin{equation}
p(N_{\textrm{ap}}) = \int {\rm d} \delta_{\mathrm{m},U} \;p(\delta_{\mathrm{m},U})\, p(N_{\textrm{ap}}|\delta_{\mathrm{m},U}) \, .
\label{eq:p_of_Nap}
\end{equation}

As seen in the derivation above, we are left with three ingredients in order to calculate the tangential shear profiles given a quantile $Q$ of the aperture number $\langle \gamma_\textrm{t}(\vartheta) |  N_{\textrm{ap}} \rangle$:
\begin{enumerate}[(I)]
\item the PDF of the matter density contrast smoothed with the filter function $U$ (used in Eqs.\,\ref{Bayes}, \ref{eq:p_of_Nap})
\begin{equation}
p(\delta_{\mathrm{m},U}) \; ;
\end{equation}
\item the expectation value of the convergence inside a radius $\vartheta$ given the smoothed density contrast (used in Eq.\,\ref{Eq:Kappa_given_Nap})
\begin{equation}
\langle \kappa_{<\vartheta}|\delta_{\mathrm{m},U}\rangle \; ;
\end{equation}
\item the distribution of $N_{\textrm{ap}}$ for the given filter function $U$ given the smoothed density contrast (used in Eqs.\,\ref{Bayes}, \ref{eq:p_of_Nap})
\begin{equation}
p(N_{\textrm{ap}}|\delta_{\mathrm{m},U}) \,.
\end{equation}
\end{enumerate}
Since all three ingredients are sensitive to the filter $U$, we need to adjust all of them coherently with respect to the top-hat case.

\subsection{$(\nt{I}): p(\delta_{{\rm m},U})$}
\label{subsec:I}

As shown by F18 the full LDT computation of the matter density PDF can accurately approximated by a shifted log-normal distribution with vanishing mean \citep{Hilbert:2011}, which is fully characterised by two parameters, $\sigma$ and $\delta_0$, as
\begin{equation}
p(\delta_{\mathrm{m},U}) = \frac{1}{\sqrt{2\pi}\sigma(\delta_{\mathrm{m},U}+\delta_0)} \exp(-\frac{[\ln(\delta_{\mathrm{m},U}/\delta_0+1)+\sigma^2/2]^2}{2\sigma^2})\ .
\label{eq:pdmu}
\end{equation}
The two free parameters can be determined by specifying the variance $\langle \delta_{\mathrm{m},U}^2 \rangle $ and skewness $\langle \delta_{\mathrm{m},U}^3 \rangle$ of the PDF as \citep{Hilbert:2011}
\begin{align}
\langle \delta_{\mathrm{m},U}^2 \rangle &= \delta_0^2 \left[ \exp(\sigma^2)-1\right] \;,
\label{eq:var_dmU}\\
\langle \delta_{\mathrm{m},U}^3\rangle &= \frac{3}{\delta_0} \langle \delta_{\mathrm{m},U}^2\rangle^2 + \frac{1}{\delta_0^3} \langle \delta_{\mathrm{m},U}^2\rangle^3 \; ;
\label{eq:skew_dmU}
\end{align}
we derive the expression of $\langle \delta_{\mathrm{m},U}^2 \rangle $ and $\langle \delta_{\mathrm{m},U}^3 \rangle$ in Appendix~\ref{sec:deatiled_der} (see Eq. \ref{eq:delta_moment_proj}).

As we show later, this approximation works well for non-negative filter functions like top-hat or Gaussian filters. However, the log-normal PDF approximation becomes less accurate for compensated filters that include negative weights. In these cases we instead divide $U$ into  $U_>(\vartheta)=U(\vartheta)\mathcal{H}(\vartheta_\textrm{ts}-\vartheta)$ and $U_<(\vartheta)=-U(\vartheta)\mathcal{H}(\vartheta-\vartheta_\textrm{ts})$, where $\vartheta_\textrm{ts}$ is the transition scale from positive to negative filter weights. As a consequence, we obtain two correlated log-normal random variables, $\delta_{\mathrm{m},U_>}$ and $\delta_{\mathrm{m},U_<}$, whose
joint distribution can be represented by a bi-variate log-normal distribution as
\begin{align}
\begin{split}
&p(\delta_{\mathrm{m},U_>},\delta_{\mathrm{m},U_<}) = \frac{1}{2\pi\sigma_>(\delta_{\mathrm{m},U_>}+\delta_{0,>})\,\sigma_<(\delta_{\mathrm{m},U_<}+\delta_{0,<})\sqrt{1-\rho^2}}  \\ 
& \hspace{1cm} \times \exp\left( -\frac{1}{2(1-\rho^2)} \left[  \Delta_> ^2 + \Delta_<^2  -2\rho\Delta_>\Delta_< \right] \right) \;,
\label{bivariate_lognormal}
\end{split}
\end{align}
where we defined 
\begin{equation}
    \Delta_> = \frac{\ln(\delta_{\mathrm{m},U_>}/\delta_{0,>}+1)+\sigma_>^2/2}{\sigma_>}
\end{equation}
and similarly for $\Delta_<$. The correlation coefficient $\rho$ is determined by
\begin{align}
    \rho & = \ln\left(\frac{\langle\delta_{\mathrm{m},U_>}\delta_{\mathrm{m},U_<}\rangle}{\delta_{0,>}\delta_{0,<}}+1\right)\frac{1}{\sigma_>\sigma_<} \, ,
\end{align}
and in order to calculate the difference of two independent random variables $\delta_{\mathrm{m},U}=\delta_{\mathrm{m},U_>} - \delta_{\mathrm{m},U_<}$ we can use the convolution theorem \citep{Arfken:2008} to get
\begin{equation}
    p(\delta_{\mathrm{m},U})=\int\limits_{-\infty}^{\infty} \dd \delta_{\mathrm{m},U_>}\,p(\delta_{\mathrm{m},U_>},\delta_{\mathrm{m},U_>}-\delta_{\mathrm{m},U})\;.
    \label{conv_dist}
\end{equation}

\subsection{$(\mathrm{II}): \langle \kappa_{<\vartheta}|\delta_{\mathrm{m},U}\rangle$}
\label{subsec:II}
In order to calculate the expectation value of the mean convergence inside an angular radius $\vartheta$, $\kappa_{<\vartheta}$, given the matter density contrast $\delta_{\mathrm{m},U}$, we assume that both follow a joint log-normal distribution (see e.g. the discussion in Appendix B of G18). In this case, the expectation value can be written as
\begin{equation}
\frac{\langle \kappa_{<\vartheta}|\delta_{\mathrm{m},U}\rangle}{\kappa_0} = \exp\left( \frac{C\,[2\ln(\delta_{\mathrm{m},U}/\delta_0+1)+V-C]}{2V}\right)-1 \;,
\label{eq:kappa_delta}
\end{equation}
where $\delta_0$ is determined with Eq.~\eqref{eq:skew_dmU} and the three variables $C$, $V$, and $\kappa_0$ can be calculated from the moments $\langle \delta_{\mathrm{m},U}^2 \rangle$, $\langle \kappa_{<\vartheta}\, \delta_{\mathrm{m},U} \rangle$, and $\langle \kappa_{<\vartheta}\, \delta_{\mathrm{m},U}^2 \rangle$, which follow from the derivation in Appendix~\ref{sec:deatiled_der} (see Eq.~\ref{eq:kappa_delta_moment_proj}):
\begin{align}
V &= \ln(1+\frac{\langle \delta_{\mathrm{m},U}^2 \rangle}{\delta_0^2})\;, \label{eq:V}\\
C &= \ln(1+\frac{\langle \kappa_{<\vartheta} \,\delta_{\mathrm{m},U} \rangle}{\delta_0\kappa_0})\;,\\
\kappa_0 &= \frac{\langle \kappa_{<\vartheta}\, \delta_{\mathrm{m},U} \rangle^2\mathrm{e}^V}{\langle \kappa_{<\vartheta}\, \delta_{\mathrm{m},U}^2 \rangle - 2\langle \kappa_{<\vartheta}\, \delta_{\mathrm{m},U} \rangle \langle \delta_{\mathrm{m},U}^2 \rangle /\delta_0 } \label{eq:kappa_0} \; .
\end{align}

We note that the assumption that $\delta_{\mathrm{m},U}$ is log-normal distributed is not well justified for filters with negative weights as we mentioned in the previous section. A possible improvement could be done for instance by assuming again that $\delta_{\mathrm{m},U}$ is made up of two log-normal random variables, and we would need to calculate conditional moments like $\langle \kappa_{<\vartheta}|\delta_{\mathrm{m},U_>}-\delta_{\mathrm{m},U_<}\rangle$. This would significantly increase the amount of joint moments needed in our calculation and would render fast modelling unfeasible. However, an improved modelling is also unnecessary at present, given the statistical uncertainties we expect for Stage III weak lensing surveys such as KiDS-1000. We demonstrate this empirically in Sect.~\ref{sec_model_test} by comparison to N-body simulated data, but we also want to give a brief theoretical motivation. The average value of $\kappa_{<\vartheta}$, given that $\delta_{\mathrm{m},U}$ lies within the range $[\delta_{\min}, \delta_{\max}]$, is given by
\begin{equation}
    \langle \kappa_{<\vartheta} | \delta_{\mathrm{m},U} \in [\delta_{\min}, \delta_{\max}]\rangle = 
    {\int_{\delta_{\min}}^{\delta_{\max}} \dd\delta_{\mathrm{m},U}\ p(\delta_{\mathrm{m},U}) \langle \kappa_{<\vartheta}|\delta_{\mathrm{m},U}\rangle\ \over \int_{\delta_{\min}}^{\delta_{\max}} \dd\delta_{\mathrm{m},U}\ p(\delta_{\mathrm{m},U})}\, .
\end{equation}
If $\kappa_{<\vartheta}$ and $\delta_{\mathrm{m},U}$ were joint Gaussian random variables, then $p(\delta_{\mathrm{m},U})$ would be a Gaussian PDF and we would have $\langle \kappa_{<\vartheta}|\delta_{\mathrm{m},U}\rangle = \delta_{\mathrm{m},U}\ \langle \delta_{\mathrm{m},U} \kappa_{<\vartheta}\rangle / \langle \delta_{\mathrm{m},U}^2 \rangle$. We now argue that the leading-order correction to this Gaussian approximation consists of replacing $p(\delta_{\mathrm{m},U})$ by our full non-Gaussian model, without changing $\langle \kappa_{<\vartheta}|\delta_{\mathrm{m},U}\rangle$, since this would be exactly correct in the limit of strong correlation between the two variables. Our log-normal approximation to $\langle \kappa_{<\vartheta}|\delta_{\mathrm{m},U}\rangle$ is then already a next-to-leading-order correction and a bi-variate log-normal approximation for $\langle \kappa_{<\vartheta}|\delta_{\mathrm{m},U}\rangle$ would be of even higher order. While this reasoning is admittedly only heuristic, it is proven correct by the accuracy of our model predictions for the lensing signals in Sect.~\ref{sec_model_test}.

\subsection{$(\mathrm{III}): p(N_{\mathrm{ap}}|\delta_{\mathrm{m},U})$}
\label{Sec:Characteriscfunction}

The third basic ingredient is the PDF of $N_{\textrm{ap}}$ given the projected matter density contrast smoothed with the filter $U$. Assuming a Poisson distribution for $N_{\textrm{ap}}$, which is the most straightforward ansatz, is unfortunately not possible because negative values are expected with a compensated filter (i.e. in some of the $U_{<}$ contributions). We use instead a completely new approach compared to F18, and derive an expression for $p(N_{\textrm{ap}}|\delta_{\mathrm{m},U})$ by use of the characteristic function \citep[][hereafter CF]{Papoulis1991}, which is an alternative representation of a probability distribution, similar to the moment generating functions, but based on the Fourier transform of the PDF. Of interest to us, the $n$-th derivative of the CFs can be used to calculate the $n$-th moment of the PDF. The CF corresponding to $p(N_{\textrm{ap}}|\delta_{\mathrm{m},U})$ is defined as 
\begin{equation}
\Psi(t) = \langle\mathrm{e}^{i \textrm{t}N_\textrm{ap}} \rangle_{\delta_{\mathrm{m},U}} = \int_\mathbb{R} \dd N_\textrm{ap} p(N_{\textrm{ap}}|\delta_{\mathrm{m},U})\mathrm{e}^{\mathrm{i} tN_\textrm{ap}} \, ,
\label{eq:CF_defintion}
\end{equation}
where in our particular case, we derive in Appendix \ref{sec:CF} a closed expression as
\begin{equation}
\Psi(t)=\exp\left(2\pi n_0 \int_0^\infty d \vartheta\;\vartheta\; \left(1+b\, \langle w_\vartheta|\delta_{\mathrm{m},U}\rangle\right)  \left[\mathrm{e}^{\mathrm{i} t U(\vartheta)} -1\right]\right) \label{eq:CF}
\end{equation}
with $n_0$ being the mean number density of foreground galaxies on the sky. The assumption of linear galaxy bias enters here by the term $b\, \langle w_\vartheta|\delta_{\mathrm{m},U}\rangle$, with 
\begin{equation}
    w_\vartheta = \frac{1}{2\pi} \int_0^{2\pi} \dd \phi \,\delta_{\rm m, 2D}(\vartheta,\phi) \, .
    \label{eq:w_theta}
\end{equation}
Hence, $n_0(1+b\, \langle w_\vartheta|\delta_{\mathrm{m},U}\rangle)$ is the effective number density at $\vartheta$ given $\delta_{\mathrm{m},U}$. The conditional expectation value $\langle w_\vartheta|\delta_{\mathrm{m},U}\rangle$ is given in analogy to Eq.~\eqref{eq:kappa_delta}, but replacing $\langle \kappa_{<\vartheta}\, \delta_{\mathrm{m},U}^k \rangle \rightarrow \langle w_{<\vartheta}\, \delta_{\mathrm{m},U}^k \rangle$ in Eqs.~(\ref{eq:V}--\ref{eq:kappa_0}) for $k=1,2$ and using that
\begin{equation}
    \langle w_\vartheta|\delta_{\mathrm{m},U}\rangle = \langle w_{<\vartheta}|\delta_{\mathrm{m},U}\rangle+ \frac{\vartheta}{2}\frac{\dd}{\dd \vartheta} \langle w_{<\vartheta}|\delta_{\mathrm{m},U}\rangle \, ,
    \label{eq:w_delta}
\end{equation}
where the joint moments $\langle w_{<\vartheta}\, \delta_{\mathrm{m},U}^k \rangle$ are also derived in Appendix~\ref{sec:deatiled_der} (see Eq.~\ref{eq:w_delta_moment_proj}).
Next, we re-express Eq.~\eqref{eq:CF} as the product of two terms,
\begin{equation}
\Psi(t) = \exp[p(t)]\,\exp[\mathrm{i}q(t)] \;, 
\end{equation}
where
\begin{align}
    p(t)&=2\pi n_0 \int_0^{R_{\rm max}} \dd \vartheta\;\vartheta\; \left(1+b\, \langle w_\vartheta|\delta_{\mathrm{m},U}\rangle\right) \left(\cos[t U(\vartheta)]-1\right) \,,\\
    q(t)&=2\pi n_0 \int_0^{R_{\rm max}} \dd \vartheta\;\vartheta\; \left(1+b\, \langle w_\vartheta|\delta_{\mathrm{m},U}\rangle\right) \sin[t U(\vartheta)]\,,
\end{align}
and $R_\mathrm{max}$ is the angular radius beyond which $U$ vanishes. We note that G18 and F18 found super-Poisson shot-noise in their work. They interpret these deviations from Poisson noise as having a number $\neq 1$ of galaxies per Poisson halo. This would suggest that we could incorporate non-Poissonian behaviour by replacing $n_0$ with an effective density of Poisson halos and making this a free parameter of our model. However, more recent investigations (e.g.\ Friedrich et al. in prep.) cast doubt on the simplified interpretation of F18 and G18. A proper investigation of the problem of non-Poissonian shot-noise is beyond the scope of this work, and we will address it in future investigations.

Finally, the probability density function $p(N_{\textrm{ap}}|\delta_{\mathrm{m},U})$ follows from the inverse Fourier transform of the CF
\begin{align}
p(N_{\textrm{ap}}|\delta_{\mathrm{m},U}) &= \frac{1}{2\pi} \int_\mathbb{R} \mathrm{d} t \exp(-\mathrm{i}t N_{\textrm{ap}}) \Psi(t) \nonumber \\
&=\frac{1}{2\pi} \int_\mathbb{R} \mathrm{d} t \cos( q(t)-t N_{\textrm{ap}}) \exp[p(t)] \,,
\label{eq:True_PDF_charac}
\end{align}
where the second step follows from the fact that the imaginary part cancels out.

In Appendix\,\ref{sec:CF} we discuss a similar approach, where we assume that $p(N_{\textrm{ap}}|\delta_{\mathrm{m},U})$ is log-normal distributed. In that case, to specify the PDF, only the first three moments are needed, which follow from derivatives of the CF. As shown in Appendix\,\ref{sec:CF} both methods yield almost identical results, and since the log-normal approach is significantly faster, we use it hereafter, unless otherwise stated.

To summarise, the major changes compared to the F18 model are the following:
\begin{enumerate}
    \item To determine $p(\delta_{\mathrm{m},U})$ we
    \begin{itemize}
        \item updated the calculation of the variance $\langle \delta_{\mathrm{m},U}^2 \rangle$
        and of the skewness $\langle \delta_{\mathrm{m},U}^3
        \rangle$ in Appendix \ref{sec:deatiled_der} to general filter functions;
        \item combine in Eqs.~(\ref{bivariate_lognormal}--\ref{conv_dist}) two log-normal random variables for the positive and negative parts for compensated filters to obtain the final expression for any filter shape.
    \end{itemize}  
    \item To determine $p(N_{\textrm{ap}}|\delta_{\mathrm{m},U})$ we
    \begin{itemize}
        \item calculate the characteristic function of galaxy shot-noise around a given matter density profile via Eq.~\eqref{eq:CF_defintion};
        \item use log-normal approximation or inverse Fourier transform Eq.~\eqref{eq:True_PDF_charac} to obtain the PDF of shot-noise from its characteristic function.
    \end{itemize}
    \item To determine $\langle\kappa_{<\vartheta}|\delta_{\mathrm{m},U}\rangle$ we
    \begin{itemize}
        \item  updated the calculations of $\langle \kappa_{<\vartheta}\, \delta_{\mathrm{m},U} \rangle$ 
        and $\langle \kappa_{<\vartheta}\, \delta_{\mathrm{m},U}^2 \rangle$ to general filter functions (see Appendix \ref{sec:deatiled_der}).
    \end{itemize}
\end{enumerate}

\section{Simulation data}
\label{Sect:Data}

Before using our revised model in data analyses, it is mandatory to quantify its  precision and range of validity. We use for this validation exercise three simulations suites: 
\begin{itemize}
    \item the full-sky gravitational lensing simulations described in \citet[][hereafter T17]{Takahashi2017}, with which we carry out a detailed investigation of the model in a simple survey configuration;
    \item the cosmo-SLICS simulations, described in \citet{Harnois-Deraps:2019}, with which we validate our model on a independent simulation suite;
    \item the SLICS simulations, described in \citet{Harnois-Deraps:2018}, with which we construct a KiDS-1000 like covariance matrix.
\end{itemize}

\subsection{T17 simulations}
\label{sec:T17_description}

The T17 simulations are constructed from a series of nested cubic boxes with side lengths of $L,2L,3L...$ placed around a fixed vertex representing the observer’s position, with $L=450\,\mathrm{Mpc}/h$. Each box is replicated eight times and placed around the observer using periodic boundary conditions. The number of particles per box is fixed to $2048^3$, which results in higher mass and spatial resolutions at lower redshifts. Within each box three spherical lens shells are constructed, each with a width of $150\,\mathrm{Mpc}/h$, which are then used by the public code \textsc{GRayTrix}\footnote{\url{http://th.nao.ac.jp/MEMBER/hamanatk/GRayTrix/}} to trace the light-ray trajectories from the observer to the last scattering surface\footnote{These maps are freely available for download at \url{http://cosmo.phys.hirosaki-u.ac.jp/takahasi/allsky_raytracing/}}. With the $N$-body code \textsc{gadget2} \citep{Springel2001} the gravitational evolution of dark matter particles without baryonic processes are followed from the initial conditions, which in turn are determined by use of second-order Lagrangian perturbation theory. The initial linear power spectrum followed from the Code for Anisotropies in the Microwave Background \citep[CAMB;][]{Lewis2000} with $\Omega_{\rm m}=1-\Omega_\Lambda=0.279$, $\Omega_{\rm b}=0.046$, $h=0.7$, $\sigma_8=0.82$, and $n_{\rm s}=0.97$. The matter power spectrum agrees with theoretical predictions of the revised Halofit \citep{Takahashi2012} within $5\%(10\%)$ for $k<5 (6)\,h\,\mathrm{Mpc}^{-1}$ at $z<1$. In order to account for the finite shell thickness and angular resolution, T17 provide correction formulae, which we repeat in Appendix~\ref{Sect:power_correction}. Although various resolution options are available, for our purpose the realisations with a resolution of $\textsc{nside}=4096$ are sufficient. 

\begin{figure}[!htbp]
\centering
\includegraphics[width=\columnwidth]{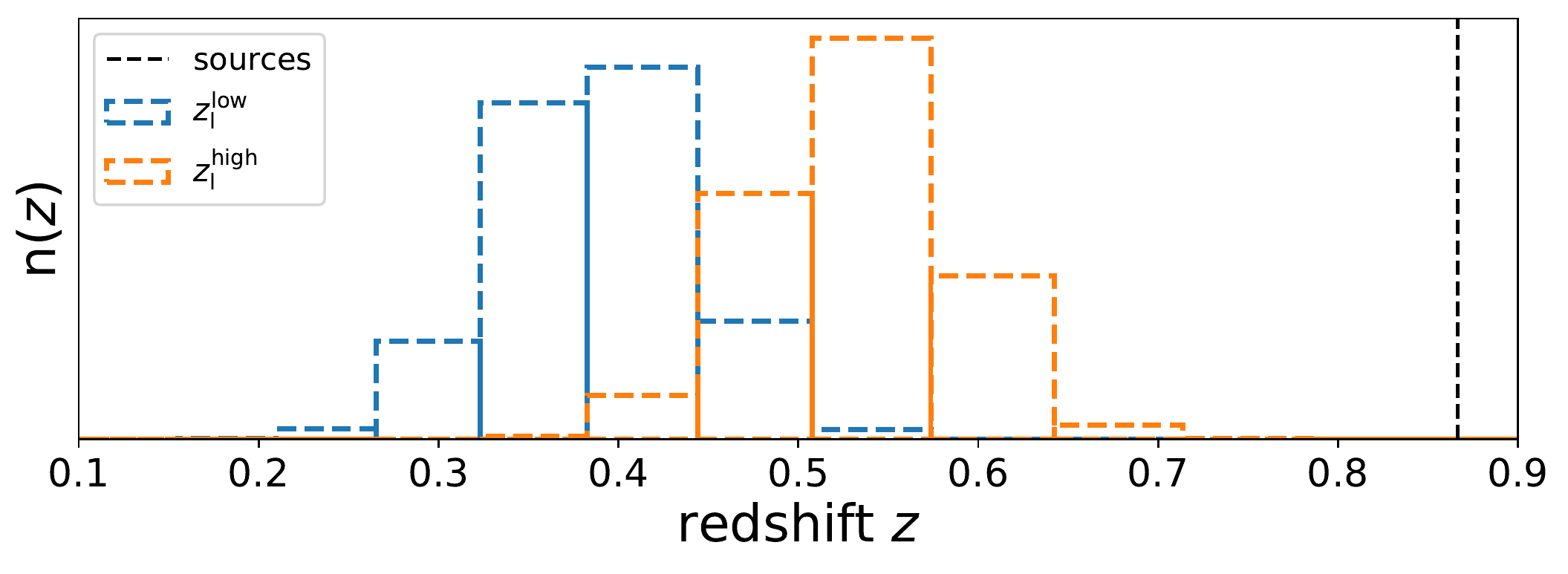}
    \caption{Lens galaxy redshift distribution constructed from the T17 simulation given the true $n(z)$ of the second $z_{\rm l}^\mathrm{low}$ and third $z_{\rm l}^\mathrm{high}$ LRG bin \citep{MJ:2019}. The black dashed line shows the redshift of the source galaxies.}
    \label{fig:nofz}
\end{figure}

We use the publicly available matter density contrast maps to create a realistic lens galaxy catalogue that  mimics the second and third redshift bins of the luminous red galaxies sample constructed from the KiDS-1000 data \citep{MJ:2019}, as shown by the solid lines in Fig.\,\ref{fig:nofz}. The reason to mock the LRG sample is that the galaxy bias for this kind of galaxies can be roughly described with a constant linear bias, which is needed for the analytical model. We excluded the lowest-redshift lens bin, first because of its low galaxy number density ($n_0=0.012$\,gal/arcmin$^2$) in which the shot-noise level is significant, and second because the density field is more non-linear, and hence we expect the log-normal approximation to break down. Since there is a significant overlap between the KiDS-1000 sources and the lenses in the fourth LRG redshift bin, we reject it as well. To create our lens galaxy samples we first project the T17 3D density maps $\delta_{\mathrm{m,3D}}$ following the $n(z)$ shown as the step functions in Fig.\,\ref{fig:nofz} to get two $\delta_{\mathrm{m,2D}}$ maps. For both maps we then distribute galaxies following a Poisson distribution with parameter $\lambda=n(1+b\,\delta_{\mathrm{m,2D}})$, where $b$ is a constant linear galaxy bias and $n$ is chosen such that the galaxy number density is $n_0=0.028$\,gal/arcmin$^2$ for the second bin (hereafter the low-redshift bin $z_{\rm l}^\mathrm{low}$) and $n_0=0.046$\,gal/arcmin$^2$ for the third lens bin (hereafter the high-redshift bin \smash{$z_{\rm l}^\mathrm{high}$}). Since our method requires a constant linear galaxy bias, we specify a bias of $1.72$ for lens bin two and $1.74$ for lens bin three, similar to those reported in \citet{MJ:2019}. F18 found this linear bias assumption to be accurate enough for year 1 data of the Dark Energy Survey, which is similar in constraining power to our target KiDS data (but we note that an investigation of higher-order biasing is underway in Friedrich et al., in prep.).

In our validation test, we use a shear grid at a single source plane located at $z=0.8664$, indicated by the black dashed line in Fig.\,\ref{fig:nofz}. F18 showed that the model works for realistic redshift distributions, and this choice simplifies the generation of our source catalogues. Furthermore, in order to determine a realistic covariance matrix, we transform the shear field into an observed ellipticity field by adding shape noise to the shear grid as  
\begin{equation}
\epsilon^{\nt{obs}} = \frac{\epsilon^{\nt{s}}+g}{1+\epsilon^{\nt{s}}g^{*}} \, ,
\label{eq:ebos}
\end{equation}
where $\epsilon^{\nt{obs}}$,
$\epsilon^{\nt{s}}$, and $\gamma$ are complex numbers, and the
asterisk $(*)$ indicates complex conjugation. The source ellipticities $\epsilon^{\nt{s}}$ per pixel are generated by drawing random numbers from a Gaussian distribution with width 
\begin{equation}
    \sigma_\mathrm{pix} = \frac{\sigma_\epsilon}{\sqrt{n_\mathrm{gal}A_\mathrm{pix}}} \approx 0.29\,,
\end{equation}
where $A_\mathrm{pix}$ is the pixel area of the shear grid, and the effective number density $n_\mathrm{gal}$ and $\sigma_\epsilon$ are chosen such that they are consistent with the KiDS data. While this transformation is valid in terms of the reduced shear $g=\gamma/(1-\kappa)$, we use throughout this paper the approximation $\gamma\approx g$, as the typical values for the convergence are small, $|\kappa|\ll 1$. 
We neglect the intrinsic alignment of galaxies in this work.

\subsection{Extracting the model components from the T17 simulations}
In order to validate the different components of our model, we need to extract $p(\delta_{\mathrm{m},U})$, $p(N_{\textrm{ap}})$, and $\langle \gamma_\textrm{t} |  \mathcal{Q} \rangle$  from the simulation. The first two follow directly by smoothing the maps of the projected density contrast and the lens galaxy with the corresponding filters. This smoothing can be performed in two different ways. The first is to use the \textsc{healpy} function \textsc{query\_disc}, which finds all pixel centres that are located within a given radius, whereas the second approach uses the \textsc{healpy} function \textsc{smoothing}, with a given beam window function created by the function \textsc{beam$2$bl}. The two approaches result in PDFs that differ slightly, since the \textsc{query\_disc} does not reproduce an exact top-hat, while the \textsc{smoothing} approach is only over a finite $\ell$-range. Nevertheless, we found that both approaches are consistent for $\textsc{nside}=4096$ well within the uncertainty we estimate from 48 sub-patches (see discussion below), and hence we use the second approach which is significantly faster.

The tangential shear information $\langle \gamma_\textrm{t} |  \mathcal{Q} \rangle$ is measured for each quantile $Q$ by the software \textsc{treecorr} \citep{Jarvis:Bernstein:2004} in 15 log-spaced bins with angular separation $\Theta/20 < \vartheta < \Theta$, where $\Theta$ is the size of the filter being used. For the top-hat filter we measured the shear profiles from $6' < \vartheta < 120'$, corresponding to a filter with a size of $120'$. We note here that for all measured shear profiles the shear around random points is always subtracted, which ensures that the shear averaged over all quantiles for one realisation vanishes by definition. 

In order to have an uncertainty for all three model quantities, we divide the full-sky map into 48 sub-patches, such that each patch has a size of approximately $859.4$\,deg$^2$. For $p(\delta_{\mathrm{m},U})$ and $p(N_{\textrm{ap}})$ we determined for each sub-patch one distribution, such that we were able to calculate a standard deviation from 48 values for each bin in the PDF. For the covariance matrix we use 10 out of the 108 realisations and divide each full-sky map in 48 sub-patches, which then results in a covariance matrix measured from 480 fields. Furthermore, both for the covariance and for the error bars in the plotted shear profiles we use Eq.~\eqref{eq:ebos} to create noisy shear profiles for each sub-patch, which are then re-scaled to the effective KiDS-1000 area \citep[see][]{Giblin:2020}.

\subsection{Cosmo-SLICS}
\label{Cosmo-SLICS}

We use the cosmo-SLICS simulations described in \citet{Harnois-Deraps:2019} to determine the validity regime of our revised model for different cosmologies. These are a suite of weak lensing simulations sampling 26 points (listed in Table~\ref{cos_overview}) in a broad cold dark matter (CDM) parameter space, distributed in a Latin hypercube to minimise interpolation errors. Specifically, the matter density $\Omega_{\nt{m}}$, the dimensionless Hubble parameter $h$, the normalisation of the matter power spectrum $\sigma_8$, and the time-independent equation-of-state parameter of dark energy $w_0$ are varied over a range that is large enough to complement the analysis of current weak lensing data \citep[see e.g.][]{Harnois-Deraps2020}. Each simulation follows 1536$^3$ particles inside a  cube of co-moving side length $L_{\nt{box}}=505\,h^{-1}\nt{Mpc}$ and $n_\nt{c}=3072$ grid cells on the side, starting with initial conditions produced with the Zel'dovich approximation. Moreover, the cosmo-SLICS evolve a pair of simulations at each node, designed to suppress the sampling variance \citep[see][for more details]{Harnois-Deraps:2019}. Each cosmological model is ray-traced multiple times to produce 50 pseudo-independent light cones of size 100\,deg$^2$. 

For each realisation, we create KiDS-1000-like sources and KiDS-LRG-like lens catalogues, following the pipeline described in \citet{Harnois-Deraps:2018};  notably, we reproduce exactly the source galaxy number density and $n(z)$ that is used in \citet{Asgari2021},  who report a total number density $n_{\nt{gal}}=6.93/$arcmin$^2$ and a redshift distribution estimated from self-organising maps \citep[see][]{Wright:2020}. These mock galaxies are then placed at random angular coordinates on 100\,deg$^2$ light cones. In contrast to the T17 mocks, we test our model with two source redshift bins, corresponding to the KiDS-1000 fourth and fifth tomographic bins (hereafter $z_{\rm s}^\mathrm{low}$ and $z_{\rm s}^\mathrm{high}$). The source galaxies are assigned a shear signal $\gamma$ from a series of lensing maps, following the linear interpolation algorithm described in Sect.~2 in \citet{Harnois-Deraps:2018}. For our lens sample we opted to include the second and third tomographic bin of the LRG galaxies described in \cite{MJ:2019} ($z_{\rm l}^\mathrm{low}$ and $z_{\rm l}^\mathrm{high}$). Compared to the T17 values, the $n(z)$ of the cosmo-SLICS LRG mocks have a coarser redshift resolution of the simulations. Moreover, the $n(z)$ vary slightly for different underlying cosmologies, due to variations in the relation between co-moving distance and redshift. Following \citet{MJ:2019}, we generate our LRG catalogues assuming a constant linear galaxy bias of 1.72 and 1.74, with a galaxy number density of $n_0=0.028$\,gal/arcmin$^2$ and $n_0=0.046$\,gal/arcmin$^2$.

\begin{figure}[!htbp]
\centering
\includegraphics[width=\columnwidth]{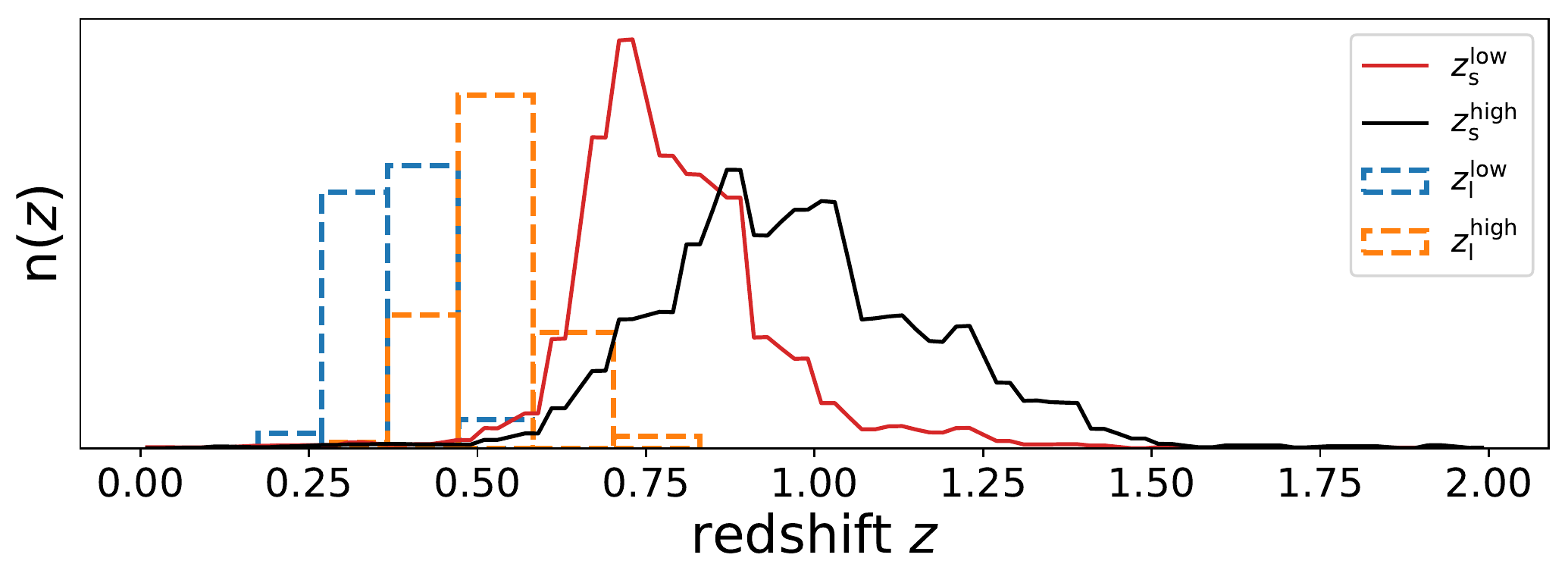}
    \caption{Redshift distributions of the second and third LRG (lens) bins and the last two KiDS-1000 (source) bins of the SLICS simulations. The $n(z)$ are scaled such that a comparison is possible.}
    \label{fig:n_of_z}
\end{figure}

\subsection{SLICS}
In total the SLICS\footnote{The SLICS are made  publicly available on the SLICS portal at \url{https://slics.roe.ac.uk/}.} are a set of over 800 fully independent realisations  similar to the fiducial $\Lambda$CDM cosmo-SLICS model. The underlying cosmological parameters for each run are the same, fixed to $\Omega_{\nt{m}}=0.2905$, $\Omega_{\Lambda}=0.7095$,
$\Omega_{\nt{b}}=0.0473$, $h=0.6898$, $\sigma_8=0.826$ and $n_{\rm s}=0.969$ \citep[see][]{Hinshaw:2012}. For Fourier modes $k<2.0\,h\,\nt{Mpc}^{-1}$, the SLICS and cosmo-SLICS  three-dimensional dark matter power spectrum $P(k)$ agrees within 2$\%$ with the predictions from the Extended Cosmic Emulator \citep[see][]{Heitmann:Lawrence:2014}, followed by a progressive deviation for higher $k$-modes \citep{Harnois-Deraps:2018}. We use the SLICS to estimate a reliable covariance matrix, which, combined with the cosmo-SLICS, allows us to test our model for a simulation that is independent of T17. Similar to the T17 simulations, the signal of the SLICS is combined with the randomly oriented intrinsic shapes $\epsilon^{\nt{s}}$ to create ellipticities, whereas $\epsilon^{\nt{s}}$ is drawn from a Gaussian distribution with width $\sigma_\epsilon$ directly since the shear information are given here per galaxy. We added an additional layer of realism and used a redshift-dependent shape noise that better reproduces the data properties. Specifically, we used $\sigma_\epsilon = 0.25$ and $0.27$ for the source bins, as reported in \citet{Giblin:2020}.

\subsection{Extracting the SLICS and cosmo-SLICS data vector}
The extraction of the data vector for the SLICS and cosmo-SLICS analyses is similar to the T17 case, where shape noise was not included for the cosmo-SLICS data vector to better capture the cosmological signal. Another slight difference is that the light cones are now square, which accentuates the edge effects when the aperture filter overlaps with the light-cone boundaries. In principle, it is possible to weight the outer rims for each $N_\mathrm{ap}$ map, so that the whole map can be used. Although this would increase our statistical power, it could also introduce a systematic offset. We opted instead to exclude the outer rim for each realisation resulting in an effective area of $(10-2\Theta)^2\,\mathrm{deg}^2$ with $\Theta$ the size of the corresponding filter. This procedure also ensures that roughly the same number of background galaxies are used to calculate the shear profile around each pixel.
\begin{figure}[!htbp]
\centering
\includegraphics[width=\columnwidth]{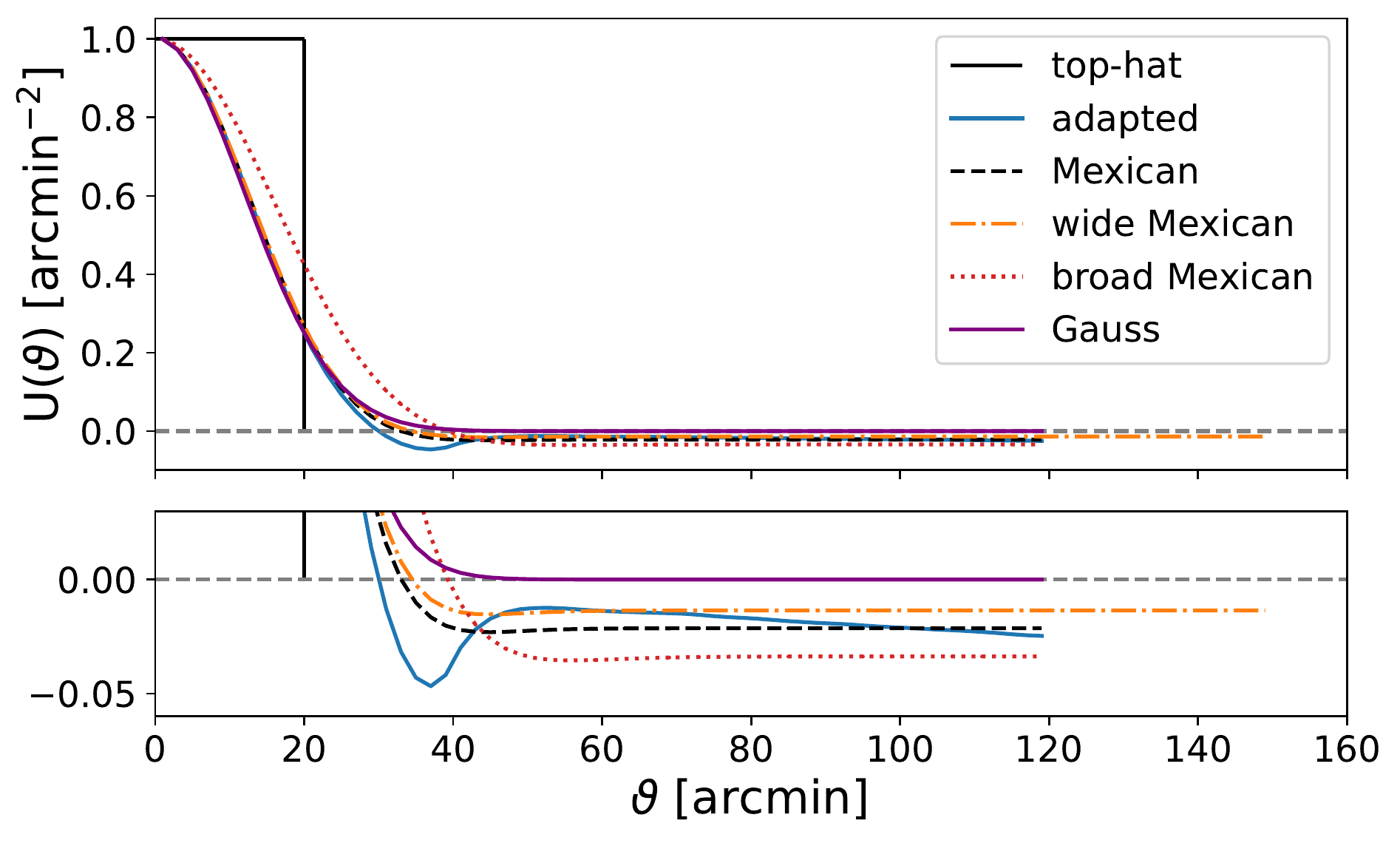}
    \caption{Different filters $U$ used in this work to verify the new model. For all filters we scaled the first bin value to $1/$arcmin$^{-2}$ for comparison. The corresponding $Q$-filters are shown in Fig.~\ref{fig:filter_Q}. The wide Mexican filter extends up to $150’$.}
    \label{fig:filter_U}
\end{figure}

\begin{figure*}[!htbp]
\begin{subfigure}{\columnwidth}
 \includegraphics[width=\columnwidth]{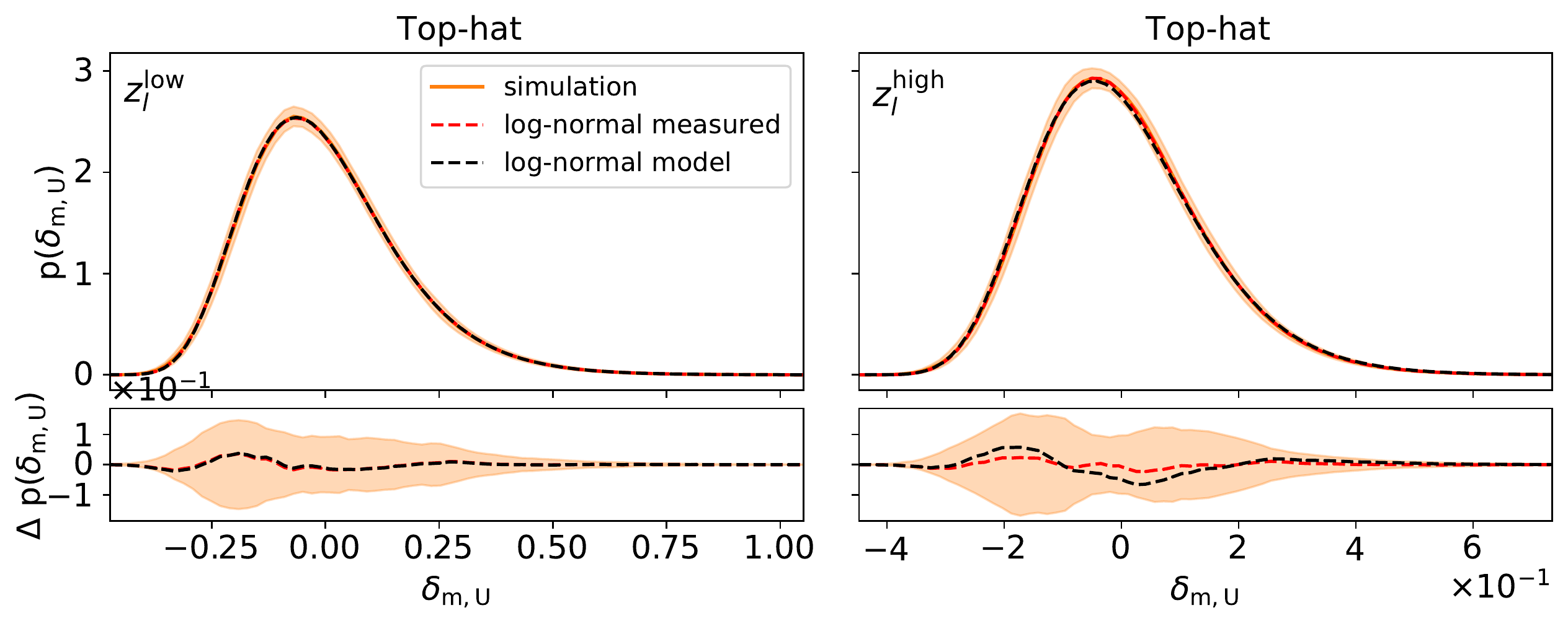}
\end{subfigure}
\begin{subfigure}{\columnwidth}
 \includegraphics[width=\columnwidth]{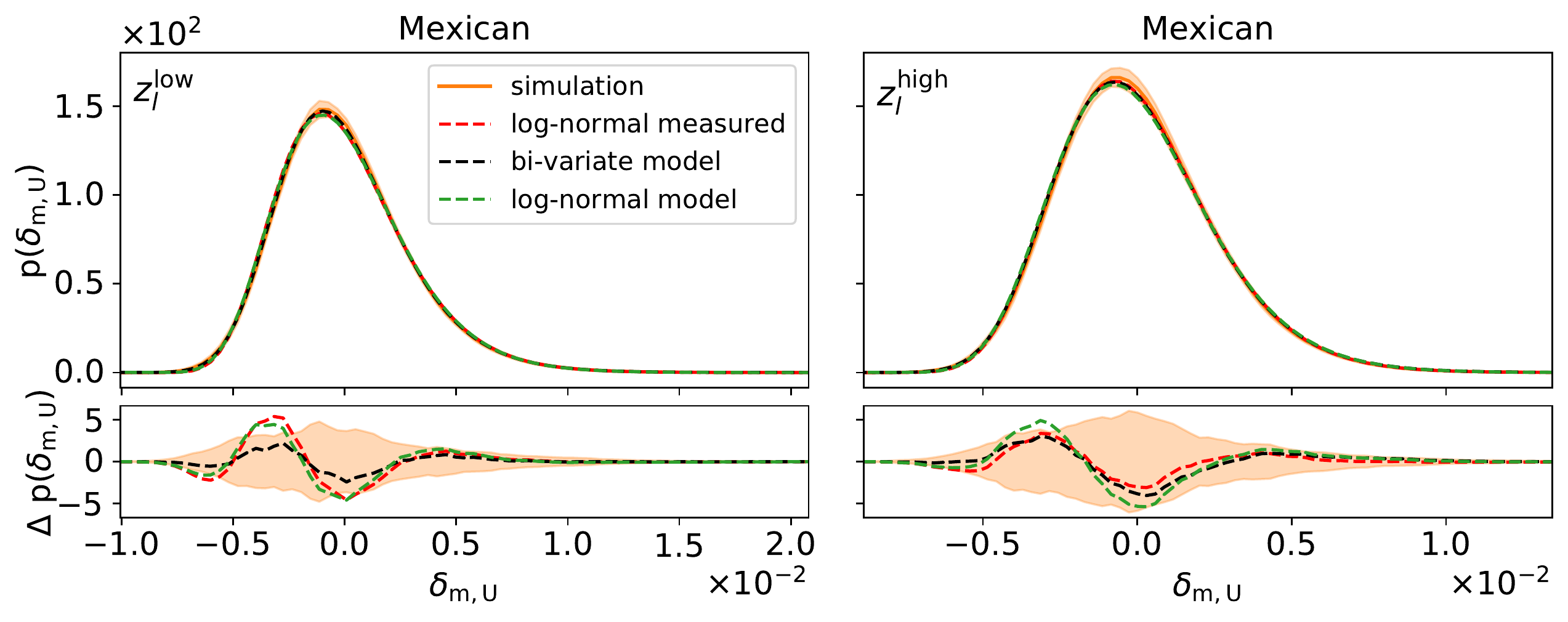}
\end{subfigure}
\begin{subfigure}{\columnwidth}
 \includegraphics[width=\columnwidth]{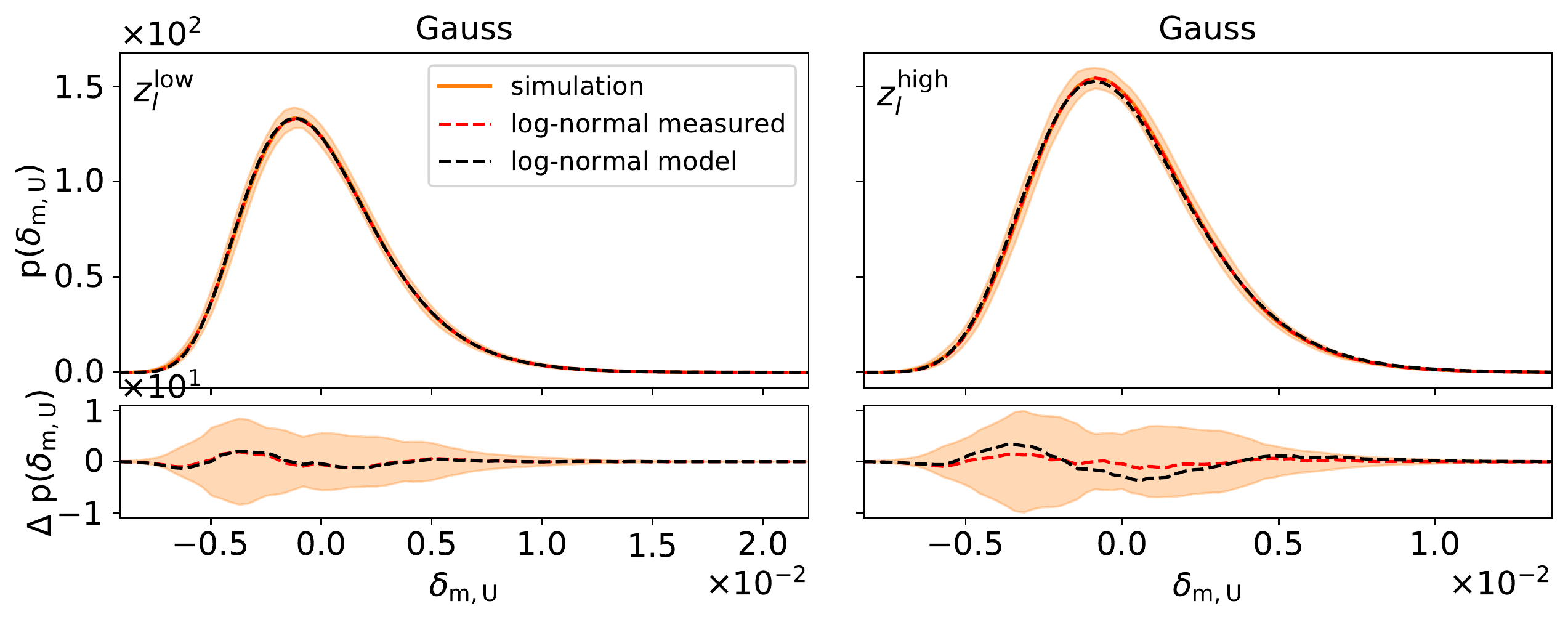}
\end{subfigure}
\begin{subfigure}{\columnwidth}
 \includegraphics[width=\columnwidth]{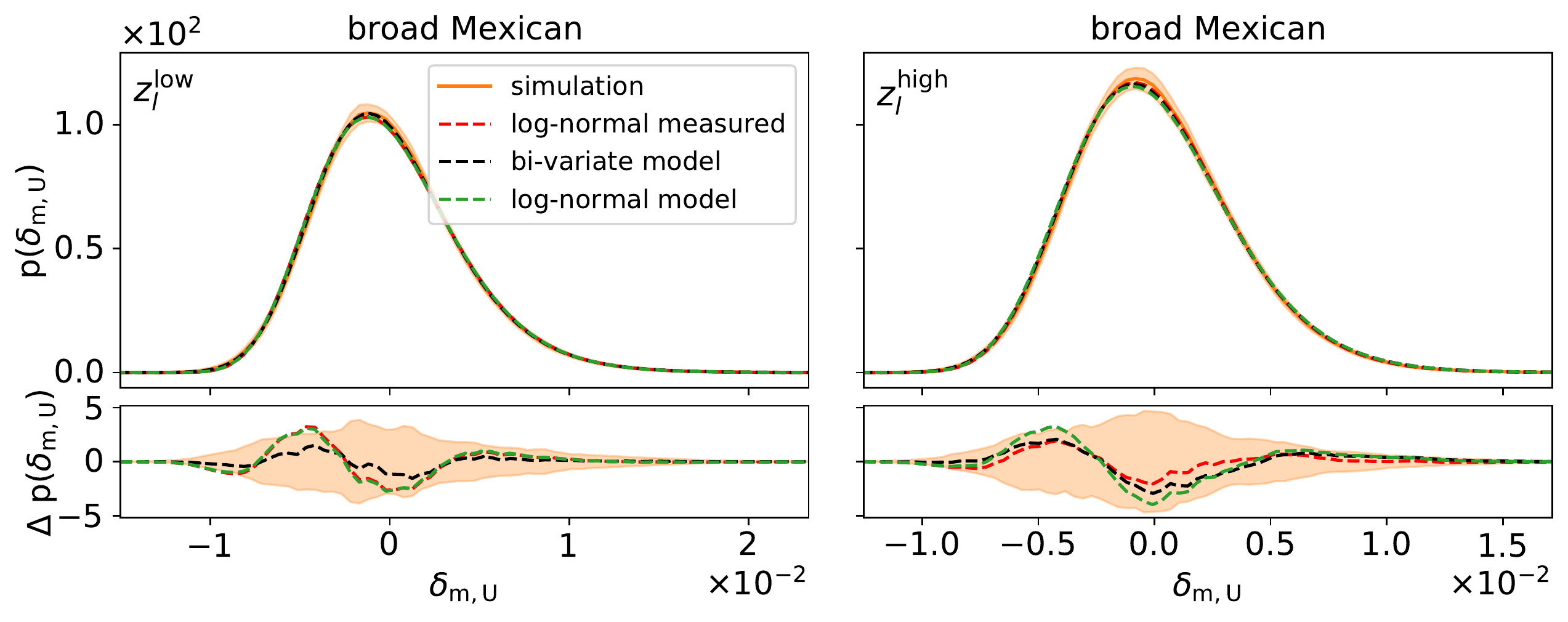}
\end{subfigure}
\begin{subfigure}{\columnwidth}
 \includegraphics[width=\columnwidth]{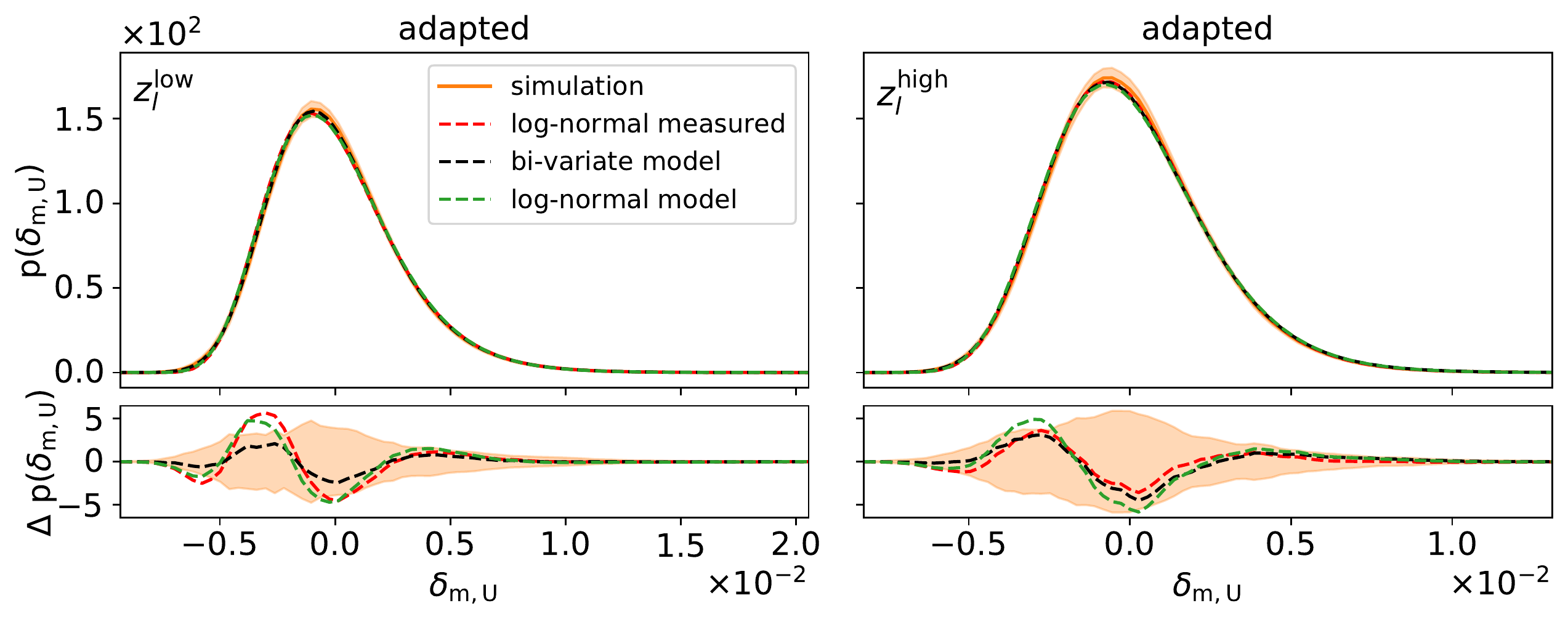}
\end{subfigure}
\begin{subfigure}{\columnwidth}
\hspace{0.3cm}\includegraphics[width=\columnwidth]{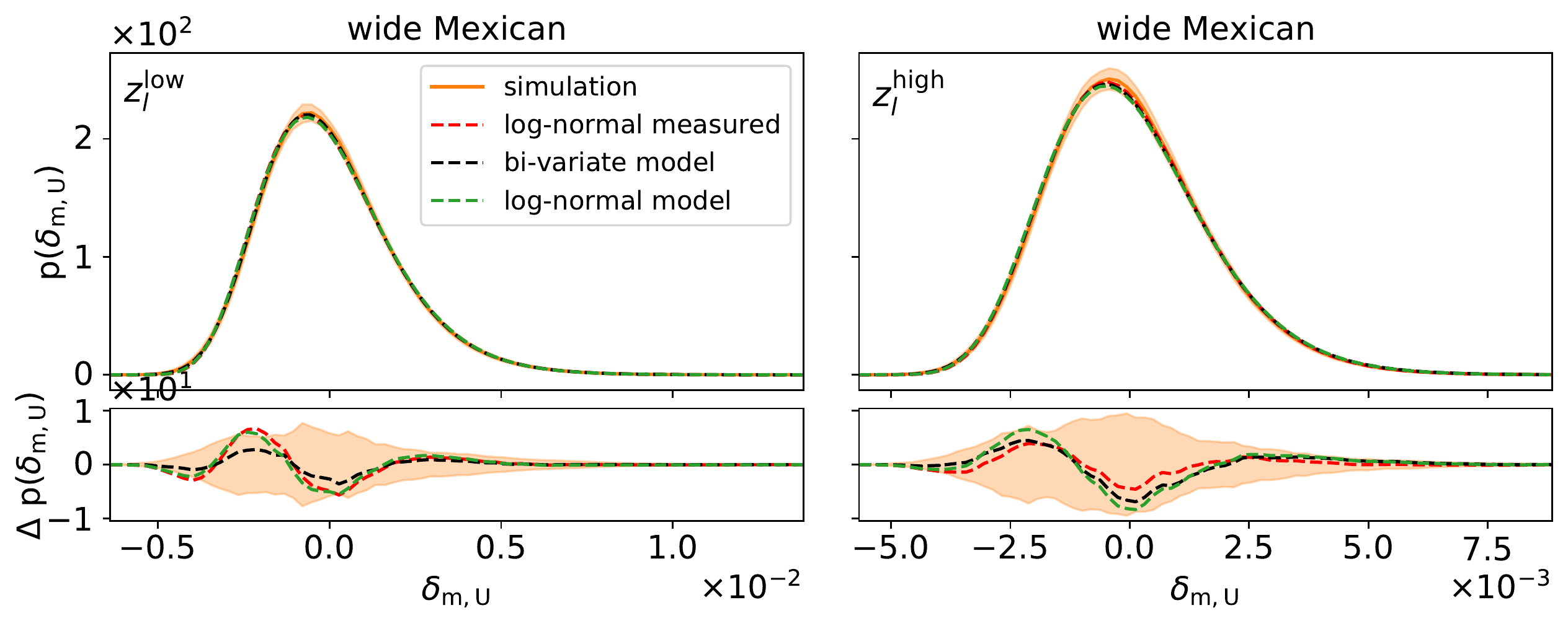}
\end{subfigure}
    \caption{PDF of $\delta_{\mathrm{m},U}$ smoothed with the filters shown in Fig.~\ref{fig:filter_U}. The orange shaded region is the standard deviation of 48 sub-patches scaled by a $777.4/859.4$, where $777.4$\,deg$^2$ is the effective survey area of KiDS-1000 \citep[see][]{Giblin:2020} and $859.4$\,deg$^2$ is the area of one sub-patch. The red dashed curve corresponds to a log-normal PDF with the measured moments $\langle \delta_{\mathrm{m},U}^2\rangle$ and $\langle \delta_{\mathrm{m},U}^3\rangle$ from the smoothed T17 density maps, and indicates the accuracy using a log-normal PDF. The green and the black dashed lines are both from the model; the green corresponds to the PDF of $\delta_{\mathrm{m},U}$ when using log-normal and the black using the bi-variate approach described in Eq.~\eqref{bivariate_lognormal}. The lower panels show the residuals $\Delta p(\delta_{\mathrm{m},U})$ of all lines with respect to the simulations.}
    \label{fig:pofdelta_mU}
\end{figure*}
\begin{figure*}[!htbp]
\begin{subfigure}{\columnwidth}
\includegraphics[width=\columnwidth]{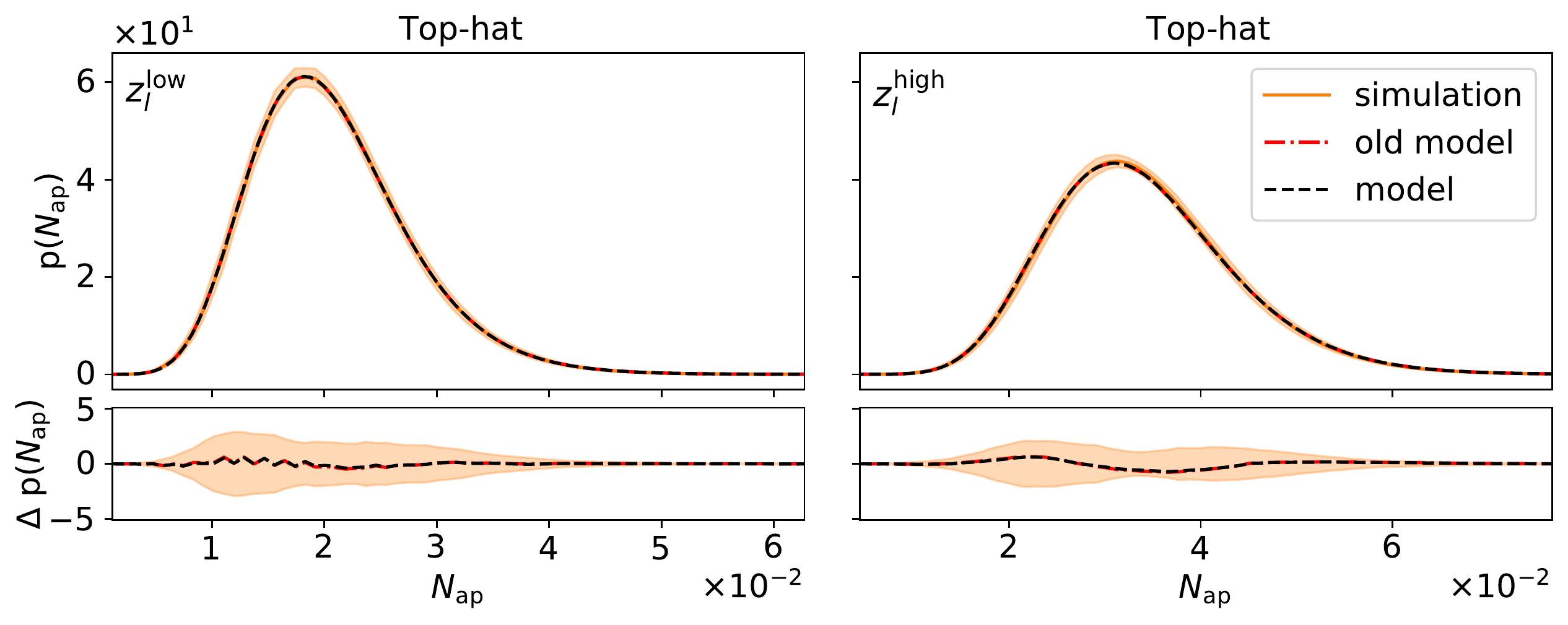}
\end{subfigure}
\begin{subfigure}{\columnwidth}
\includegraphics[width=\columnwidth]{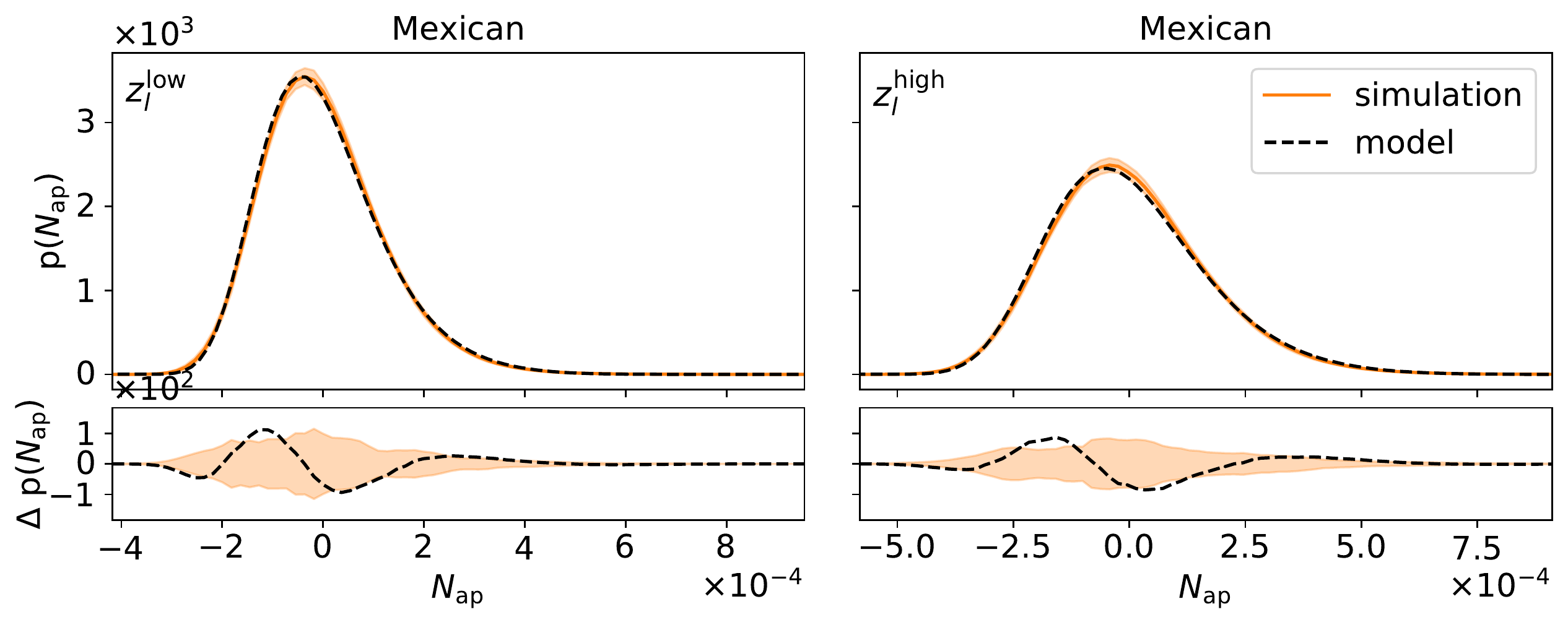}
\end{subfigure}
\begin{subfigure}{\columnwidth}
\includegraphics[width=\columnwidth]{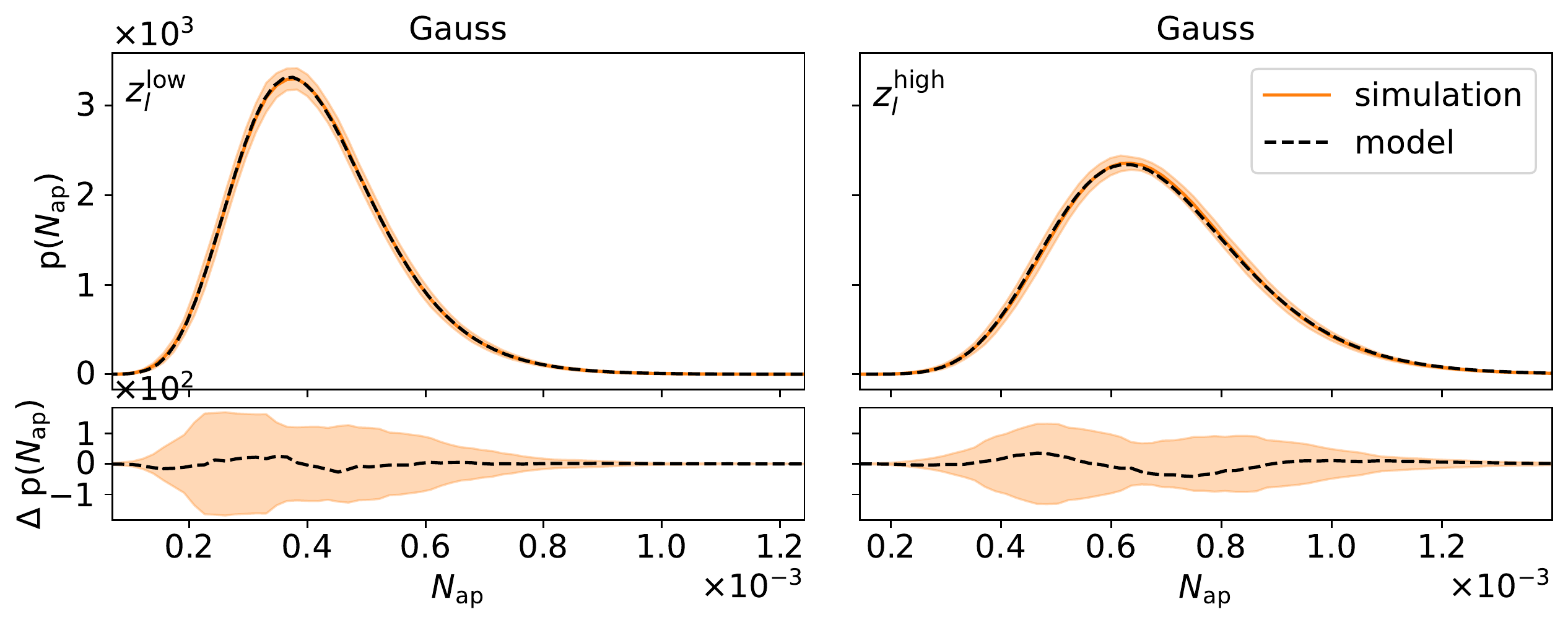}
\end{subfigure}
\begin{subfigure}{\columnwidth}
\includegraphics[width=\columnwidth]{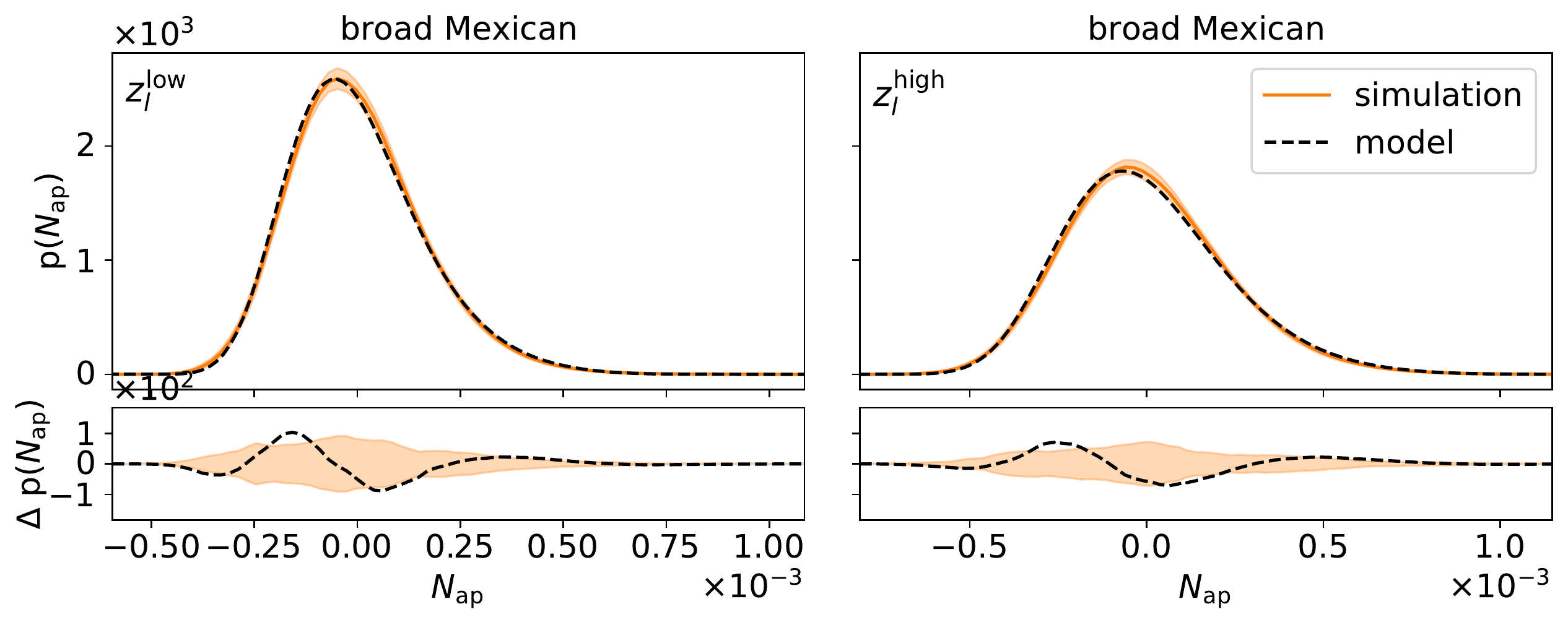}
\end{subfigure}
\begin{subfigure}{\columnwidth}
\includegraphics[width=\columnwidth]{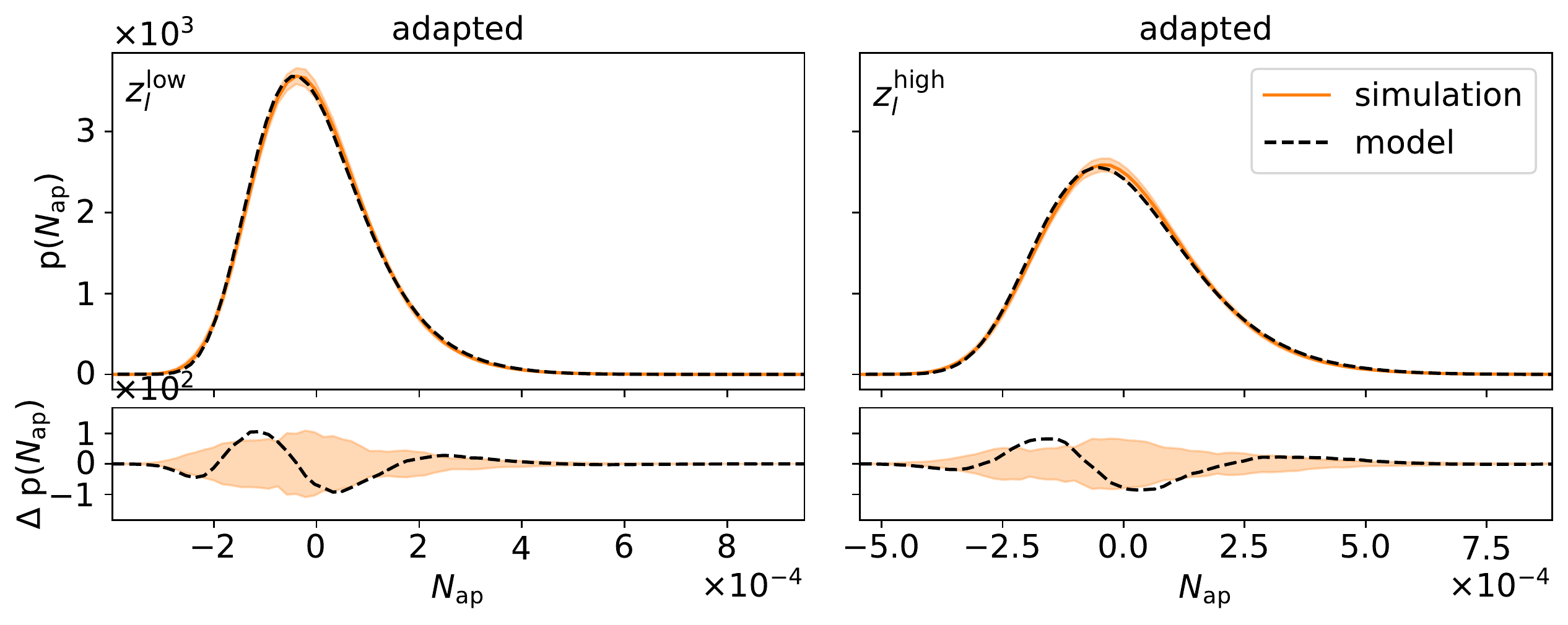}
\end{subfigure}
\begin{subfigure}{\columnwidth}
\hspace{0.3cm}\includegraphics[width=\columnwidth]{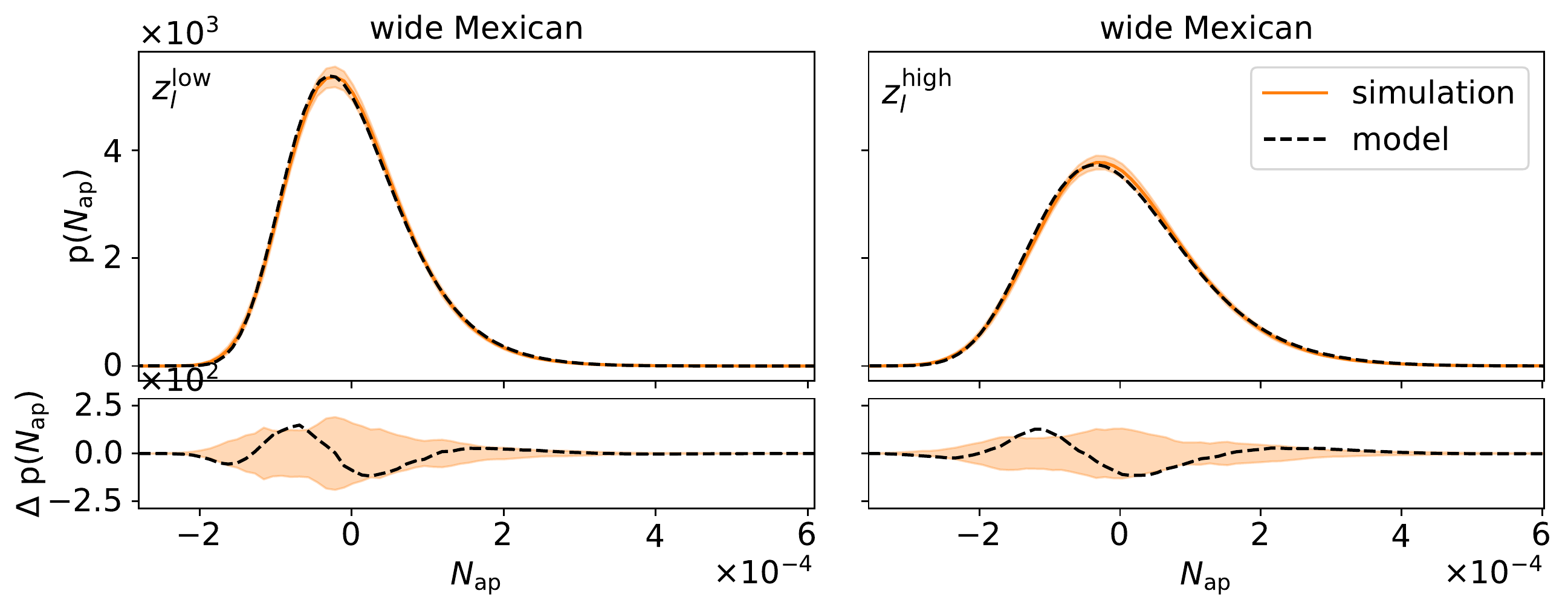}
\end{subfigure}
\caption{PDF of $ N_{\textrm{ap}}$ calculated with the filters $U$ in Fig.~\ref{fig:filter_U}. The orange lines are determined with the simulations and the orange shaded region is the standard deviation from 48 sub-patches. The black dashed lines correspond to the results from the new model, and for comparison the red dashed line in the upper left panel is from the old model. The lower panels show the residuals $\Delta p(N_{\textrm{ap}})$ of all lines with respect to the simulations.}
\label{fig:pofNap}
\end{figure*}

\begin{figure*}
\begin{subfigure}{\columnwidth}
\includegraphics[width=\columnwidth]{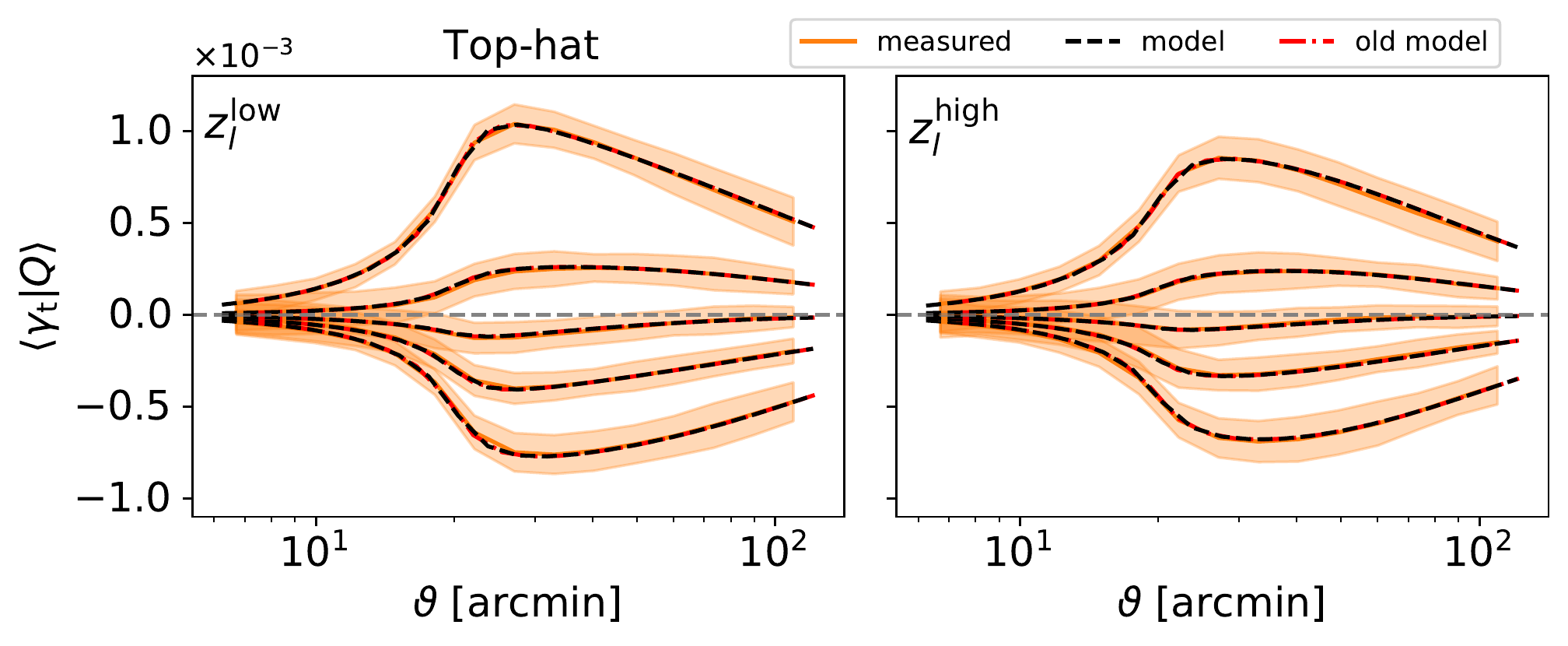}
\end{subfigure}
\begin{subfigure}{\columnwidth}
\includegraphics[width=\columnwidth]{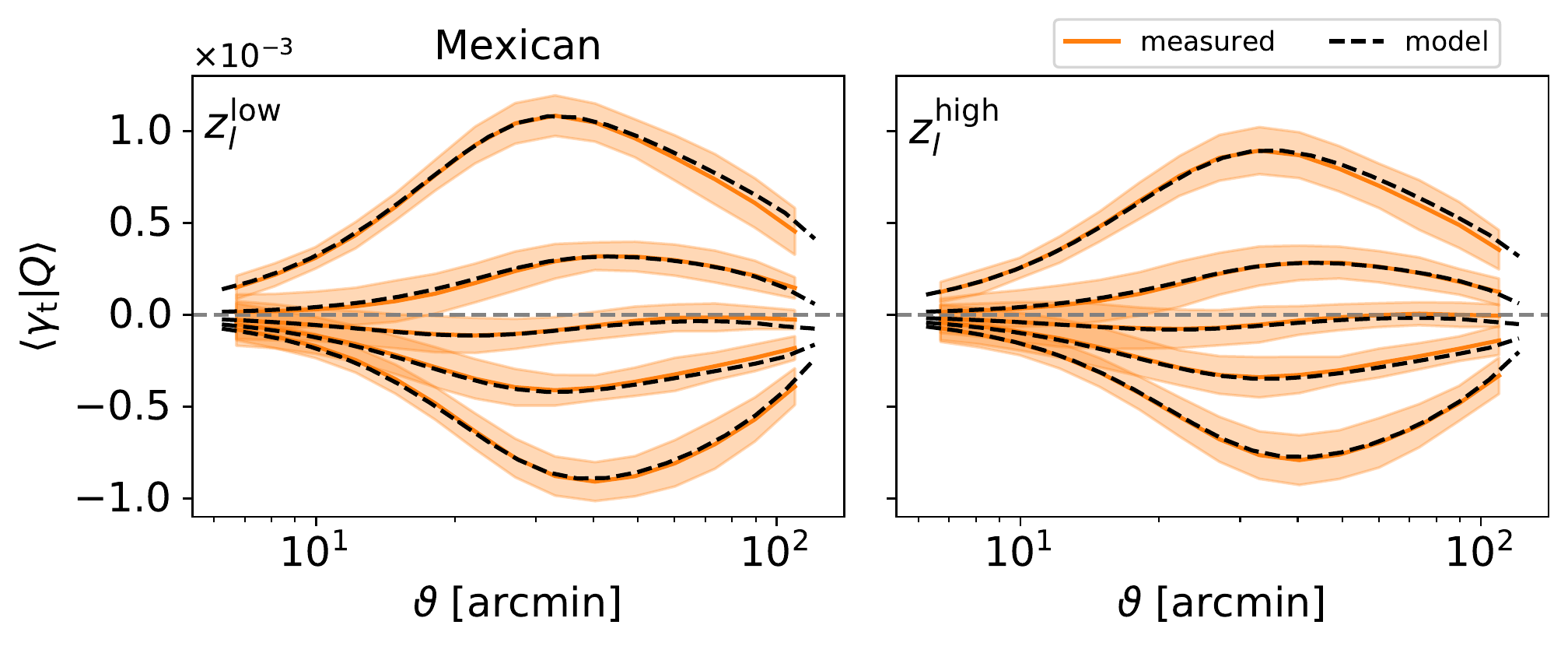}
\end{subfigure}
\begin{subfigure}{\columnwidth}
\includegraphics[width=\columnwidth]{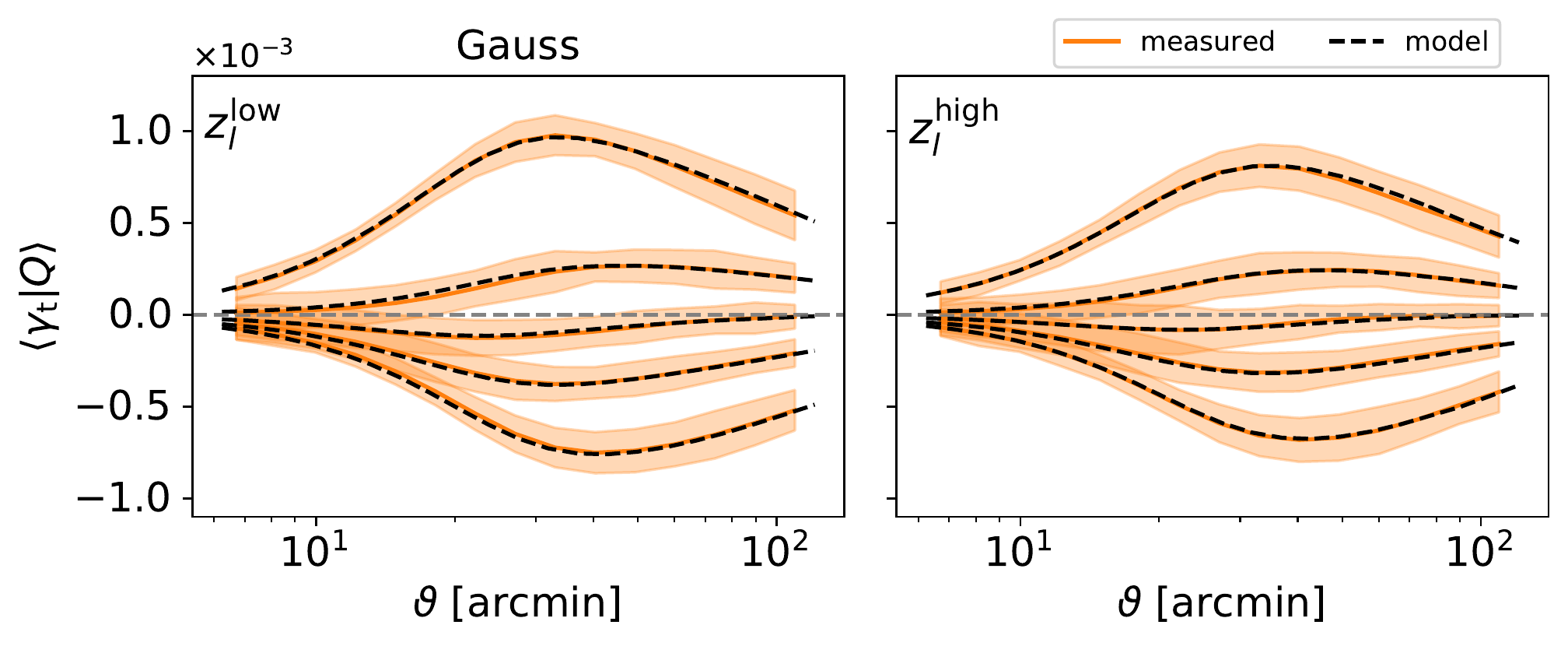}
\end{subfigure}
\begin{subfigure}{\columnwidth}
\includegraphics[width=\columnwidth]{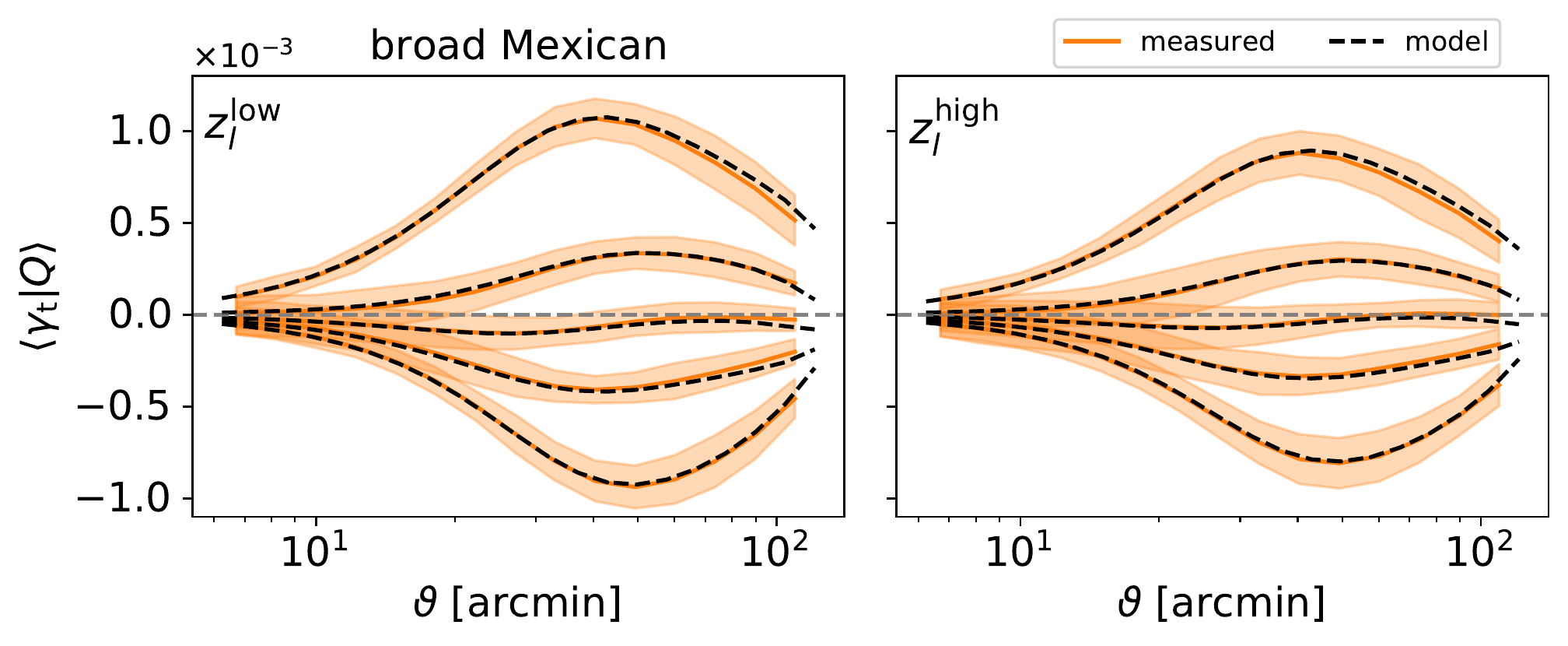}
\end{subfigure}
\begin{subfigure}{\columnwidth}
\includegraphics[width=\columnwidth]{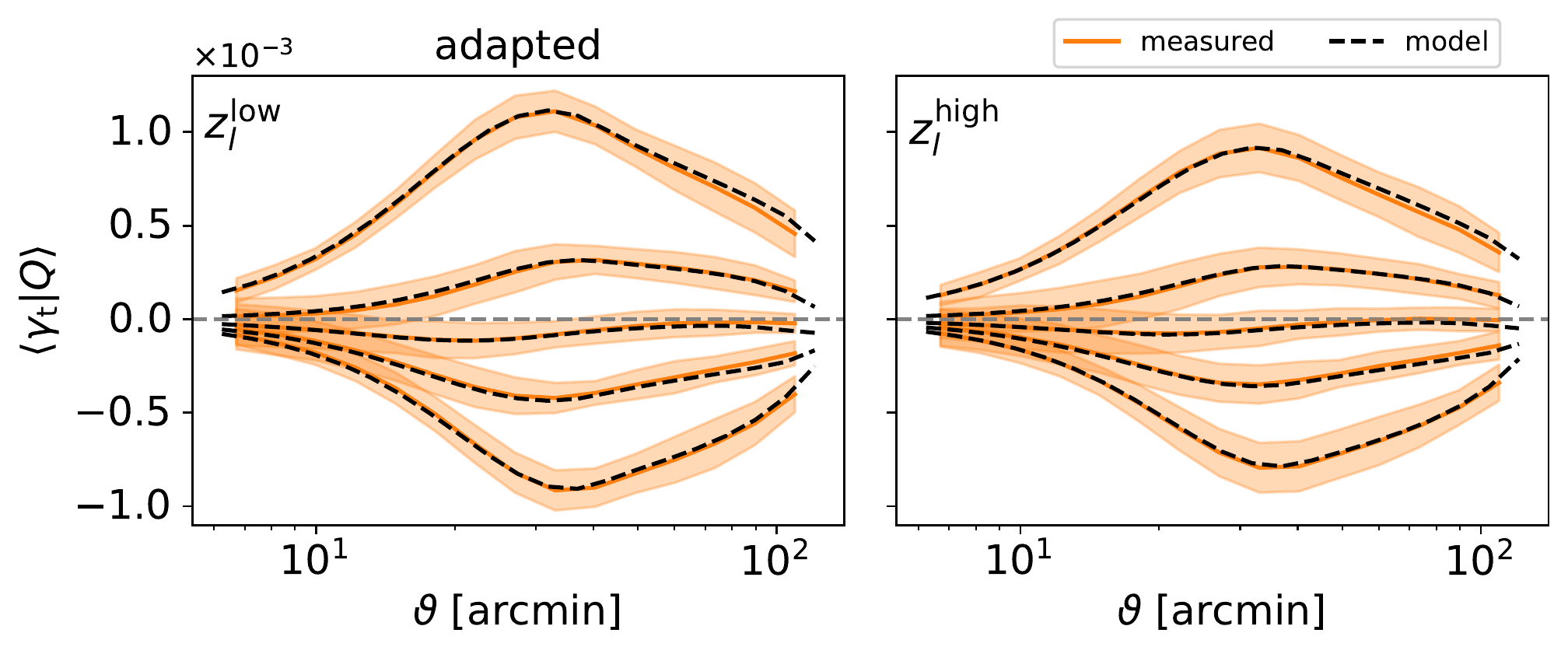}
\end{subfigure}
\begin{subfigure}{\columnwidth}
\hspace{0.3cm}\includegraphics[width=\columnwidth]{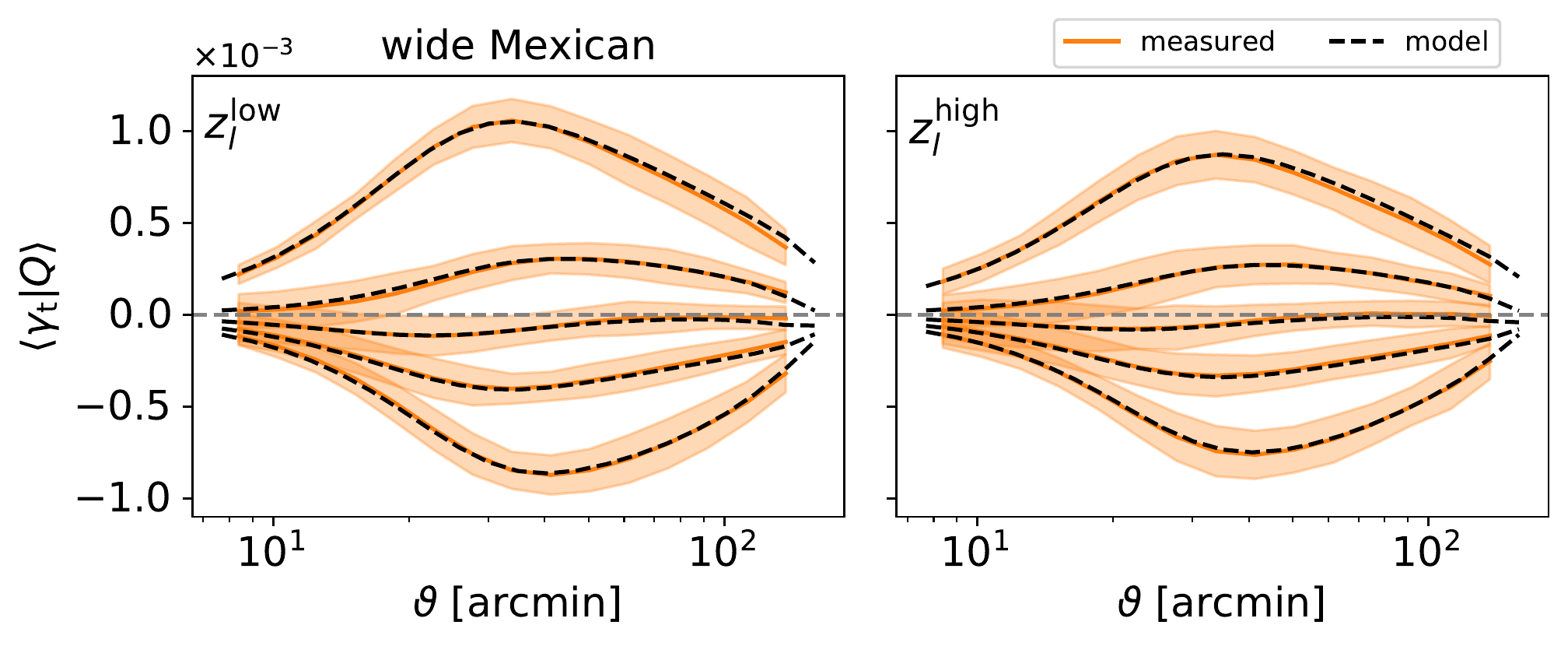}
\end{subfigure}
\caption{Predicted shear profiles for the two lens samples (dashed black line) and measured shear profiles (in orange) for the new model with filter $U$. The orange shaded region is the standard deviation on the mean from 48 sub-patches, scaled to the KiDS-1000 area. The residuals between model and simulations were tested to determine whether they can be erased when the PDF of the aperture number is fixed to the measured value from T17, but the same discrepancies were present.}
\label{fig:shear}
\end{figure*}

\begin{figure}
\includegraphics[width=\columnwidth]{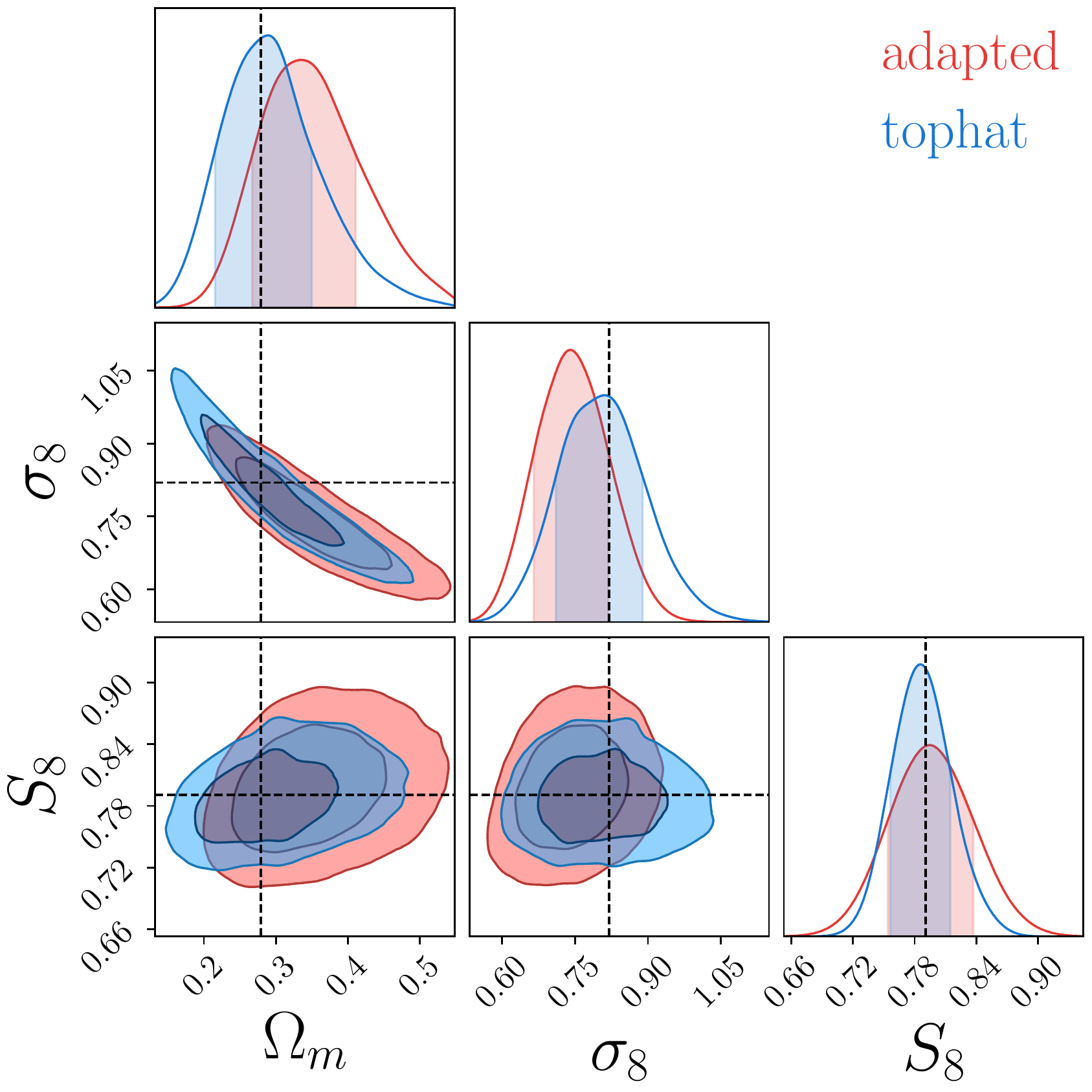}
\caption{MCMC results for the top-hat and adapted filter using the model and the T17 simulations as our data vector and a covariance matrix calculated from ten T17 realisations each divided into 48 sub-patches. For the adapted filter a systematic bias for $\sigma_8$ and $\Omega_\mathrm{m}$ is found, although it cancels out for the $S_8 = \sigma_8 \sqrt{\Omega_\nt{m}/0.3}$ parameter. The contours here are marginalised over the lens galaxy bias parameters.} 
    \label{fig:MCMC_orginal}
\end{figure}

\section{Testing the revised model}
\label{sec_model_test}
We used the simulations described in Sect.\,\ref{Sect:Data} to test our revised model and its accuracy in predicting shear profiles. Following the results of F18 we chose a top-hat filter of $20'$ as our starting point and we considered a number of more general filters with a similar angular extent, shown in Fig.~\ref{fig:filter_U}. Our motivation for studying these filters is as follows: We use a Gaussian filter to test whether the model performs well for non-constant but positive filters; the `adapted' filter is the filter that results from B20; the `Mexican' filter removes the local minimum at $\vartheta\sim 40'$; the `broad Mexican' has a larger width; finally, the `wide Mexican' suppresses the negative tail. In order to lower the amplitude of the negative part while keeping a similar width, we adjusted the upper bound of the wide-Mexican filter to conserve the compensation to $150'$, which makes it better suited to large contiguous survey areas.

Before comparing our model to the simulations, we note that we are using here the revised model even for the top-hat filter, for which we could instead use the F18 model directly. Notably, the derivations of $\langle\delta_{\mathrm{m},U}^2\rangle$ and $\langle \delta_{\mathrm{m},U}^3\rangle$ are identical in the revised model, and we show in the following plots for the top-hat filters that both models yield almost identical results in predicting the shear profiles with a top-hat filter. Therefore, from here on, we only show results from the revised model. In the following three sections, we validate the key model ingredients introduced in Sect. \ref{subsec:I}- \ref{Sec:Characteriscfunction}.

\subsection{Validating $p(\delta_{\mathrm{m},U})$}
We show in Fig.~\ref{fig:pofdelta_mU} the PDF of the smoothed two-dimensional density contrast for all six filters, and for the two lens bins. We see by inspecting the different panels that the predictions agree with the simulations for the two lens bins within 1\,$\sigma$ cosmic variance expected for KiDS-1000. We note here that this PDF cannot be measured in real data, and that the real test for the accuracy of our model are the shear signals, with larger uncertainties due to shape noise. Nevertheless, for the top-hat and the Gaussian we have an agreement between model and simulation well within the 1\,$\sigma$, which indicates that the log-normal approximation for theses filters is good. The other filters show stronger deviations when using a log-normal approximation, but these are weaker when the negative part of the filter approaches zero (wide Mexican) or when the width of the filter increases (broad Mexican), although the negative part of the broad Mexican is stronger than for the Mexican filter. This indicates that probing on larger scales either with a broader or wider filter the log-normal approximation is more accurate. Furthermore, when using the bi-variate log-normal approach discussed in Sect.~\ref{subsec:I}, the residuals are even more suppressed, and thus we cannot recognise differences in the match between predicted and measured PDF for all compensated filters. Although the model for the compensated filters is not as good as for the non-negative filters (top-hat and Gaussian), the revised model remains consistent throughout with the T17 simulations. 

\subsection{Validating $p(N_{\mathrm{ap}})$}
We show in Fig.~\ref{fig:pofNap} how well the model can predict $p(N_{\textrm{ap}})$ given the galaxy distributions described in Sect.\,\ref{Sect:Data}. As for $p(\delta_{\mathrm{m},U})$, the best matches are observed for the non-negative filters, where the simple log-normal PDF is used. For the compensated filters with the bi-variate log-normal $p(\delta_{\mathrm{m},U})$ we note a slight deviation in the skewness of $p(N_{\textrm{ap}})$. These discrepancies are not seen when placing galaxies at random positions regardless of any underlying matter density field as shown in Fig.\,\ref{fig:pofNap_uniform}, which indicates that they must originate either from $p(\delta_{\mathrm{m},U})$ or from the $\langle w_\vartheta|\delta_{\mathrm{m},U}\rangle$ term (we set the latter to $0$ for uniform random fields). It might be that the deviations seen in $p(N_{\textrm{ap}})$ are exclusively caused by the deviations in $p(\delta_{\mathrm{m},U})$, but since they are much smaller, we expect that the assumptions made in computing $\langle w_\vartheta|\delta_{\mathrm{m},U}\rangle$ induce additional inaccuracies. Nevertheless, we show next that these deviations result in shear signals whose residuals are well within the statistical uncertainties of Stage III weak lensing surveys such as KiDS-1000. However the accuracy of the $\langle w_\vartheta|\delta_{\mathrm{m},U}\rangle$ term will likely need to be improved for future surveys like {\it Euclid}, as discussed in Sect.~\ref{subsec:II}.

\subsection{Validating $\langle \gamma_\textrm{t} |  \mathcal{Q} \rangle$}
Having quantified the accuracy of the basic ingredients of our model, we are now in a position to compare the predicted and measured shear profiles. This is a major result of our paper, which is shown in Fig.~\ref{fig:shear}. Following G18, we used five quantiles and we measured the shear profiles up to $120'$ (or $150'$ for the wide Mexican case). For the top-hat, Gaussian, and wide Mexican filters we see no significant deviations between the model and the simulations. For the adapted and the smaller Mexicans the shear profiles show minor discrepancies in some quantiles and at large angular scales, but are always consistent within the KiDS-1000 accuracy. The shapes of the signals are affected by the choice of the filter. We can observe shifts in the peak positions and changes in the slope of the signals especially at small scales. This will allow us in the future to select filters that optimise the signal-to-noise ratio of the measurement, while being clean of systematics related to small-scale inaccuracies. Finally, we show in Fig.\,\ref{fig:lognomal_shear_comp} that for the compensated filter the difference in using the proposed bi-variate log-normal approach is slightly more accurate than using a plain log-normal. Although the difference does not change the final results noticeably, and although it introduces some inconsistency in the sense that we use a bi-variate approach for $p(\delta_\mathrm{m,U})$ but not for $\langle \kappa_{<\theta}|\delta_\mathrm{m,U}\rangle$\footnote{Since the impact is already quite small when adjusting $p(N_\mathrm{ap})$, we are confident that also using a bi-variate approach for $\langle \kappa_{<\theta}|\delta_\mathrm{m,U}\rangle$ would result in even smaller improvements as discussed in greater detail at the end of Sect.\,\ref{subsec:II}.}, we decided to stay with the proposed ansatz because it is slightly more accurate, and we plan to use $p(N_\mathrm{ap})$ in future analysis.

In order to check whether the discrepancies seen for some compensated filters yield biased results, we performed an MCMC analysis. As our data vector we used the T17 shear profiles shown in Fig.~\ref{fig:shear}, where we made a conservative cut and included only scales above $14'$, since as shown in F18 the model is not fully accurate for small angular scales. For the comparison we decided to use the adapted filter and the top-hat filter to have one analysis with and one without these discrepancies. Furthermore, since the mean aperture mass summed over all quantiles vanishes per definition, one of the five shear signals is fully determined by the others, and so we discarded for all cases the middle quantile with the lowest signal. Thus, we ended up with data and model vectors of size 88. As explained previously, we measured our covariance matrix from ten T17 simulations, each divided into 48 sub-patches, for a total of 480 sub-patches. We note here that the galaxy number density can slightly deviate between the different realisations due to the Poisson sampling. Given the amplitude of these small fluctuations, these can be safely neglected.
Next we de-biased the inverse covariance matrix $C^{-1}$ following \cite{Hartlap2007},
\begin{equation}
    C^{-1} = \frac{n-p-2}{n-1} \hat{C}^{-1}\, ,
    \label{eq:hartlap}
\end{equation}
where $n$ is the number of simulations (480) and $p$ the size of the data vector (88). Finally, given our data $\vb{d}$ measured from only one noise-free T17 realisation, and our model vector $\vb{m}$, we measured the $\chi^2$ statistics as
\begin{equation}
\chi^2 =  \left[\vb{m}-\vb{d}\right]^T C^{-1} \left[\vb{m}-\vb{d}\right] \, .
\label{eq:chi2}
\end{equation}
Given this set-up we ran an MCMC varying the matter density parameter $\Omega_\nt{m}$ and normalisation of the power spectrum $\sigma_8$ for the adapted and the top-hat filters, where we marginalised over the biases of the lens samples. As shown in Fig.\,\ref{fig:MCMC_orginal} the analysis with the adapted filter results in a biased inference for the $\Omega_\nt{m}$-$\sigma_8$-plane (although still within $1\,\sigma$); this is not the case for the top-hat filter. We note here that this bias is due to the systematic offset in the slope of the highest quantile, which in turn is is highly sensitive to $\Omega_\mathrm{m}$. Since the amplitude of the shear profiles are correct and these are highly correlated with the $S_8 = \sigma_8 \sqrt{\Omega_\nt{m}/0.3}$ parameter, the contours shift to smaller $\sigma_8$ values in order to compensate for the bigger $\Omega_\mathrm{m}$ value\footnote{The calibration of the residual in the highest quantile alone led to an unbiased result.}. In the next section we calibrate the model to investigate whether this systematic bias can be corrected.

\subsection{Calibrating the model}
\label{sec:calibrate_model_with_cosmoSlics}
In this section we calibrate the remaining small inaccuracies of the analytical model seen in Fig. \ref{fig:shear} which result in the systematic bias we had observed in the parameter constraints shown in Fig.~\ref{fig:MCMC_orginal}. For this we decided to divide out for each quantile the residuals between the model, $\gamma_{\textrm{M}_\textrm{T}}$, at the T17 cosmological parameters, $p_\textrm{T17}$, and the noiseless shear profiles measured from the T17 simulations, $\gamma_\textrm{T17}$, such that the calibrated model at parameters $p$ is defined as
\begin{equation}
    \gamma_\textrm{M,cal}(p) = \gamma_{\textrm{M}}(p) \frac{\gamma_\textrm{T17}}{\gamma_{\textrm{M}}(p_\textrm{T17})} \, .
    \label{eq:new_model}
\end{equation}
Since we used the $n(z)$ combinations of the fiducial cosmo-SLICS shown in Fig.~\ref{fig:n_of_z} to validate the calibration, we decided to use the $n(z)$ shown in Fig.~\ref{fig:nofz}, and in order to have the source $n(z)$ as close as possible to the one of the cosmo-SLICS we averaged several T17 shear grids at different redshifts for the same realisation weighted by the source $n(z)$ shown in Fig.~\ref{fig:n_of_z}. In Fig.~\ref{fig:calibrated_vector} we show the calibration vectors for the highest and lowest quantile for the top-hat and adapted filter, where it can be seen that the different lens $n(z)$ is more important than the source $n(z)$.

\begin{figure*}
\begin{subfigure}{\columnwidth}
\includegraphics[width=\columnwidth]{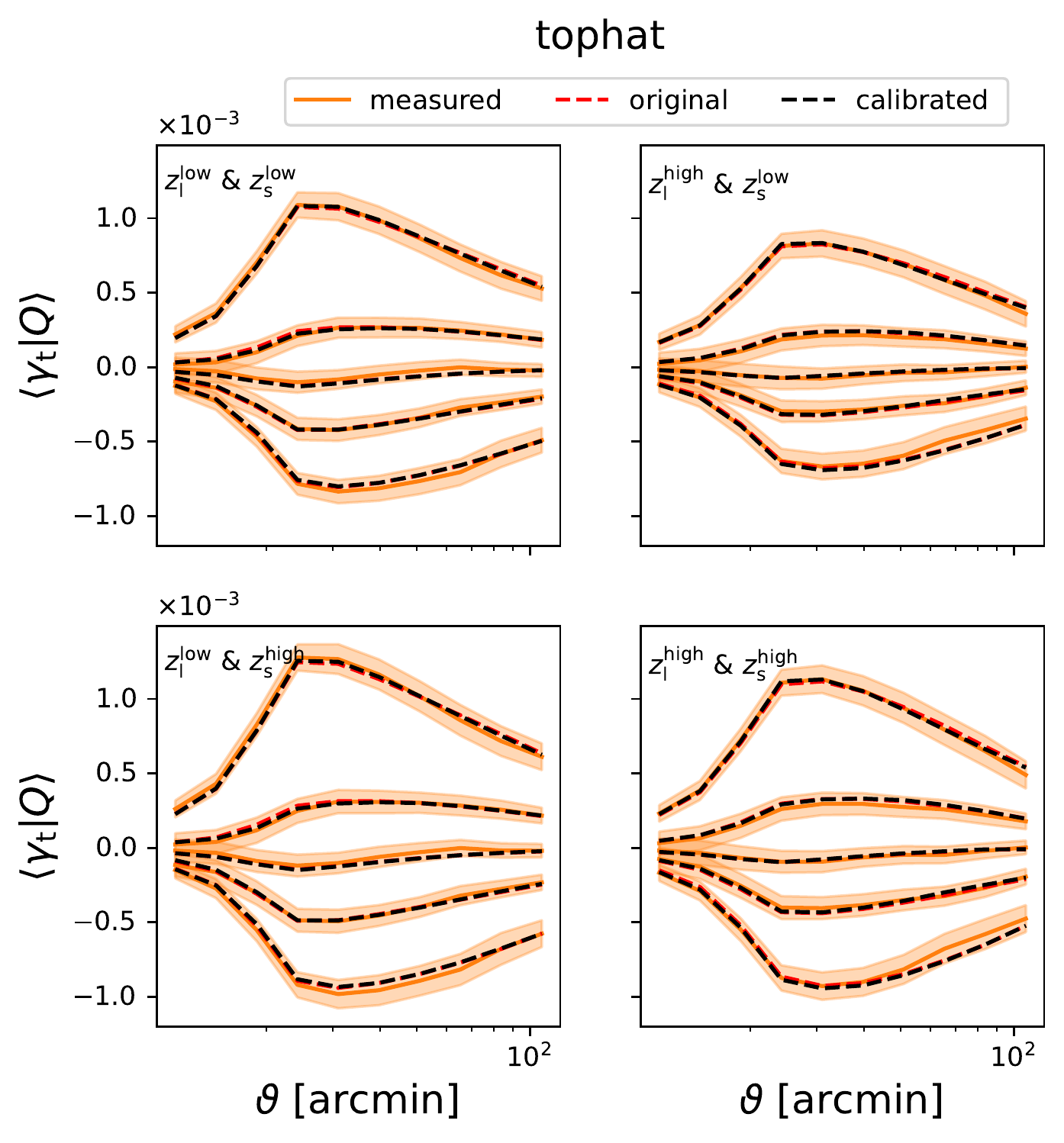}
\end{subfigure}
\begin{subfigure}{\columnwidth}
\includegraphics[width=\columnwidth]{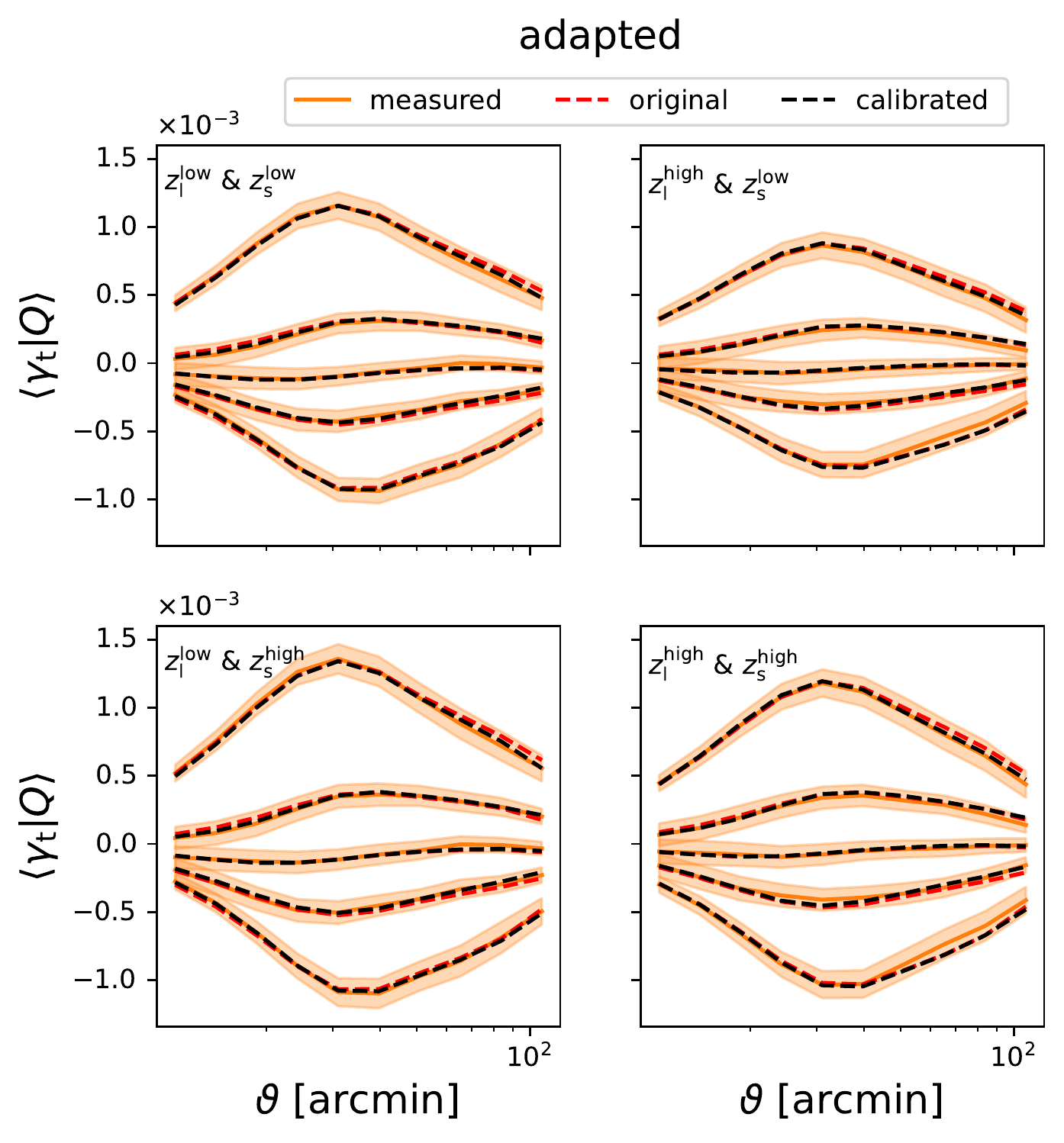}
\end{subfigure}
\caption{Shear profiles for the top-hat filter (left) and for the adapted filter (right) for the fiducial cosmology of cosmo-SLICS. The orange lines are the mean shear profiles and the orange shaded region is the expected KiDS-1000 uncertainty. The red dashed line corresponds to the original model and the black to the calibrated model.}
\label{fig:shear_cosmo_SLICS}
\end{figure*}
Next, in order to investigate whether the calibration decreases the systematic biases we performed another MCMC analysis on independent simulations, where our data vector is the fiducial cosmology from the cosmo-SLICS shear profiles shown in Fig.~\ref{fig:shear_cosmo_SLICS}, with the original model in red and the calibrated one in black. As before we used the adapted filter and the top-hat filter. The match between the predicted and measured shear profiles is slightly degraded compared to the T17 simulations, which could be caused by edge effects (contiguous full-sky vs 100\,deg$^2$ patches), the smaller statistic (41\,253\,deg$^2$ vs 5000\,deg$^2$), or differences in the underlying matter power spectrum $P(k)$ that is used in the model. We use the \citet{Takahashi2012} {\sc Halofit} function throughout this paper, which is calibrated on the same $N$-body code that is used to create the T17 simulations \citep[][\sc{Gadget2}]{Springel2001}, and which is known to have an excess power of 5--8\% in the mildly non-linear regime \citep[][]{Heitmann:Lawrence:2014}. The cosmo-SLICS, in contrast, are produced from {\sc cubep$^3$m} \citep{Joachim2013}, whose $P(k)$ agrees better with the Cosmic Emulator of \citet{Heitmann:Lawrence:2014}. We tested different choices of power spectra models calculated with the \textsc{pyccl} package \citep{pyccl}, but found differences in the predicted shear profiles that are negligible compared to the expected KiDS-1000 uncertainties.

\begin{figure}
\includegraphics[width=\columnwidth]{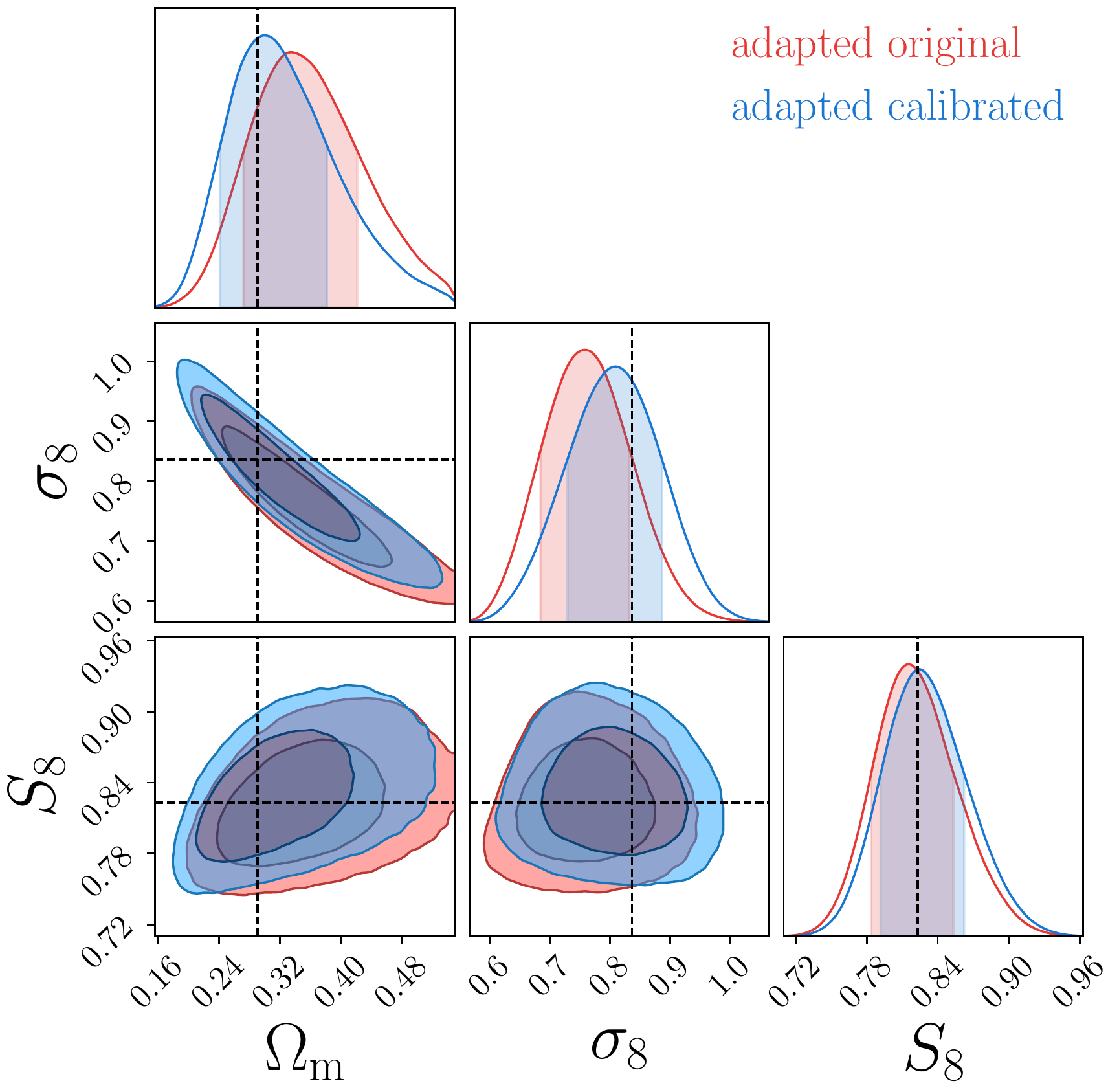}
\caption{MCMC results for the adapted filter using the original and calibrated model. The data vector is calculated from the fiducial cosmology of cosmo-SLICS and a covariance matrix from 614 SLICS realisations. It is clearly seen that the calibrated model is less biased than the original one. The contours are marginalised over the lens galaxy bias parameters.}
    \label{fig:MCMC_calibrated}
\end{figure}

Discarding again the middle quantile with the lowest signal, using four different redshift combinations (Sect.~\ref{Cosmo-SLICS}) and the signal at all scales because the model is calibrated at all scales, we have data and model vectors of size 160. In this scenario we calculated our covariance matrix from 614 SLICS simulations\footnote{For the remaining $\sim 200$ realisations we have no corresponding lens galaxy mocks.} with shape noise that mimics the KiDS-1000 data. After de-biasing the inverse covariance matrix $C^{-1}$ with Eq.~\eqref{eq:hartlap} we calculated the $\chi^2$ with Eq.~\eqref{eq:chi2}. Given this set-up we ran multiple MCMC, where we used the original model and the calibrated model. As shown in Fig.\,\ref{fig:MCMC_calibrated} the calibrated model for the adapted filter results compared to the original model in less a biased inference. Interestingly, the results for the top-hat filter seen in Fig.\,\ref{fig:MCMC_calibrated_tophat} are slightly more biased than the calibrated model for the adapted filter. Since this offset is still inside $1\,\sigma$, it is likely to be only a statistical fluke due to the remaining residual between model and cosmo-SLICS simulations. The constraining power between top-hat and adapted filter are different because the smoothing scales of the two filters were not adjusted as in \citet{Burger:2020}, and are here sensitive to different physical scales. Nevertheless, we show in Table \ref{Table_constrains} the resulting constrains for both filters, where it is seen that the calibration moves the results, also for the top-hat filter, closer to the truth.
\begin{table}[h!]
\centering
\caption{Overview of the maximum-posterior cosmologies with the constraining power that we obtain for the original and calibrated model. For all results we marginalise over the biases $b$ of the lenses. The input cosmology of the fiducial cosmo-SLICS is $\Omega_\nt{m}=0.2905$, $\sigma_8=0.8364$ and thus ${S_8 = \sigma_8 \sqrt{\Omega_\nt{m}/0.3}=0.8231}$. We fixed the time-independent equation-of-state parameter of dark energy $w_0=-1.0$ and Hubble parameter $h=0.6868$ to their true values. We note that the parameter uncertainties increase slightly if we also vary parameters like $w_0$, $h$, or the scalar spectral index $n_{\rm s}$.}
\begin{tabular}{cccc}
& $\Omega_\nt{m}$ & $\sigma_8$ & $S_8$ \\
\hline
\vspace{0.1cm}
ad. original & $0.332^{+0.089}_{-0.060}$ & $0.760^{+0.072}_{-0.077}$   &  $0.814^{+0.039}_{-0.031}$ \\
\vspace{0.1cm}
th. original & $0.320^{+0.068}_{-0.048}$ & $0.774^{+0.069}_{-0.060}$ &  $0.812^{+0.028}_{-0.022}$   \\
\vspace{0.1cm}
ad. calibrated & $0.298^{+0.084}_{-0.056}$ & $0.809^{+0.078}_{-0.080}$  & $0.824^{+0.038}_{-0.033}$  \\
\vspace{0.1cm}
th. calibrated & $0.309^{+0.058}_{-0.051}$ & $0.794^{+0.067}_{-0.062}$  &  $0.812^{+0.027}_{-0.021}$ \\
\end{tabular}
\label{Table_constrains}
\end{table}

In order to compare our results with those of G18, who derived constraints of $\Omega_\nt{m}= 0.26^{+0.04}_{-0.03}$ and $S_8= 0.90^{+0.10}_{-0.08}$ with their fiducial analysis, we need to multiply our uncertainty intervals by $\sqrt{777.4/1321}$ to account for the smaller area of KiDS-1000 ($777.4$\,deg$^2$) compared to the DES Y1 area ($1321$\,deg$^2$). Furthermore, we exclusively used information about the shear profiles, whereas G18 also used the mean aperture number in each quantile. For this work we were a bit sceptical about using the aperture number here for the compensated filters because we have significant residual discrepancies between model and simulation, which would affect our analysis. The match of the shear profiles in turn is very accurate in our simulations, which shows that they are robust against uncertainties in $P(N_\mathrm{ap})$\footnote{The predicted shear profiles do not change significantly even if the predicted $P(N_\mathrm{ap})$ is substituted with the measured $P(N_\mathrm{ap})$}. For instance, if one monotonically transforms the $N_\mathrm{ap}$ values, the predicted $P(N_\mathrm{ap})$ changes, but the segmentation into quantiles is not affected, hence the shear profiles would remain the same. In order to use $P(N_\mathrm{ap})$ in future analysis we need to model shot noise in the galaxy distribution and investigate if the residuals between model and simulations result in systematic biases, but we will keep this for future work. Nevertheless, we see that our constraints from using only information about the shear profiles can be similar to the ones in G18. In addition, due to the calibration method used here, smaller smoothing scales are available than those recommended in F18, where even the top-hat filter has significant deviations. This could allow us to further improve the significance for future DSS analyses or to investigate effects such as baryonic feedback and intrinsic alignments, which are typically relevant on scales $< 10 \mathrm{Mpc}/h$.

\section{Summary and conclusion}
\label{sec:Conclusion}
In our previous work \citep{Burger:2020} we showed that using compensated filters in the density split statistic (DSS) to quantify over- and underdense regions on the sky have advantages compared to the top-hat filter, both in terms of the overall S/N and of recovering accurately the galaxy bias term. Furthermore, we expect that compensated filters are less influenced by baryonic effect, since they are more confined in Fourier space and therefore are better in smoothing out large $\ell$-modes where baryonic effects play an important role. This will be investigated in more detail in a follow-up paper, when we start dealing with real data. \cite{Gruen:2015} demonstrated that the DSS is a powerful cosmological tool by constraining cosmological parameters with DSS measurements from the Dark Energy Survey (DES) First Year and Sloan Digital Sky Survey (SDSS) data, using the DSS model derived in \cite{Friedrich:Gruen:2018} which uses a top-hat filter.  They found for the matter density parameter $\Omega_\nt{m}= 0.26^{+0.04}_{-0.03}$, a constraint that agrees with and is competitive with the DES analysis of galaxy and shear two-point functions \citep[see][]{DES:2018}.

Following these works, we modify the model of \cite{Friedrich:Gruen:2018} in such a way that it can predict the shear profiles $\langle \gamma_\textrm{t} |  \mathcal{Q} \rangle$ for a given quantile $Q$ of the aperture number $N_{\textrm{ap}}$ for general filters (Gaussian and also compensated filters). This is achieved by recalculating the three basic ingredients, which are the PDF of the projected matter density contrast smoothed with the filter function, $p(\delta_{\mathrm{m},U})$; the expectation value of the convergence inside a radius $\vartheta$ for a fixed smoothed matter density contrast, $\langle \kappa_{<\vartheta}|\delta_{\mathrm{m},U}\rangle$; and the distribution of $N_{\textrm{ap}}$ for the given filter function $U$ given the smoothed matter density contrast, $p(N_{\textrm{ap}}|\delta_{\mathrm{m},U})$. For $\langle \kappa_{<\vartheta}|\delta_{\mathrm{m},U}\rangle$ we modified the calculation of the moments for general filters, while we introduced new approaches to calculate $p(N_{\textrm{ap}}|\delta_{\mathrm{m},U})$ and $p(\delta_{\mathrm{m},U})$ for compensated filters. For non-negative filters, $\delta_{\mathrm{m},U}$ is well described by a log-normal PDF, although we found significant deviations for compensated filters. To solve this issue we used a bi-variate log-normal ansatz, where we assumed that $\delta_{\mathrm{m},U}$ can be divided into two log-normal random variables with each separately following a log-normal distribution. For the calculation of $p(N_{\textrm{ap}}|\delta_{\mathrm{m},U})$ we derived an expression for the corresponding characteristic function, which can be used either directly to calculate $p(N_{\textrm{ap}}|\delta_{\mathrm{m},U})$ by inverse Fourier transformation or by calculating the first three moments, which then specify a log-normal distribution for $p(N_{\textrm{ap}}|\delta_{\mathrm{m},U})$. The differences between these two approaches are considerably smaller than the statistical uncertainty, and so we used the latter approach because of its smaller computational time. 

In order to validate the revised model, we compared it to the \cite{Takahashi2017} simulations. For non-negative filters like a top-hat or a Gaussian, no significant difference between the model and simulations for the PDF or the tangential shear profiles were detected. For compensated filters, however, we found some discrepancies in the predicted PDF of $N_{\textrm{ap}}$ and shear signals, which results in a biased inference, although still inside $1\,\sigma$. To correct this biased result, we calibrated the model to match the noiseless \cite{Takahashi2017} and tested the calibrated model with the independent fiducial cosmology of cosmo-SLICS \citep{Harnois-Deraps:2019}. With the calibration applied, all systematic biases are removed, so we are confident that we can apply the model to Stage III surveys such as KiDS-1000. Although this calibration is less important for the top-hat and Gaussian filter, it is still an interesting approach because it allows even smaller scales to be used for both the shear profiles and the filter scales. The use of smaller scales, where the original models fail, makes it possible to increase the constraining power or to study baryonic effects that normally play an important role only at small scales.

After passing all these tests, we are confident that the revised model can be readily applied to Stage III lensing data. We note that a number of systematic effects related to weak lensing analyses will require external simulations, notably regarding the inclusion of secondary signal from the intrinsic alignments of galaxies, or from the impact of baryonic feedback on the matter distribution. However, our model is able to capture the uncertainty on the lens and source redshift distribution, the shape calibration bias, or the galaxy bias at a low computational cost, and is therefore ideally suited to perform competitive weak lensing analyses in the future.

\begin{acknowledgement}
We thank the anonymous referee for the very constructive and fruitful comments. This paper went through the whole KiDS review process, where we especially want to thank the KiDS-internal referee Benjamin Joachimi for his fruitful comments to improve this work. Further, we would like to thank Mike Jarvis for maintaining \textsc{treecorr} and Ryuichi Takahashi for making his simulation suite publicly available. PB acknowledges support by the Deutsche Forschungsgemeinschaft, project SCHN342-13. OF gratefully acknowledges support by the Kavli Foundation and the International Newton Trust through a Newton-Kavli-Junior Fellowship and by Churchill College Cambridge through a postdoctoral By-Fellowship. JHD is supported by a STFC Ernest Rutherford Fellowship (project reference ST/S004858/1). Author contributions: all authors contributed to the development and writing of this paper. 
\end{acknowledgement}

\bibliographystyle{aa}
\bibliography{cite}

\begin{appendix}
\onecolumn
\section{Detailed derivations for the new model}
\label{sec:deatiled_der}

In this appendix we show more detailed derivations of the results than in the main text. We start with the calculation of the variances or covariances in the flat-sky approximation, continue with calculations of the third-order moments and finish with the derivation of the PDF of the aperture number given a smoothed density contrast by use of the characteristic function.

\subsection{Variance and skewness for general filters at leading order in perturbation theory}
\label{sec:pofdeltamU}
Although analytical possible we decided against using the bi-spectrum to calculate third-order moments like the skewness. Instead we use a formalism where we calculate the second- and third-order moments of the smoothed density contrasts within cylinder of physical radius $R$ and physical length $L$ using the flat-sky approximation shown in Appendix B in F18 for a top-hat filter, and apply it to our case with a general filter $U$. Numerically, this approach is faster since, as we will see below, it is possible to express the third-order moments in terms of second-order moments. Another advantage is that the projection is only along one dimension (radius of the cylinder) compared to the bi-spectrum, where the projection is at least along a 2D grid. Following F18 we start by considering a cylinder of radius $R$ and length $L$. In Fourier space the top-hat filter for such a cylinder is given by
\begin{equation}
    W_{R,L}(\vb{k}) = \frac{1}{(2\pi)^3}\frac{\sin(Lk_{||}/2)}{Lk_{||}/2} \frac{2J_1(k_\perp R)}{k_\perp R} \equiv \frac{1}{(2\pi)^3}\frac{\sin(Lk_{||}/2)}{Lk_{||}/2} W_R^{\mathrm{th}}(k_\perp) \, ,
\end{equation}
where $J_1$ is the first Bessel function, and $k_{||}$ and $k_\perp$ are the components of $\vb{k}$ parallel and orthogonal to the cylinder, respectively. The variance of the matter contrast within such a cylinder is given at leading order by
\begin{equation}
\langle \delta_{R,L}^2 \rangle(\chi) = D_+^2 \int \dd k_{||} \, \dd^2 k_\perp \, \frac{\sin^2(Lk_{||}/2)}{\left(Lk_{||}/2\right)^2}\, \left[W_R^{\mathrm{th}}(k_\perp)\right]^2 P_{\mathrm{lin},0}(k_\perp) \approx \frac{2\pi D_+^2}{L} \int \dd k ~k \left[W_R^{\mathrm{th}}(k)\right]^2 P_{\mathrm{lin},0}(k) \, ,
\end{equation}
where the last expression follows from $L \gg R $, and since the integration depends from now on only on $k_\perp$ we write the orthogonal component as $k$. The linear matter power spectrum of $P_{\mathrm{lin},0}$ is calculated using \citet{Eisenstein1998}, and $D_+$ is the growth factor which depends on the conformal time. We note that the factor $1/L$ cancels out when projecting the moments in Eq.~(\ref{eq:delta_moment_proj}-\ref{eq:w_delta_moment_proj}) using the Limber approximation \citep{Limber1953}. According to this derivation for a top-hat filter we get for a general filter $U$ that
\begin{align}
W_{U_\chi}(k) &= \int\limits_0^{2\pi} \int\limits_0^{\infty}\dd r\,\dd\vartheta \, U_\chi(r)\,\mathrm{e}^{-{\rm i}kr\cos\vartheta}= 2\pi\int\limits_0^{\infty}\dd r \, J_0(kr)\,r\,U_\chi(r)\,,
\label{eq:W_U}
\end{align}
where $U_\chi(r) = U(r/\chi) = U(\vartheta)/\chi^2$, with $U(\vartheta)$ being a filter measured in angular coordinates (see Fig.~\ref{fig:filter_U}). Correspondingly, the variance of the matter density contrast for a general filter $U$ in the flat-sky approximation is 
\begin{equation}
\langle \delta_{U,L}^2 \rangle(\chi)  = \frac{2\pi D_+^2}{L} \int \dd k ~k\ W_{U_\chi}^2(k)\,, P_{\mathrm{lin},0}(k) \, .
\label{eq:variance_tau}
\end{equation}
Following lines similar to those of Appendix B.4 of F18, the leading-order contribution to the skewness of matter density contrast for the general filter $U$ can be calculated as
\begin{align}
\langle \delta_{U,L}^3 \rangle(\chi)  & = 3\hat{c} \pi^{-1} \int \int \dd q_1 \,\dd q_2 \, q_1 \, q_2 \, W_{U_\chi}(q_1) \, W_{U_\chi}(q_2) \,P_{\mathrm{lin},0}(q_1)\,P_{\mathrm{lin},0}(q_2) \int \dd \phi~ W_{U_\chi}\left(\sqrt{q_1^2+q_2^2+2q_1q_2\cos\phi}\right) \, F_2(q_1,q_2,\phi) \nonumber \\
&\equiv 3\hat{c}\pi^{-1} \int \int \dd q_1\, \dd q_2 \, q_1 \,q_2 \, W_{U_\chi}(q_1) \,W_{U_\chi}(q_2) \, P_{\mathrm{lin},0}(q_1)\,P_{\mathrm{lin},0}(q_2) \, \Phi_{U_\chi}(q_1,q_2) \, ,
\label{eq:skewness2}
\end{align}
 where $\hat{c} = \frac{4\pi^2 D_+^4}{L^2}$. The function $F_2$ in a general $\Lambda$CDM universe is given by 
\begin{align}
F_2(q_1,q_2,\phi) &= \frac{1}{2} \left(2+\frac{q_1}{q_2}\cos\phi+\frac{q_2}{q_1}\cos\phi\right) +  (1+\mu)\,(\cos^2\phi-1) = 1+\frac{1}{2}\cos\phi\left( \frac{q_1}{q_2}+\frac{q_2}{q_1}\right) -   (1-\mu)\,\sin^2\phi \, ,
\end{align}
where $\mu$ results from perturbation theory and is a function of the growth factor $D_+$ (see Appendix B.1 in F18 for more details)\footnote{For an Einstein-de Sitter universe $\mu=5/7$.}, and $\phi$ is the angle between the vectors with absolute values $q_1$ and $q_2$.
Given the definition of $W_{U_\chi}$ in Eq.~\eqref{eq:W_U},  $\Phi_{U_\chi}$ can be written as
\begin{align}
    \Phi_{U_\chi}(q_1,q_2)=2\pi\int\limits_0^{\infty} \dd r \, r\, U_\chi(r) \int \dd \phi~ J_0\left(r\sqrt{q_1^2+q_2^2+2q_1q_2\cos\phi}\right) F_2(q_1,q_2,\phi)  \, .
\end{align}
Next, we use Graf's addition theorem \citep[see e.g.][]{Abramowitz1972}, which states that 
\begin{align}
    J_0\left(\sqrt{q_1^2+q_2^2+2q_1q_2\cos\phi}\right)= \sum_{m=-\infty}^\infty (-1)^m  J_m(q_1)\,J_m(q_2)\,\mathrm{e}^{\mathrm{i}m\phi}=J_0(q_1)\,J_0(q_2) + 2\sum_{m=1}^\infty (-1)^m  J_m(q_1)\,J_m(q_2)\,\cos(m\phi) \, ,
\end{align}
such that $\Phi_{U_\chi}(q_1,q_2)$ becomes
\begin{align}
 &2\pi\int\limits_0^{\infty} r\,U_\chi(r) \dd r \int_{0}^{2\pi} \dd \phi~ \left[J_0(rq_1)\,J_0(rq_2) + 2\sum_{m=1}^\infty (-1)^m   J_m(rq_1)\,J_m(rq_2)\,\cos(m\phi)\right] \left[1+\frac{1}{2}\cos(\phi)\left( \frac{q_1}{q_2}+\frac{q_2}{q_1}\right) -   (1-\mu)\,\sin^2\phi\right] \nonumber\\
&= \underbrace{ 2\pi^2  (1+\mu)\,\int\limits_0^{\infty} r\,U_\chi(r) \dd r \, J_0(rq_1)\,J_0(rq_2)}_{A} - \underbrace{ 2\pi^2\int\limits_0^{\infty} r\,U_\chi(r) \dd r \,  J_1(rq_1)\,J_1(rq_2)\left( \frac{q_1}{q_2}+\frac{q_2}{q_1}\right)}_{B} + \underbrace{2\pi^2   (1-\mu)\,\int\limits_0^{\infty} r\,U_\chi(r) \dd r \,  J_2(rq_1)\,J_2(rq_2)}_{C}\, ,
\end{align}
where we made use of the orthogonality of the trigonometric functions. Plugging $\Phi_{U_\chi}(q_1,q_2)$ back into Eq.~\eqref{eq:skewness2} and considering each term separately we get
\begin{align}
 A:\quad & 3\hat{c} \pi^{-1} \int \int \dd q_1 \,\dd q_2 \, q_1\, q_2 \, W_{U_\chi}(q_1) \,W_{U_\chi}(q_2) \,P_{\mathrm{lin},0}(q_1)\,P_{\mathrm{lin},0}(q_2) \, 2\pi\int\limits_0^{\infty} \dd r \, rU_\chi(r) \, \pi  (1+\mu)\, J_0(rq_1)\,J_0(rq_2) \nonumber \\
 &= 6 \pi \hat{c}  (1+\mu)\, \int\limits_0^{\infty} \dd r \, rU_\chi(r)  \left[\int \dd q \,q \,W_{U_\chi}(q)\, P_{\mathrm{lin},0}(q)\,J_0(rq)\right]^2 \, ,
\end{align}
and by analogy 
\begin{align}
 C:\quad &  3\hat{c} \pi^{-1} \int \int \dd q_1 \dd q_2 \, q_1 q_2 \, W_{U_\chi}(q_1) W_{U_\chi}(q_2) \,P_{\mathrm{lin},0}(q_1)P_{\mathrm{lin},0}(q_2) \, 2\pi\int\limits_0^{\infty}\dd r \, rU_\chi(r) \, \pi   (1-\mu)\, J_2(rq_1)J_2(rq_2) \nonumber \\
 &= 6 \pi \hat{c}   (1-\mu)\, \int\limits_0^{\infty} \dd r \,  rU_\chi(r)  \left[ \int \dd q \,q \,W_{U_\chi}(q)\, P_{\mathrm{lin},0}(q)\, J_2(rq)\right]^2 \, ,
\end{align}
and finally  
\begin{align}
 B:\quad &  -3\hat{c} \pi^{-1} \int \int \dd q_1 \, \dd q_2 \, q_1 \,q_2 \, W_{U_\chi}(q_1) \,W_{U_\chi}(q_2) \,P_{\mathrm{lin},0}(q_1)\,P_{\mathrm{lin},0}(q_2) \, 2\pi \int\limits_0^{\infty} \dd r \, rU_\chi(r) \, \pi J_1(rq_1)\,J_1(rq_2)\left[ \frac{q_1}{q_2}+\frac{q_2}{q_1}\right] \nonumber \\
 &=-12 \pi \hat{c} \int\limits_0^{\infty} \dd r \, rU_\chi(r)  \int \dd q_1 \, q_1^2 \,W_{U_\chi}(q_1)\, P_{\mathrm{lin},0}(q_1)\,J_1(rq_1)\, \int \dd q_2 \,W_{U_\chi}(q_2)\, P_{\mathrm{lin},0}(q_2)\,J_1(rq_2) \, .
\end{align}

The following transformations provide a more compressed expression for $\langle \delta_{U_\chi,L}^3 \rangle(\chi)$, which can then be used to verify our derivation by comparing it with the result from F18 for a top-hat filter. For this, we rewrite the expression of Bessel functions in terms of $W_r^\mathrm{th}(q)$ as
\begin{align}
    J_2(rq)&=\frac{1}{rq}J_1(rq)-\frac{1}{q}\frac{\dd }{\dd r}J_1(rq) = \frac{1}{rq}J_1(rq)-\left[ r\frac{\dd }{\dd r} \frac{J_1(rq)}{rq}+\frac{1}{rq}J_1(rq)\right] = -\frac{1}{2}\frac{\dd}{\dd \ln(r)} W_r^\mathrm{th}(q) \, ,\\
    J_0(rq)&=\frac{1}{rq}J_1(rq)+\frac{1}{q}\frac{\dd }{\dd r}J_1(rq)= \frac{1}{rq}J_1(rq)+\left[ r\frac{\dd }{\dd r} \frac{J_1(rq)}{rq}+\frac{1}{rq}J_1(rq)\right] = W_r^\mathrm{th}(q)+\frac{1}{2}\frac{\dd}{\dd \ln(r)} W_r^\mathrm{th}(q) \, ,
\end{align}
and with 
\begin{align}
    \frac{1}{rq} \frac{\dd^2}{\dd r^2} J_1(rq) = \frac{1}{2}\frac{\dd^2}{\dd r^2} W_r^\mathrm{th}(q) - \frac{2}{r^2}\left[ \frac{1}{rq}J_1(rq)- \frac{1}{q}\frac{\dd }{\dd r}J_1(rq)  \right] =\frac{1}{2}\frac{\dd^2}{\dd r^2} W_r^\mathrm{th}(q) + \frac{1}{r^2}  \frac{\dd}{\dd \ln(r)} W_r^\mathrm{th}(q) \, ,
\end{align}
we get
\begin{align}
    rqJ_1(rq) =  J_2(rq)-rq\frac{1}{q^2}\frac{\dd^2}{\dd r^2} J_1(rq) = J_2(rq)-r^2\frac{1}{rq}\frac{\dd^2}{\dd r^2} J_1(rq) =-\frac{3}{2}\frac{\dd}{\dd \ln(r)} W_r^\mathrm{th}(q) - \frac{r^2}{2}\frac{\dd^2}{\dd r^2} W_r^\mathrm{th}(q) \, .
\end{align}
Using these relations together with the following notation 
\begin{align}
     Q_1(r,\chi) &= \frac{2\pi D_+^2}{L} \int \dd k \, k \, W_{U_\chi}(k)\, W_r^\mathrm{th}(k) \, P_{\mathrm{lin},0}(k) \, ,\\
    Q_2(r,\chi) &= \frac{2\pi D_+^2}{L} \int \dd k \,k \, W_{U_\chi}(k)\, \frac{\dd}{\dd \ln(r)} W_r^\mathrm{th}(k) \, P_{\mathrm{lin},0}(k) \, , \\
     Q_3(r,\chi) &= \frac{2\pi D_+^2}{L} \int \dd k \, k \,W_{U_\chi}(k)\, \frac{\dd^2}{\dd r^2} W_r^\mathrm{th}(k)\,  P_{\mathrm{lin},0}(k).
\end{align}
we find that 
\begin{align}
    A:\quad 6 \pi \hat{c}  (1+\mu)\, \int\limits_0^{\infty} \dd r \, r\,U_\chi(r)  \left[\int \dd q \,q \,W_{U_\chi}(q)\, P_{\mathrm{lin},0}(q)\,J_0(rq)\right]^2 = 6 \pi  (1+\mu)\, \int\limits_0^{\infty} \dd r \, r\,U_\chi(r) \left[Q_1(r,\chi)+\frac{1}{2}Q_2(r,\chi) \right]^2 \;,
\end{align}
\begin{align}
   C:\quad  6 \pi \hat{c}   (1-\mu)\, \int\limits_0^{\infty} \dd r \,  r\,U_\chi(r)  \left[ \int \dd q \,q \,W_{U_\chi}(q)\, P_{\mathrm{lin},0}(q)\,J_2(rq)\right]^2 = 6 \pi   (1-\mu)\, \int\limits_0^{\infty} \dd r \,  r\,U_\chi(r)  \left[-\frac{1}{2}Q_2(r,\chi) \right]^2 \;,
\end{align}
and 
\begin{align}
 B:\quad & -12 \pi \hat{c} \int\limits_0^{\infty} \dd r \, r\,U_\chi(r)   \int \dd q_1 \, q_1 \, W_{U_\chi}(q_1)\, P_{\mathrm{lin},0}(q_1) \,rq_1 J_1(rq_1) \int \dd q_2 \, q_2  \,W_{U_\chi}(q_2) \, P_{\mathrm{lin},0}(q_2) \,\frac{1}{rq_2} J_1(rq_2) \nonumber\\
    &= -12 \pi \int\limits_0^{\infty} \dd r \, r\,U_\chi(r)  \left[ -\frac{3}{2}Q_2(r,\chi) - \frac{r^2}{2}Q_3(r,\chi)\right] \frac{1}{2}Q_1(r,\chi) \, .
\end{align}
Finally, combining $A$, $B$, and $C$, the skewness of $ \delta_{U_\chi,L}$ simplifies to
\begin{align}
    \langle \delta_{U,L}^3 \rangle(\chi)&=6 \pi \int\limits_0^{\infty} \dd r \,  r\,U_\chi(r) \bigg (  (1+\mu)\, \left[ Q_1(r,\chi)+ \frac{1}{2} Q_2(r,\chi) \right]^2 +  (1-\mu)\, \frac{1}{4} Q_2^2(r,\chi)  + \frac{3}{2}  Q_1(r,\chi)  Q_2(r,\chi) + \frac{r^2}{2} Q_1(r,\chi)\, Q_3(r,\chi) \bigg ) \nonumber \\
    &=  3 \pi \int\limits_0^{\infty} \dd r \,  r\,U_\chi(r) \bigg( 2 (1+\mu)\,\left[ Q^2_1(r,\chi)+ Q_1(r,\chi) \,Q_2(r,\chi)\right] + 3 Q_1(r,\chi)\, Q_2(r,\chi) +  Q_2^2(r,\chi) +  r^2Q_1(r,\chi)\, Q_3(r,\chi) \bigg) \nonumber\\
    &=3\pi  \int \dd r \,U_\chi(r) \,\frac{\dd}{\dd r} \left( r^2 \left[  (1+\mu)\,Q^2_1(r,\chi)+Q_1(r,\chi)\,Q_2(r,\chi)\right] \right)\, ,
\end{align}
where it is seen that for a top-hat of size $\vartheta$ with $U_\chi(r)=\mathcal{H}(\vartheta-\chi r)$ the result in Eq.~(B.35) immediately follows. 

Although all necessary ingredients for specifying the PDF of $\delta_{\mathrm{m},U}$ are derived already, we still need moments like $\langle \delta_{\chi\vartheta,L}\, \delta_{U,L}^2 \rangle(\chi)$ to compute quantities like $ \langle \kappa_{<\vartheta}|\delta_{\mathrm{m},U}\rangle$ or $ \langle w_{<\vartheta}\, \delta_{\mathrm{m},U}^k \rangle$. With the definitions of two further integrals,
\begin{align}
Q_4(r,\chi\vartheta) &= \frac{2\pi D_+^2}{L}\int \dd q \,q \, W_{\chi\vartheta}^\mathrm{th}(q) \,P_\mathrm{lin}(q) \, W_r^\mathrm{th}(q)\;, \label{eq:I_3}\\
Q_5(r,\chi\vartheta) &= \frac{2\pi D_+^2}{L}\int \dd q \,q\, W_{\chi\vartheta}^\mathrm{th}(q)\,P_\mathrm{lin}(q) \,\frac{\dd}{\dd \ln r} W_{r}^\nt{th}(q) \, ,
\end{align}
and using the result of F18 that for a top-hat filter of size $R$
\begin{align}
\Phi^\mathrm{th}_{R}(q_1,q_2) = \int \dd \phi \,  W_R^{\mathrm{th}}\left(\sqrt{q_1^2+q_2^2+2q_1q_2\cos\phi}\right)\, F_2(q_1,q_2,\phi) =\pi (1+\mu)\, W_R^{\mathrm{th}}(q_1)\, W_R^{\mathrm{th}}(q_2) + \frac{\pi}{2}\frac{\dd}{\dd \ln R} \left[W_R^{\mathrm{th}}(q_1)\, W_R^{\mathrm{th}}(q_2) \right] \,,
\end{align}
the joint filter moments between the matter density contrast smoothed with the general filter and the matter density contrast smoothed with a top-hat of size $\vartheta$ follows analogously to the skewness, and is given by 
\begin{align}
\langle \delta_{\vartheta,L}\, \delta_{U,L}^2 \rangle(\chi) &=
 \frac{\hat{c}}{\pi} \int \int \dd q_1 \,\dd q_2 \, q_1\, q_2 \, W_{U_\chi}(q_1) \,W_{U_\chi}(q_2) \,P_{\mathrm{lin},0}(q_1)\,P_{\mathrm{lin},0}(q_2)\, \Phi^\nt{th}_{\chi \vartheta}(q_1,q_2) \, + \nonumber \\ & \hspace{0.5cm} \frac{2\hat{c}}{\pi} \int \int \dd q_1 \, \dd q_2\,  q_1\, q_2 \, W_{U_\chi}(q_1)\, W_{\chi \vartheta}^{\mathrm{th}}(q_2) \,P_{\mathrm{lin},0}(q_1)\,P_{\mathrm{lin},0}(q_2)\, \Phi_{U_\chi}(q_1,q_2)   \nonumber\\
& =  (1+\mu)\, Q^2_1(\chi\vartheta,\chi) + Q_1(\chi\vartheta,\chi)\,Q_2(\chi\vartheta,\chi) \, +  \nonumber \\ & \hspace{0.5cm} 2 \pi \int \dd r\, U_\chi(r) \frac{\dd}{\dd r}  \left( r^2  (1+\mu)\, Q_1(r,\chi)\,Q_4(r,\chi\vartheta) + \frac{r^2}{2}\left[ Q_1(r,\chi)\,Q_5(r,\chi\vartheta)+Q_2(r,\chi)\,Q_4(r,\chi\vartheta) \right]  \right) \, . 
\end{align}

\twocolumn
\subsection{Limber projection}

Given the moments of the smoothed density contrasts at co-moving distance $\chi$ derived in the previous section, the moments in Eqs.~(\ref{eq:var_dmU}, \ref{eq:skew_dmU}) and Eqs.~(\ref{eq:V}--\ref{eq:kappa_0}) for $k=1,2, \mathrm{~or~}3$ follow \citep[see e.g.][]{Bernardeau2000},
\begin{align}
   \langle \delta_{\mathrm{m},U}^k \rangle &= \int \dd \chi \, q_\mathrm{f}^k(\chi) \, L^{k-1} \, \langle \delta_{U,L}^k \rangle(\chi)\;, \label{eq:delta_moment_proj}\\
   \langle \kappa_{<\vartheta} \delta_{\mathrm{m},U}^k \rangle &= \int \dd \chi \, W_\mathrm{s}(\chi) \, q_\mathrm{f}^k(\chi) \, L^{k-1} \, \langle \delta_{\vartheta,L}\, \delta_{U,L}^k \rangle(\chi)\;, \label{eq:kappa_delta_moment_proj} \\
   \langle w_{<\vartheta} \delta_{\mathrm{m},U}^k \rangle &= \int \dd \chi \, q_\mathrm{f}^{k+1}(\chi) \, L^{k-1} \, \langle \delta_{\vartheta,L}\, \delta_{U,L}^k \rangle(\chi) \label{eq:w_delta_moment_proj}\, ,
\end{align}
where $q_\mathrm{f}(\chi)$ is the projection kernel defined in Eq.~\eqref{eq:projection_kernel} and $W_\mathrm{s}(\chi)$ the lensing efficiency defined in Eq.~\eqref{eq:lensing_efficiency}. We note that these three equations employ a Limber approximation, which consists of $L\rightarrow \infty$ \citep{Limber1953}, and that the physical radius $r$ of filter $U$ scales with $\chi$ as described below Eq.~\eqref{eq:W_U}. We also note that these expectation values are independent of $L$.

\subsection{Non-linear regime}

In order to go to the non-linear regime for second-order moments, we replace the linear power spectrum in the above calculations with the non-linear power spectrum, which in turn is determined with the halofit model from \citet{Takahashi2012} using an analytic approximation for the transfer function \citep{Eisenstein1998}. 

For the third-order moments we use that for a top-hat filter of size $R$ the filter simplifies to $U_\chi(r)=\frac{1}{\pi R^2}\mathcal{H}(R-r)$, such that 
\begin{align}
     Q_1(R,\chi) &= \frac{2\pi D_+^2}{L} \int \dd k \, k \, W_{U_\chi}(k)\, W_R^\mathrm{th}(k) \, P_{\mathrm{lin},0}(k) \nonumber\\
     &= \frac{2\pi D_+^2}{L} \int \dd k \, k \,W_R^\mathrm{th}(k)\, W_R^\mathrm{th}(k) \, P_{\mathrm{lin},0}(k)  \nonumber\\
     &=\langle \delta_{R,L}^2 \rangle(\chi) \, ,
\end{align}
and
\begin{align}
Q_2(R,\chi) &= \frac{2\pi D_+^2}{L} \int \dd k \,k \, W_{U_\chi}(k)\, \frac{\dd}{\dd \ln(r)} W_R^\mathrm{th}(k) \, P_{\mathrm{lin},0}(k) \nonumber \\
      &=  \frac{2\pi D_+^2}{L} \int \dd k \, k \, W_R^\mathrm{th}(k) \frac{\dd}{\dd \ln(R)} W_R^\mathrm{th}(k)  P_{\mathrm{lin},0}(k) \nonumber \\ & = \frac{1}{2} \frac{\dd}{\dd \ln(R)} \langle \delta_{R,L}^2 \rangle(\chi) \, .
\end{align}
Furthermore, the skewness simplifies in the this case to:
\begin{align}
    \langle \delta_{R,L}^3 \rangle(\chi) = 3   (1+\mu)\,\langle \delta_{R,L}^2 \rangle^2 (\chi) + \frac{3}{2}  \langle \delta_{R,L}^2 \rangle (\chi) \frac{\dd\langle \delta_{R,L}^2 \rangle (\chi)}{\dd \ln(R)} \,. 
\end{align}
This then helps to define
\begin{equation}
      S_3 \equiv \frac{\langle \delta_{R,L}^3 \rangle(\chi)}{\langle \delta_{R,L}^2 \rangle^2 (\chi)}  = 3 (1+\mu)\, + \frac{3}{2}  \frac{\dd}{\dd \ln(R)} \ln(\langle \delta_{R,L}^2 \rangle)  \, ,
      \label{eq:S_3}
\end{equation}
which in the linear and the non-linear regime is approximately the same \citep{Bernardeau2002}, meaning that in order to get the skewness in the non-linear regime we approximate 
\begin{equation}
     \langle \delta_{R,L}^3 \rangle_\mathrm{non-linear}(\chi) \approx S_3 \, \langle \delta_{R,L}^2 \rangle^2_\mathrm{non-linear} (\chi)\, .
\end{equation}
For the general filter we use that the numerical integration of $r$ in $\langle \delta_{U_\chi,L}^3 \rangle(\chi)$ results basically in a sum of top-hat filters, such that we make use of $S_3$ to scale each term individual to the non-linear regime. For the joint filter moment $\langle \delta_{\chi \vartheta,L}\, \delta_{U_\chi,L}^2 \rangle(\chi)$ we use a generalised version of $S_3$, which states that for two different top-hat filters of size $R_1$ and $R_2$
\begin{equation}
\langle\delta_{R_1,L}^2 \delta_{R_2,L} \rangle(\chi) \propto  \langle\delta_{R_1,L} \delta_{R_2,L} \rangle(\chi)\, \langle\delta_{R_1,L}^2 \rangle(\chi) \, .
\end{equation}
Using again that the $r$-integration results in a sum of top-hat filters and factoring out the non-derivative terms similar to Eq.~\eqref{eq:S_3}, we scale individually all the non-derivative terms to the non-linear regime.

\subsection{Characteristic function}
\label{sec:CF}
\begin{figure}[!htbp]
\centering
\includegraphics[width=0.9\columnwidth]{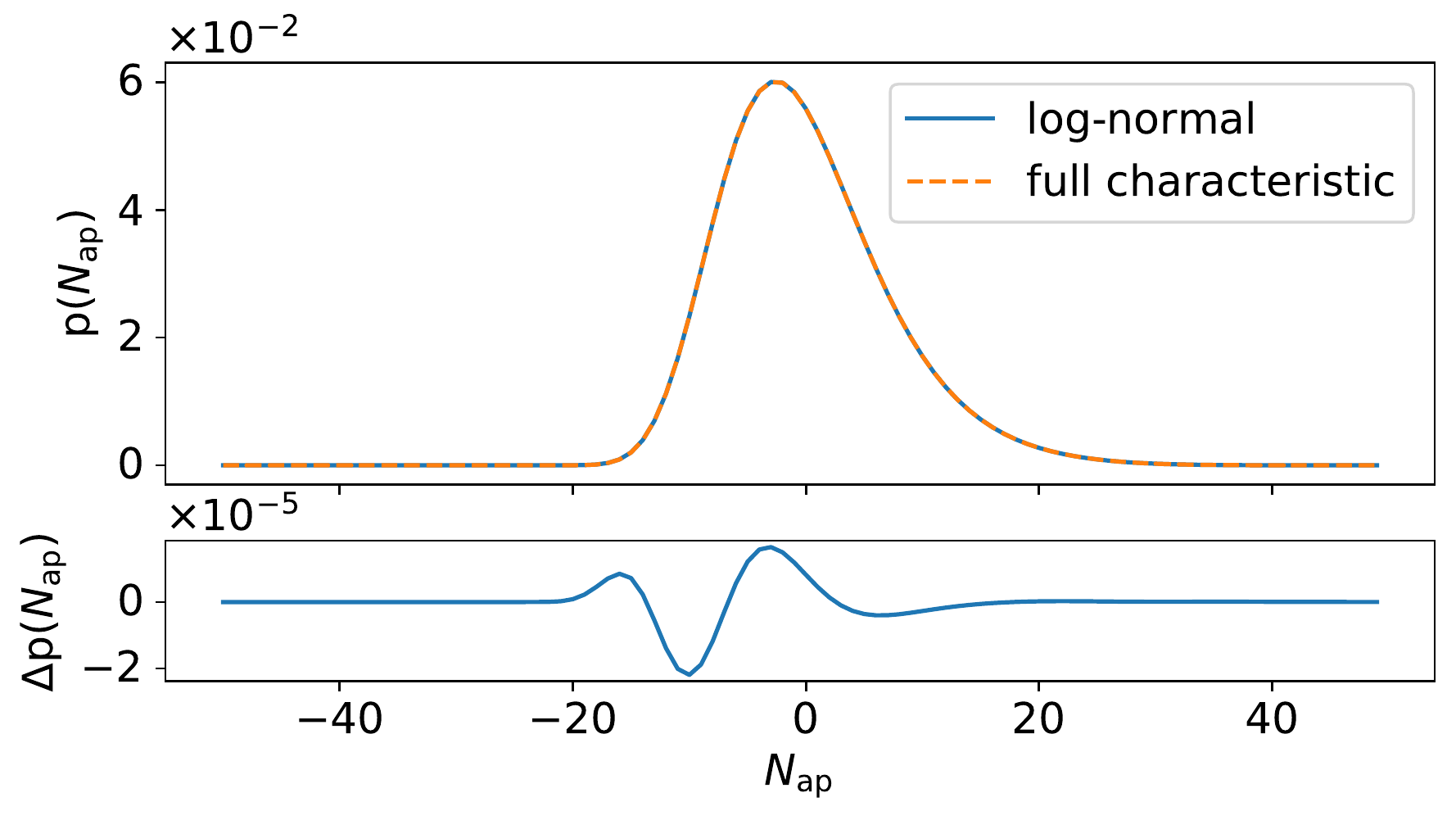}
    \caption{Comparison between the two approaches to calculate $p(N_{\textrm{ap}})$. It is clearly seen that both methods yield almost the same result.}
    \label{fig:lognormal_vs_characteristic}
\end{figure}

We consider a large circle of radius $R$, inside of which there are $N=n_0 \pi R^2$ galaxies, where $n_0$ is the galaxy number density. The probability of finding a galaxy at separation $\vartheta$ is 
\begin{equation}
     p(\vartheta;\delta_{\mathrm{m},U})=\frac{2\vartheta }{R^2\eta}(1+b\, \langle w_\vartheta|\delta_{\mathrm{m},U}\rangle) \, ,
\end{equation}
where $\langle w_\vartheta|\delta_{\mathrm{m},U}\rangle$ is the expectation of the mean 2D density contrast on a circle at $\vartheta$ (see Eq.~\ref{eq:w_theta}) given the smoothed density contrast defined in Eq.~\eqref{eq:w_delta}. The assumption of linear galaxy bias enters here by the term $b\, \langle w_\vartheta|\delta_{\mathrm{m},U}\rangle$. The normalisation is
\begin{equation}
    \eta = \int_0^R \frac{2\vartheta }{R^2}(1+b\, \langle w_\vartheta|\delta_{\mathrm{m},U}\rangle)\, \dd \vartheta  \, ,
\end{equation}
which goes to unity for $R\rightarrow\infty$. The characteristic function (CF) of the aperture number $N_\textrm{ap}$, given the smoothed 2D density contrast $\delta_{\mathrm{m},U}$, is given by
\begin{align*}
\Psi(t) &= \langle\mathrm{e}^{{\rm i} t N_\textrm{ap}} \rangle_{\delta_{\mathrm{m},U}} = \int_\mathbb{R} \dd N_\textrm{ap}\, p(N_{\textrm{ap}}|\delta_{\mathrm{m},U})\mathrm{e}^{\mathrm{i} tN_\textrm{ap}} \\
&= \left[ \prod_{i=1}^N \int_0^R \dd \vartheta_i \, p(\vartheta_i;\delta_{\mathrm{m},U}) \right]\mathrm{e}^{\mathrm{i} t \sum_j U(\vartheta_j)} \\
&= \left[ \int_0^R \dd \vartheta \,\frac{2\vartheta}{R^2\eta}(1+b\, \langle w_\vartheta|\delta_{\mathrm{m},U}\rangle) \mathrm{e}^{\mathrm{i} t U(\vartheta)} \right]^N \\
&= \left[ \int_0^R \dd \vartheta\,\frac{2\vartheta}{R^2 \eta}(1+b\, \langle w_\vartheta|\delta_{\mathrm{m},U}\rangle)  \left(\mathrm{e}^{\mathrm{i} t U(\vartheta)} -1+1\right) \right]^N \\
&= \left[ 1+ \frac{\pi n_0}{N\eta}\int_0^R \dd \vartheta \;2\vartheta\;(1+b\, \langle w_\vartheta|\delta_{\mathrm{m},U}\rangle)  \left(\mathrm{e}^{\mathrm{i} t U(\vartheta)} -1\right) \right]^N\\
&\underbrace{\longrightarrow}_{N,R\rightarrow \infty} \exp\left[ 2\pi n_0 \int_0^\infty \dd \vartheta\;\vartheta\;(1+b\, \langle w_\vartheta|\delta_{\mathrm{m},U}\rangle)  \left(\mathrm{e}^{\mathrm{i} t U(\vartheta)} -1\right)\right], 
\end{align*}
where we used in the second line that 
\begin{equation*}
N_\textrm{ap} = \int_0^\infty \dd^2 \vartheta \, U(|\boldsymbol{\vartheta}|) \, n(\boldsymbol{\vartheta}) = \sum_j U(\vartheta_j)\;,
\end{equation*}
with $n(\boldsymbol{\vartheta})=\sum_j\delta_\textrm{D}(\boldsymbol{\vartheta}-\boldsymbol{\vartheta_j})$, and that the galaxy positions $\vartheta_i$ independently trace the density profile $\langle w_\vartheta|\delta_{\mathrm{m},U}\rangle$. As discussed in Sect.\,\ref{Sec:Characteriscfunction}, the exact approach is to transform the CF to the probability density function $p(N_{\textrm{ap}}|\delta_{\mathrm{m},U})$ by use of the inverse Fourier transformation. Alternatively, we assume that the PDF is well approximated by a log-normal distribution as
\begin{equation}
p(N_{\textrm{ap}}|\delta_{\mathrm{m},U}) = \frac{1}{\sqrt{2\pi}S(N_{\textrm{ap}}+L)} \exp(\frac{-\left[\ln(N_{\textrm{ap}}+L)-M\right]^2}{2S^2}) \, ,
\label{eq:log_normal_PDF}
\end{equation}
where the parameters $S, M, L$ are fixed with the first raw moment
\begin{equation}
\mu'_1 = \langle N_{\textrm{ap}} | \delta_{\mathrm{m},U} \rangle = \langle N_{\textrm{ap}}  \rangle_{\delta_{\mathrm{m},U}}  = \exp(M+\frac{S^2}{2}) -L \, ,
\end{equation}
and the central moments 
\begin{align}
\mu_2 &= \langle (N_{\textrm{ap}}-\langle N_{\textrm{ap}} \rangle)^2 \rangle_{\delta_{\mathrm{m},U}} = \exp(2M+S^2)\left[{\rm e}^{S^2}-1\right] \, ,\\
\mu_3 &= \langle (N_{\textrm{ap}}-\langle N_{\textrm{ap}} \rangle)^3 \rangle_{\delta_{\mathrm{m},U}} =\exp(3M+\frac{3}{2}S^2)\left[{\rm e}^{S^2}-1\right]^2\left[{\rm e}^{S^2}+2\right]  \, .
\end{align}
The raw moments can be calculated from the derivatives of the CF,
\begin{equation}
\mu'_n = \left.\frac{\dd^n \Psi(t)}{\dd (\textrm{i}t)^n}\right|_{t=0} \, .
\end{equation}
With the definition 
\begin{equation}
E_n = 2\pi n_0\int_0^\infty \dd \vartheta\;\vartheta\;(1+b\, \langle w_\vartheta|\delta_{\mathrm{m},U}\rangle)  U^n(\vartheta ) \, ,
\end{equation}
it follows that
\begin{align}
\mu_0' &= 1 \;,\nonumber\\
\mu_1' &= E_1 \;,\label{eq:mu_1}\\
\mu_2' &= (E_1)^2 + E_2\;, \nonumber\\
\mu_3' &= (E_1)^3 + 3 E_1 E_2  + E_3\;, \nonumber
\end{align}
and so
\begin{align}
\mu_2 &= \mu'_2 - (\mu'_1)^2 = (E_1)^2 + E_2 - (E_1)^2 = E_2 \;,\label{eq:mu_2} \\
\mu_3 &= \mu'_3 - 3\mu'_1\mu'_2 + 2(\mu'_1)^2 = E_3 \, . \label{eq:mu_3}
\end{align}
To find the parameters of the log-normal distribution Eq.~\eqref{eq:log_normal_PDF} by use of the raw and central moments in Eqs.~(\ref{eq:mu_1}--\ref{eq:mu_3}) we define 
\begin{align}
\gamma &= \frac{\mu_3}{\mu_2^{3/2}} = \sqrt{\exp(S^2)-1}\left[2+\exp(S^2)\right] \nonumber\\ &= \sqrt{q-1}\left(2+q\right) \,,
\end{align}
where we defined in the last step $q=\exp(S^2)$. Modifying $\gamma$ we get
\begin{equation}
    0 = q^3 + 3q^2- 4 -\gamma^2 \,,
\end{equation}
which always has one real solution $q_0$, and so the parameters follow to
\begin{align}
S &= \sqrt{\ln(q_0)}\;,\label{eq:S}\\
M &= \frac{1}{2}\ln\left(\frac{\mu_2}{q_0^2-q_0}\right)\;,\label{eq:M}\\
L &= \sqrt{\frac{\mu_2}{q_0-1}}-\mu'_1 \label{eq:L}\, .
\end{align}
\begin{figure}[!htbp]
\begin{minipage}{\linewidth}
\includegraphics[width=1.0\columnwidth]{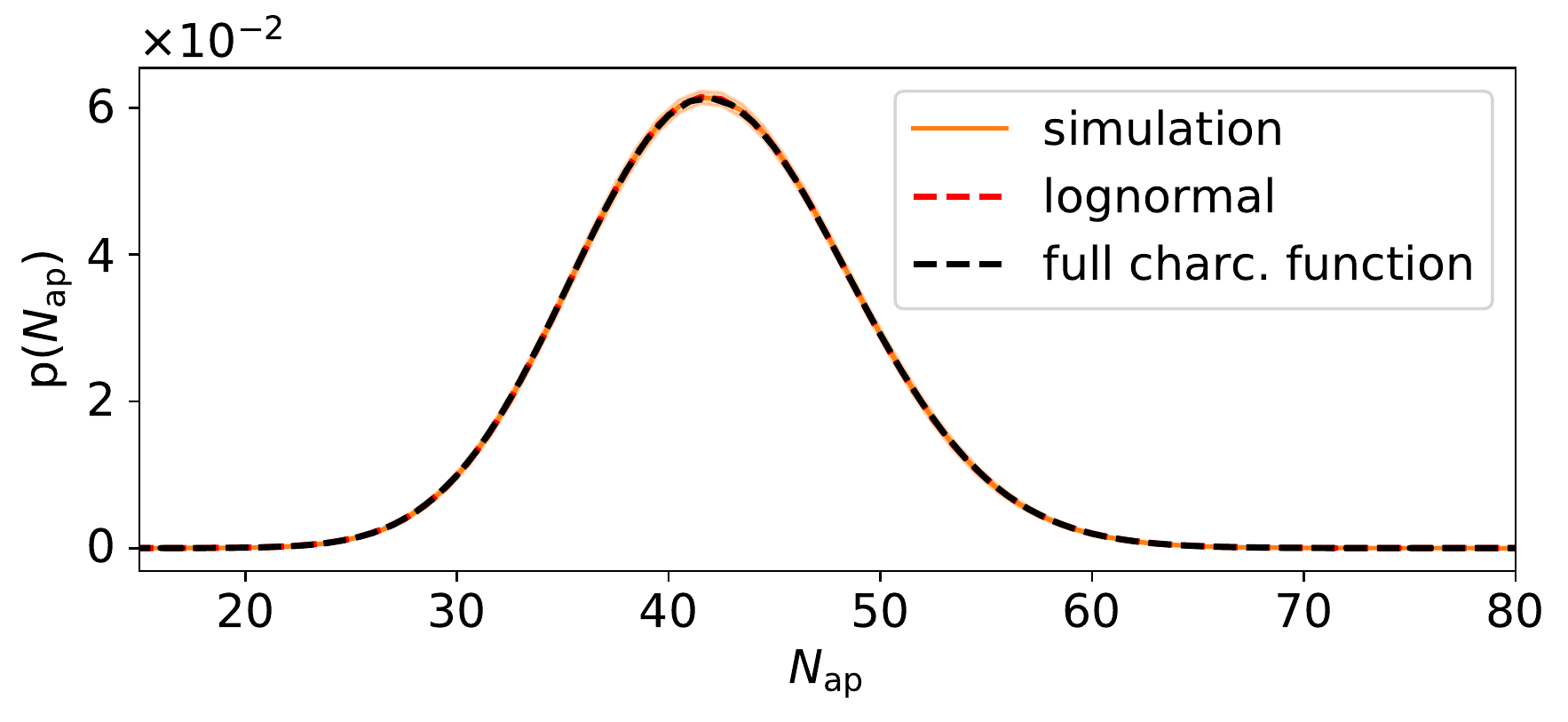}
\end{minipage}
\begin{minipage}{\linewidth}
\includegraphics[width=1.0\columnwidth]{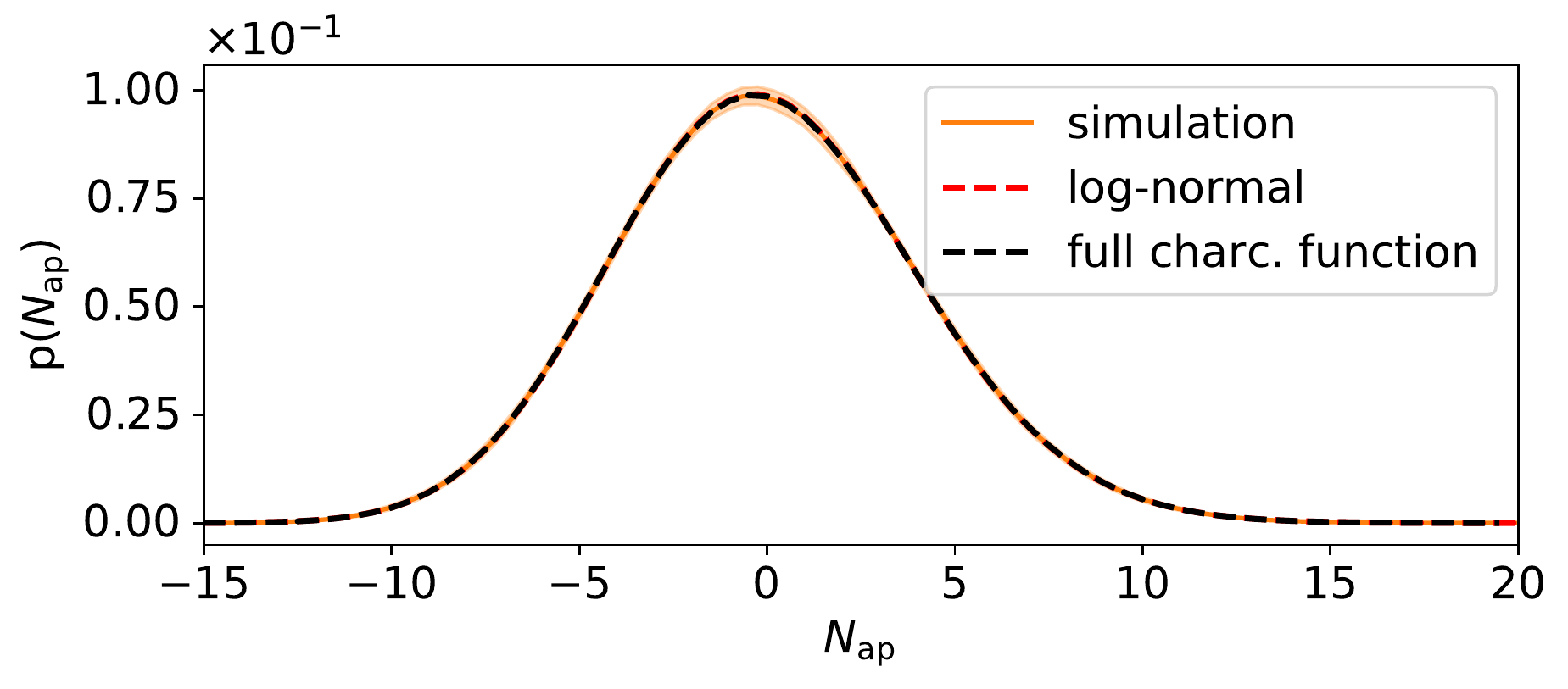}
\end{minipage}
\caption{Probability distribution of the aperture number resulting in a uniform random field smoothed with the top-hat filter of size $20'$ in the upper panel and for the adapted filter $U$ of size $120'$ in the lower panel. The orange shaded region is the standard deviation determined from 48 sub-patches.} 
\label{fig:pofNap_uniform}
\end{figure}
To check this derivation and compare it to the direct approach of using the inverse Fourier transform, we created an idealised case of a full-sky uniform random field $n_\textrm{side}=4096$ with a number density $n_0\approx0.034/\textrm{arcmin}^2$. Next we calculated by use of the \textsc{healpy} internal \textsc{smoothing} function $N_{\textrm{ap}}$ for the top-hat filter of size $20'$ and the for the adapted filter. In the determination of the predicted PDF we set, $\langle w_\vartheta|\delta_{\mathrm{m},U}\rangle=0$ so $p(N_{\textrm{ap}})$ follows immediately with Eq.~\eqref{eq:log_normal_PDF} or Eq.~\eqref{eq:True_PDF_charac}. It is clearly seen in Fig.~\ref{fig:pofNap_uniform} that the model for both filters has an excellent fit with the measured PDF of the aperture number. Additionally, we show in Fig.\,\ref{fig:lognormal_vs_characteristic} a comparison between the predicted $p(N_\mathrm{ap})$ using the full characteristic function Eq.~\eqref{eq:CF} versus the log-normal approach Eq.~\eqref{eq:log_normal_PDF} for the low-redshift bin $z_l^\mathrm{low}$ from the Takahashi set-up. In the lower panel the residual difference between the two methods is three orders of magnitude smaller than the signal itself, which shows that the two approaches are identical given the uncertainties we expect for Stage III surveys. Since the log-normal approach is faster to compute we can use this approach in future analyses where computational speed is essential. 
\section{Correction formulae for the power spectra of the T17 simulations }
\label{Sect:power_correction}
 To account for the finite angular resolution T17 suggested a simple damping factor at small scales as
\begin{equation}
    C_\ell^\kappa \rightarrow \frac{C_\ell^\kappa}{1+(\ell/\ell_\textrm{res})^2}\, ,
\end{equation}
where $\ell_\textrm{res}=1.6\times N_\textrm{side}$. Additionally, to take the shell thickness into account they conducted a simple fitting formula by which the matter power spectrum should be modified to
\begin{equation}
    P_\delta(k) \rightarrow P_\delta^W(k) = \frac{(1+c_1 k^{-\alpha_1})^{\alpha_1}}{(1+c_2 k^{-\alpha_2})^{\alpha_3}}P_\delta(k)\, ,
\end{equation}
where the parameters are simulation specific and are $c_1= 9.5171 \times 10^{-4}$, $c_2= 5.1543 \times 10^{-3}$, $\alpha_1= 1.3063$, $\alpha_2= 1.1475$, $\alpha_3= 0.62793$, and the wavenumber $k$ is in units of $h/\textrm{Mpc}$. We note that although we incorporated these corrections in the following, they have very little effect on the scales we are considering.


\begin{table}[h!]
\centering
\caption{Overview of all the different cosmological parameters for the 26 cosmo-SLICS models, which are used in Sect.~\ref{sec_model_test} for the cosmological analysis.}
\begin{tabular}{c|ccccc}
 & $\Omega_\nt{m}$ & $h$ & $w_0$ & $\sigma_8$ & S$_8$ \\
\hline
fid & 0.2905 & 0.6898 & $-1.0000$ & 0.8364 & 0.8231 \\
1 & 0.3282 & 0.6766 & $-1.2376$ & 0.6677 & 0.6984 \\
2 & 0.1019 & 0.7104 & $-1.6154$ & 1.3428 & 0.7826 \\
3 & 0.2536 & 0.6238 & $-1.7698$ & 0.6670 & 0.6133 \\
4 & 0.1734 & 0.6584 & $-0.5223$ & 0.9581 & 0.7284 \\
5 & 0.3759 & 0.6034 & $-0.9741$ & 0.8028 & 0.8986 \\
6 & 0.4758 & 0.7459 & $-1.3046$ & 0.6049 & 0.7618 \\
7 & 0.1458 & 0.8031 & $-1.4498$ & 1.1017 & 0.7680 \\
8 & 0.3099 & 0.6940 & $-1.8784$ & 0.7734 & 0.7861 \\
9 & 0.4815 & 0.6374 & $-0.7737$ & 0.5371 & 0.6804 \\
10 & 0.3425 & 0.8006 & $-1.5010$ & 0.6602 & 0.7054 \\
11 & 0.5482 & 0.7645 & $-1.9127$ & 0.4716 & 0.6375 \\
12 & 0.2898 & 0.6505 & $-0.6649$ & 0.7344 & 0.7218 \\
13 & 0.4247 & 0.6819 & $-1.1986$ & 0.6313 & 0.7511 \\
14 & 0.3979 & 0.7833 & $-1.1088$ & 0.7360 & 0.8476 \\
15 & 0.1691 & 0.7890 & $-1.6903$ & 1.1479 & 0.8618 \\
16 & 0.1255 & 0.7567 & $-0.9878$ & 0.9479 & 0.6131 \\
17 & 0.5148 & 0.6691 & $-1.3812$ & 0.6243 & 0.8178 \\
18 & 0.1928 & 0.6285 & $-0.8564$ & 1.1055 & 0.8862 \\
19 & 0.2784 & 0.7151 & $-1.0673$ & 0.6747 & 0.6500 \\
20 & 0.2106 & 0.7388 & $-0.5667$ & 1.0454 & 0.8759 \\
21 & 0.4430 & 0.6161 & $-1.7037$ & 0.6876 & 0.8356 \\
22 & 0.4062 & 0.8129 & $-1.9866$ & 0.5689 & 0.6620 \\
23 & 0.2294 & 0.7706 & $-0.8602$ & 0.9407 & 0.8226 \\
24 & 0.5095 & 0.6988 & $-0.7164$ & 0.5652 & 0.7366 \\
25 & 0.3652 & 0.7271 & $-1.5414$ & 0.5958 & 0.6574 \\
\end{tabular}
\label{cos_overview}
\end{table}

\begin{figure}[!htbp]
\centering
\includegraphics[width=0.9\columnwidth]{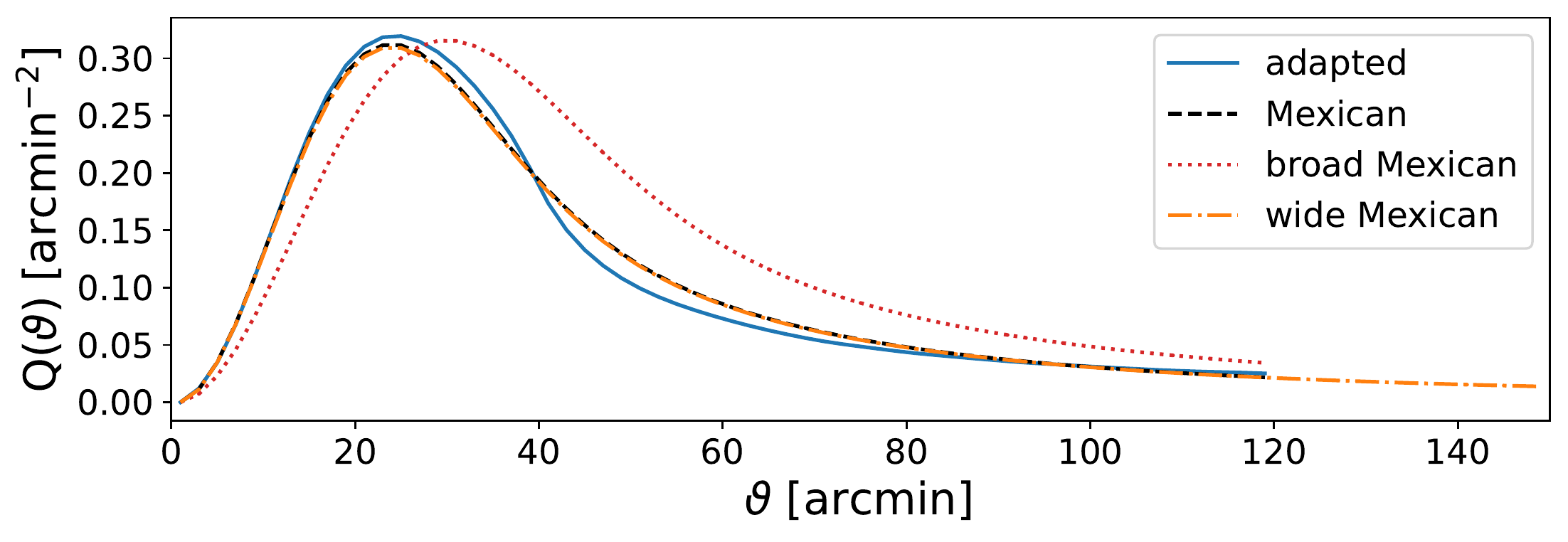}
    \caption{Different filters $Q$ resulting from the corresponding $U$ filters shown in Fig.\,\ref{fig:filter_U} used in this work to verify the new model. }
    \label{fig:filter_Q}
\end{figure}

\begin{figure}
\includegraphics[width=\columnwidth]{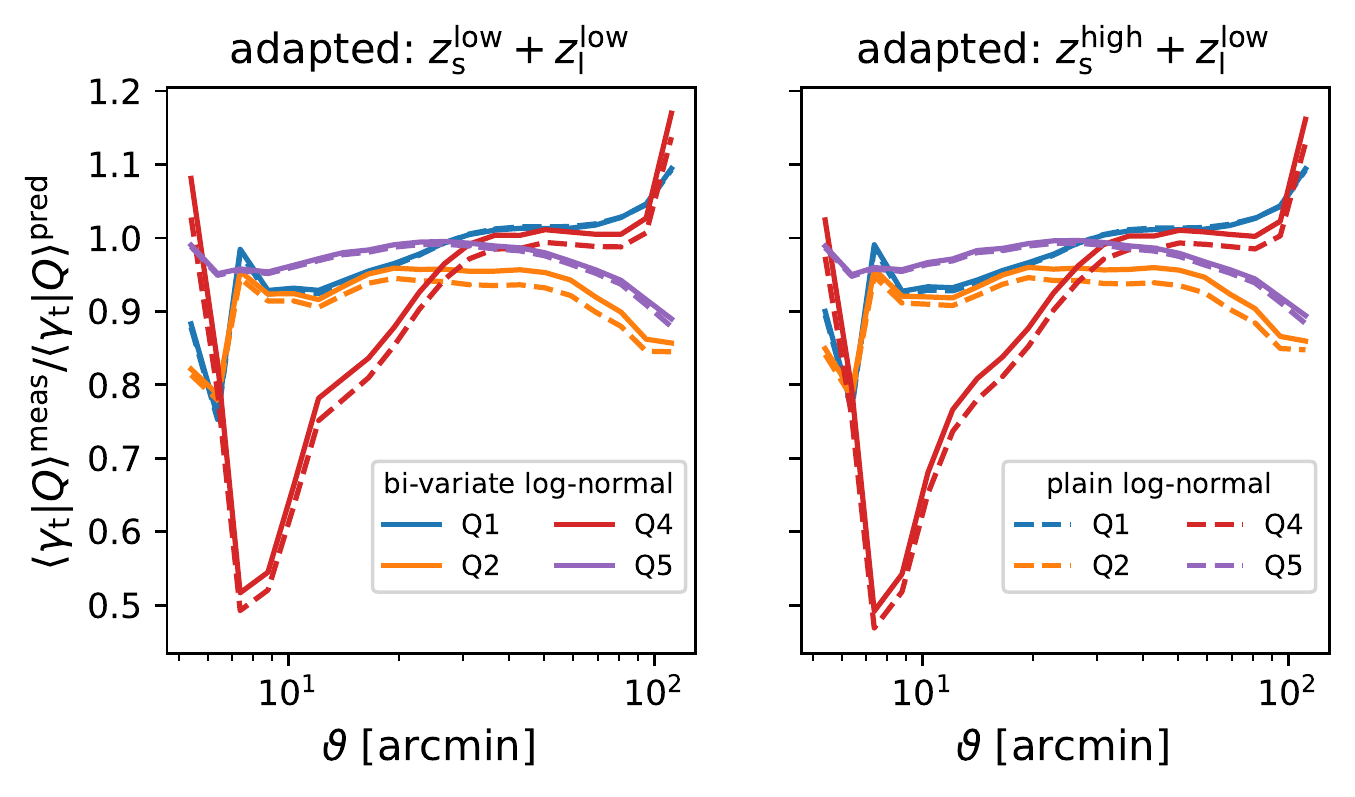}
\caption{Comparison between the uncalibrated shear profiles for the adapted filter with and without using the bi-variate log-normal approach discussed in Sect.\,\ref{subsec:I} The ratio is calculated between the measured shear profiles from T17 for the lower LRG source bin and for sources where several T17 shear grids were averaged, weighted by the $n(z)$ given in Fig.~\ref{fig:n_of_z}. The bi-variate log-normal shear profiles are more consistent with the measured shear profiles and thus (although the shear signals were calibrated) the more accurate model was chosen. Here only the highest and lowest two quantiles are shown because the middle one is to close to zero. }
\label{fig:lognomal_shear_comp}
\end{figure}

\begin{figure}
\includegraphics[width=\columnwidth]{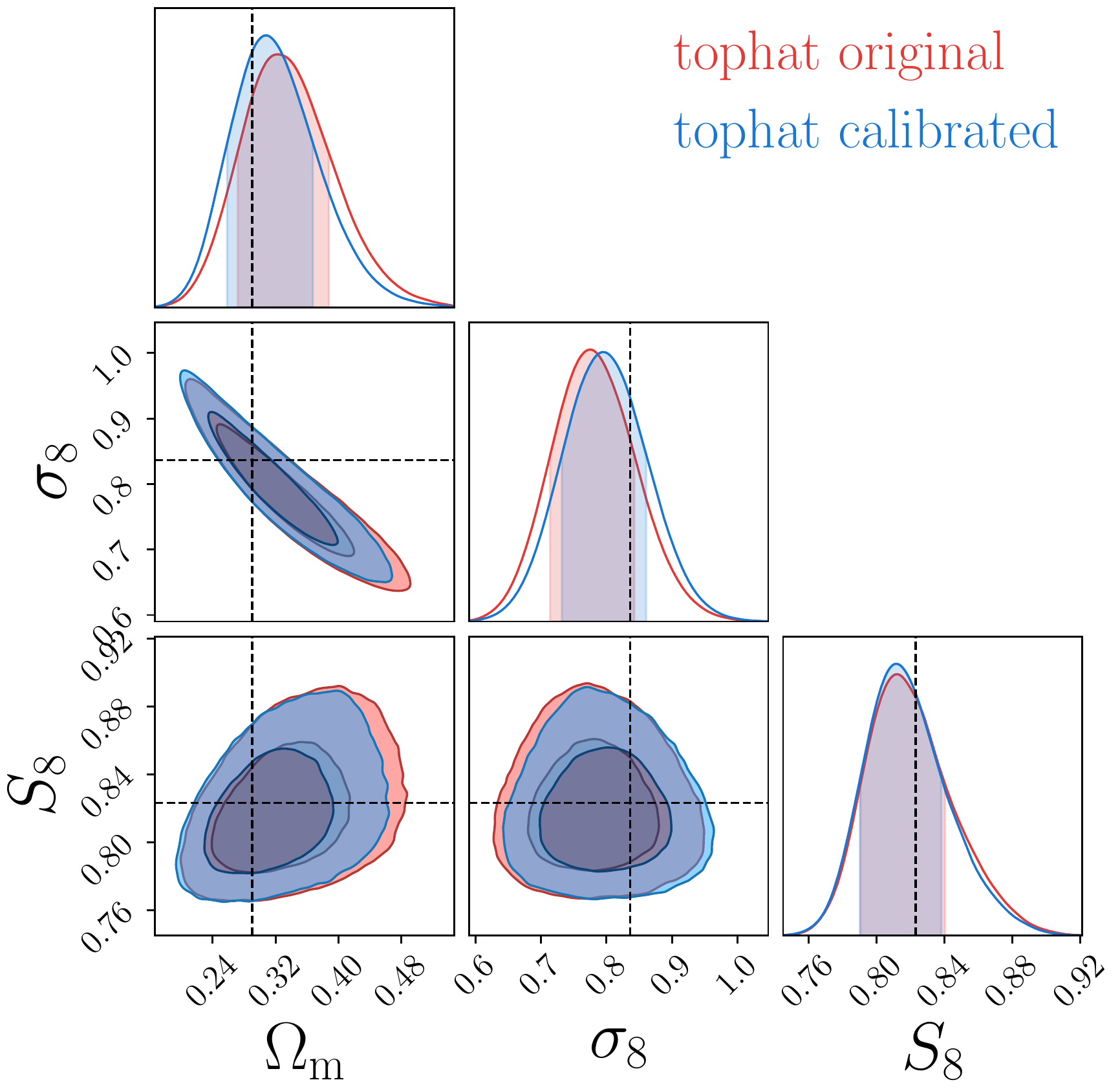}
\caption{MCMC results for the top-hat filter using the original and calibrated model. The data vector is calculated from the fiducial cosmology of cosmo-SLICS and a covariance matrix from 614 SLICS realisations. The systematic biases are likely to be statistical flukes due to the noise in the data vector. The contours are marginalised over the lens galaxy bias parameters.}
    \label{fig:MCMC_calibrated_tophat}
\end{figure}

\begin{figure}
\includegraphics[width=\columnwidth]{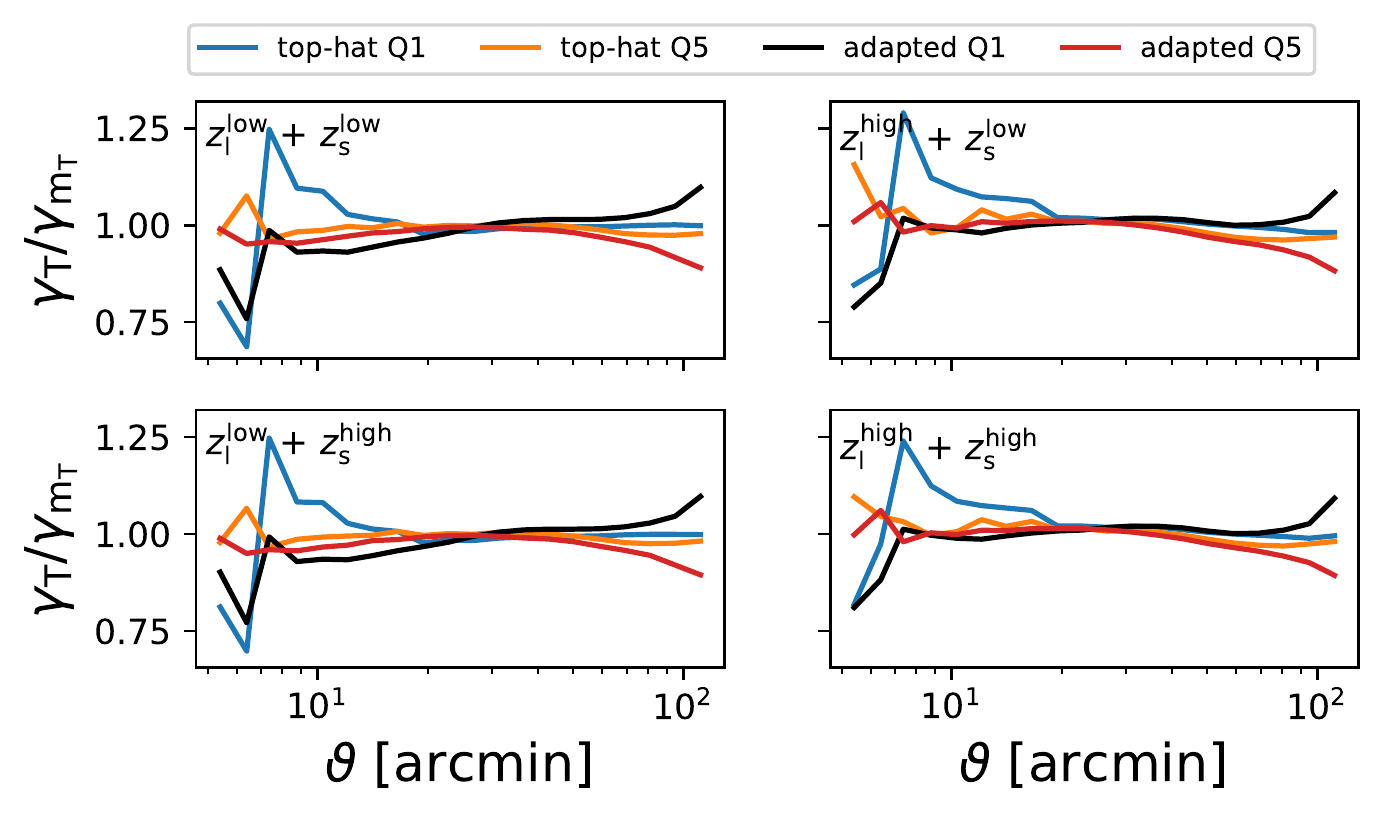}
\caption{Calibration of the model $\gamma_{\textrm{M}_\textrm{T}}(p_\textrm{T})$ by the T17 simulations $\gamma_\textrm{T}(p_\textrm{T})$ explained in Eq.~\eqref{eq:new_model}, shown for the highest and lowest quantile for the adapted and top-hat filter. The corresponding redshift distributions of the lenses are given in Fig.~\ref{fig:nofz} and for the sources several T17 shear grids are averaged, weighted by the $n(z)$ given in Fig.~\ref{fig:n_of_z}.}
\label{fig:calibrated_vector}
\end{figure}

\end{appendix}

\end{document}